\documentclass{sthesis}
\graphicspath{ {images/}{../images/}}

\usepackage[style=alphabetic,sorting=none,sortcites=true]{biblatex}
\usepackage{tocloft}
\usepackage{parskip}
\usepackage{wrapfig}
\usepackage{caption}
\usepackage{multicol}
\indexsetup{level=\chapter,toclevel=chapter}
\makeindex[title={List of definitions}]

\begin{document}

\begin{titlepage}
  \begin{center}
    \textsc{\LARGE Master thesis\\\vspace{1em}Mathematical Foundations of Computer Science}\\[1.5cm]
    \includegraphics[height=100pt]{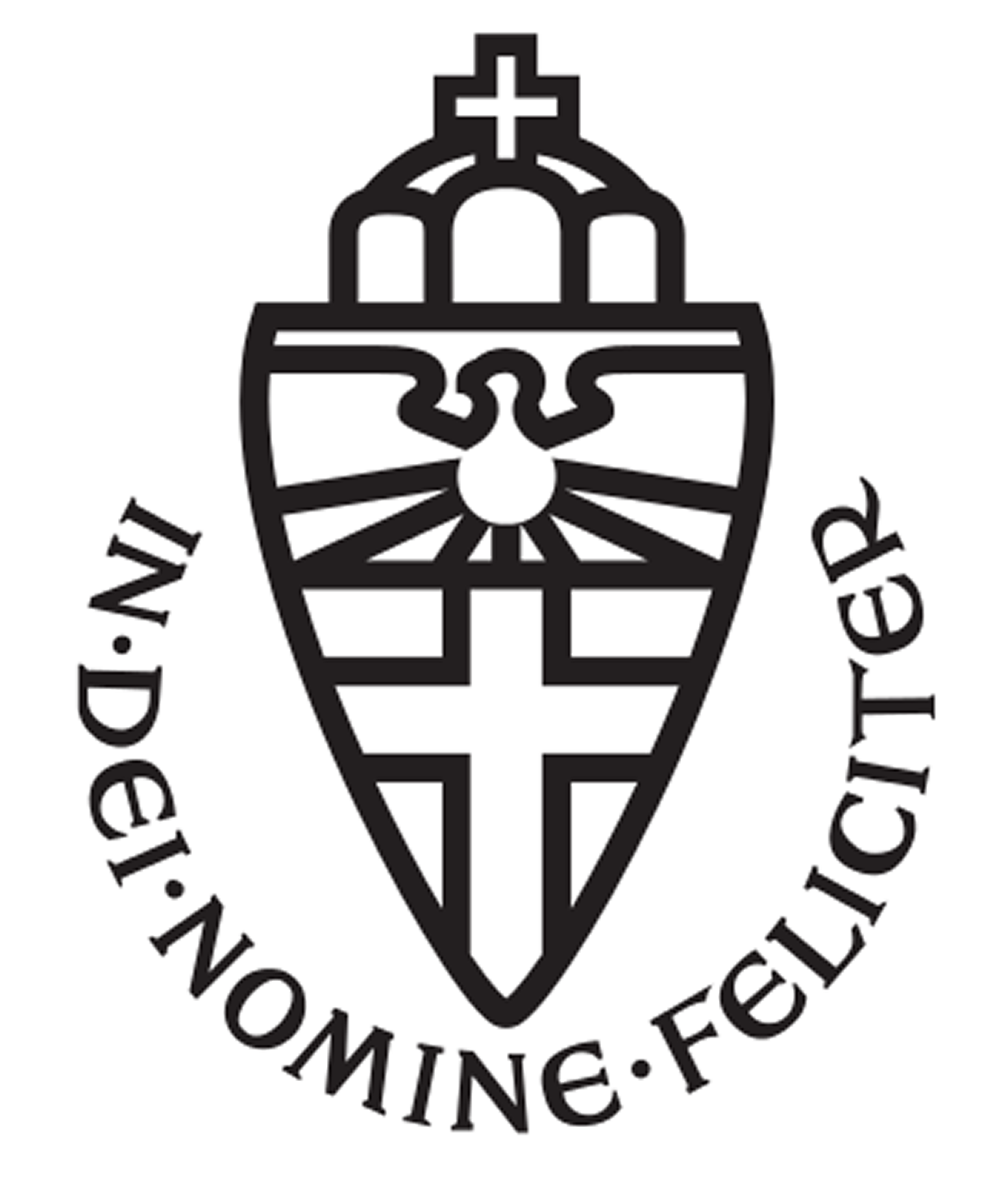}

    \vspace{0.4cm}
    \textsc{\Large Radboud University}\\[1cm]
    \hrule
    \vspace{0.4cm}
    \textbf{\huge Impredicative Encodings of Inductive and Coinductive Types}\\[0.4cm]
    \hrule
    \vspace{2cm}
    \begin{minipage}[t]{0.45\textwidth}
      \begin{flushleft} \large
        \textit{Author:}\\
        Steven Bronsveld\\
        \texttt{steven.bronsveld@ru.nl}\\
        s1020191
      \end{flushleft}
    \end{minipage}
    \begin{minipage}[t]{0.45\textwidth}
      \begin{flushright} \large
        \textit{Supervisor/assessor:}\\
        Prof. Dr. J.H. Geuvers\\
        \texttt{h.geuvers@cs.ru.nl}\\[1.3cm]
        \textit{Supervisor}\\
        Dr. N.M. van der Weide\\
        \texttt{nweide@cs.ru.nl}\\[1.3cm]
        \textit{Second assessor:}\\
        Dr. B. van den Berg\\
        \texttt{b.vandenberg3@uva.nl}
      \end{flushright}
    \end{minipage}
    \vfill
    {\large \today}
  \end{center}
\end{titlepage}

\begin{abstract}
  In the impredicative type theory of System F ($\lambda$2), it is possible to create inductive data types, such as natural numbers and lists. It is also possible to create coinductive data types such as streams. They work well in the sense that their (co)recursion principles obey the expected computation rules (the $\beta$-rules). Unfortunately, they do not yield a (co)induction principle, because the necessary uniqueness principles are missing (the $\eta$-rules). Awodey, Frey, and Speight (2018) used an extension of $\lambda C$ with sigma-types, equality-types, and functional extensionality to provide System F style inductive types with an induction principle by encoding them as a well-chosen subtype, making them initial algebras.
  In this thesis, we extend their results. We create a list and quotient type that have the desired induction principles. We show that we can use the technique for general inductive types by defining W-types with an induction principle. We also take the dual notion of their technique and create a coinductive stream type with the desired coinduction principle (also called bisimulation). We finish by showing that this dual approach can be extended to M-types, the generic notion of coinductive types, and the dual of W-types.
\end{abstract}

\tableofcontents

\chapter{Introduction}

Polymorphic lambda calculus ($\lambda 2$), also known as System F, was first introduced by Jean-Yves Girard in 1971 \cite{GIRARD197163} and John C. Reynolds in 1974 \cite{Reynolds74}. It is a very expressive system in the sense that it is possible to create both inductive data types, such as natural numbers and lists as well as coinductive data types, such as streams and infinite trees \cite{Girard_Taylor_Lafont_1993,Geuvers2014TheCR}. Under the Curry-Howard correspondence, it is isomorphic to second-order propositional logic.

One of the defining features of System F is that it is an impredicative system. This means that it allows quantification over all types in the universe. This impredicativity enables recursive definitions that are not based on well-foundedness. This allows for the creation of self-referential inductive types and infinite coinductive types. In mathematics, impredicativity is bountiful, for example, the closure of a set $A$ can be defined as $\overline{A} := \bigcap \{ C \mid A \subseteq C \land C \text{ is closed} \}$. Here, we quantify over all closed sets $C$, including $\overline{A}$ itself.

The (co)inductive data types definable in System F behave well in the sense that they adhere to the expected formation, introduction, elimination, and computation rules ($\beta$-rules). Unfortunately, they do not satisfy the expected uniqueness rules (the $\eta$-rules). These uniqueness rules enable us to prove the appropriate (co)induction principles. A concrete example is that it is not possible to derive the induction principle for the natural numbers \cite{NoInduction}.

We zoom out to place this thesis and System F in a broader context. The field of type theory finds its origins in mathematical logic and the theory of computation. A pioneer in the field is Alonzo Church, who introduced lambda-calculus as a model of computation \cite{DBLP:journals/jsyml/Church40}. He found a way to encode natural numbers by using repeated application of a function. For example, the number zero is defined as $\lambda f. \lambda x. x$ and the number three is defined as $\lambda f. \lambda x. f\ f\ x$. These numbers are currently known as the ``Church Numerals''. This encoding can be generalized to many other inductive data types with the important characteristic that they can be typed using System F. He also created a system of types to solve some paradoxes in his (untyped) lambda calculus. This system became known as simply typed lambda calculus.

Around the same time as Girard worked on System F, Martin-Löf worked on different and more powerful systems involving dependent types. Contrary to System F, he used a predicative flavor of type theory. He introduced many different systems and many new concepts such as identity types and W-types \cite{MartinLof}. His work greatly influenced the later field of Homotopy Type Theory (HoTT). This project combines the field of homotopy theory with the field of type theory to create a new foundational mathematical system built on the univalence axiom \cite{Hott}.

In 1984, Thierry Coquand and Gérard Huet introduced The Calculus of Constructions \cite{DBLP:journals/iandc/CoquandH88}. This is a very powerful system that allows for the creation of both data structures and mathematical proofs. It is an extension of System F and uses the same impredicative data type definitions. This system was used as the basis of the proof assistant Coq, named after Coquand. In the ordering of different type systems known as the lambda-cube (or Barendregt cube, after its creator Henk Barendregt), the Calculus of Constructions ($\lambda C$) is put at the top and simply typed lambda calculus ($\lambda\! \shortrightarrow$) at put at the bottom \cite{Barendregt_1991}.

The Coq system evolved away from the use of impredicative data structures in favor of adding inductive types directly \cite{DBLP:conf/mfps/PfenningP89, DBLP:conf/tlca/Paulin-Mohring93}. The main motive was the lack of induction principles in the System F style inductive types.

In this thesis, we extend the 2018 paper ``\textit{Impredicative Encodings of (Higher) Inductive Types}'' by Steve Awodey, Jonas Frey, and Sam Speight \cite{encodings}. In this paper, the authors use a modification of the HoTT system with one impredicative bottom universe to encode System F-style data types in such a way that the uniqueness rules (or $\eta$-rules) are satisfied. They take inspiration from the way inductive structures are defined in category theory: as the initial algebra of some functor. They show that these improved types satisfy their respective induction principles.

We affirmatively answer two open questions conjectured in the preceding master's thesis of Sam Speight \cite{speight}. The first question is whether the technique can be extended to general inductive types by encoding W-types. In addition, we encode a quotient type and show that it satisfies the induction principle. The second question is whether the dual notion of the technique can be extended to general coinductive types by encoding M-types. We affirmatively answer this question too. We first encode a coinductive stream type and show the bisimulation principle. We then finish by encoding an M-type that satisfies the $\eta$-rule and showing the general coinduction principle for M-types.

We give a brief overview of this thesis.

In \cref{chap:prelim}, we discuss the preliminaries of this thesis, such as a basic understanding of category theory, algebras, coalgebras, and some type theory.

In \cref{chap:system}, we introduce the type system that we shall use throughout this thesis. We elaborate on System F and the link with the system of Awodey, Frey, and Speight. We introduce $\Pi$-types, $\Sigma$-types, intensional identity types, and the axioms of function extensionality and uniqueness of identity proofs (UIP).

In \cref{chap:list}, we introduce an impredicative list type. We then follow the technique of impredicativity encoding the type to obtain a list type that satisfies the $\eta$-rule. We proceed to prove the induction principle for this encoded list type.

In \cref{chap:quotients}, we introduce an impredicative quotient type. We again encode this type such that it satisfies the required $\eta$-rule in order to prove the induction principle for quotients.

In \cref{chap:stream}, we define an coinductive stream type. We use our newly defined quotient types to create an impredicative encoding. We show that this encoding satisfies the $\eta$-rule and prove the coinduction principle (also known as bisimulation) for this stream type.

In \cref{chap:generalization}, we take a step back and look at the general pattern of the encodings of the previous chapters in order to apply the technique to general inductive and coinductive types.

In \cref{chap:wtypes}, we define impredicative W-types. We make an impredicative encoding of these types and show that this satisfies the $\eta$-rule. We prove the induction principle for W-types, showing that we can create general inductive types.

In \cref{chap:mtypes}, we define impredicative M-types. We create an impredicative encoding of these types using a quotient. We show that this encoding satisfies the $\eta$-rule. We finish by proving that the coinduction principle holds for these new M-types.

\chapter{Preliminaries}
\label{chap:prelim}

\section{Category theory}

In this thesis, we assume some basic knowledge of category theory. We make use of initial and final objects and of products and coproducts. We also assume some basic diagram-reading abilities. We denote a morphism $f$ between objects $a$ and $b$ by $a \overset{f}{\to} b$.

For products, we denote the projection functions as $(\pi_1 : A \times B \to A)$ and $(\pi_2 : A \times B \to B)$. We denote the pairing function as $\tup{\_\ , \_}$.
For coproducts, we denote the injection functions as $(\inlc : A \to A + B)$ and $(\inrc: B \to A + B)$. We denote the case function as $([\_\ , \_] : A + B \to C)$.

We assume some very basic knowledge about equalizers and coequalizers and how they are related to limits and colimits.

\section{Algebra}
\label{sec:algebras}
A categorical way to define algebraic or inductive data structures is using $F$-algebra \cite{Jacobs_Rutten_2011}.

\begin{definition}
  An $\textbf{F}$\textbf{-algebra} of an endofunctor $(F : \C \to \C)$ is a tuple $\tup{X, \a}$ where $X$, also called the \textbf{carrier}, is an object of $\C$ and $(\a : F(X) \to X)$ is a $\C$-morphism.
\end{definition}

\begin{minipage}[t]{.66\textwidth}
  \begin{definition}
    \label{def:alg-morphism}
    Given $F$-algebras $\tup{X,\a}$ and $\tup{Y,\b}$, we say that a $\C$-morphism $(f: X\to Y)$ is an $\textbf{F}$\textbf{-algebra morphism}, or $F$-morphism if it makes the diagram on the right commute. In formulas, we require that $f\,\circ\,\a = \b\,\circ\,F(f)$. If $f$ is an $F$-algebra morphism, we denote this as $(f : \tup{X,\a} \to \tup{Y,\b})$, implicitly stating that this commutativity requirement is satisfied.
  \end{definition}
\end{minipage}
\begin{minipage}[t]{.33\textwidth}
  \[\begin{tikzcd}
      {F(X)} & X \\
      {F(Y)} & Y
      \arrow["\alpha", from=1-1, to=1-2]
      \arrow["\beta"', from=2-1, to=2-2]
      \arrow["{F(f)}"', from=1-1, to=2-1]
      \arrow["f", from=1-2, to=2-2]
      \arrow["{=}"{description}, draw=none, from=1-1, to=2-2]
    \end{tikzcd}\]
\end{minipage}

\begin{lemma}
  \label{id_morph}
  For any $F$-algebra $\tup{X,\a}$, the identity function $\idc{X}$ is an $F$-morphism.
  \begin{proof}
    We have to show that $\idc{X} \circ \a = \a \circ F(\idc{X})$. This follows directly from the unfolding the definitions: $\idc{X} \circ \a = \a = \a \circ \idc{F(X)} = \a \circ F(\idc{X})$
  \end{proof}
\end{lemma}

\begin{lemma}
  \label{comp_morph}
  The composition of two $F$-morphisms is again an $F$-morphism.
  \begin{proof}
    Let $\tup{X,\a}$, $\tup{Y,\b}$ and $\tup{Z,\g}$ be $F$-algebra. Assume we have two morphisms $\bigl(f: \tup{X,\a} \to \tup{Y, \b}\bigr)$ and $\bigl(g: \tup{Y,\b} \to \tup{Z, \g}\bigr)$. We thus have:

    \begin{align}
      f \circ \a = \b \circ F(f) \label{comp1} \\
      g \circ \b = \g \circ F(g) \label{comp2}
    \end{align}
    We thus need to show that $\bigl(g \circ f: \tup{X, \a} \to \tup{Z, \g}\bigr)$. This means we need to show that $(g\circ f) \circ \a = \g \circ F(g \circ f)$.

    \begin{align*}
      (g\circ f) \circ \a   & = g\circ (f \circ \a)          \\
      \overset{\ref{comp1}} & {=} g\circ (\b \circ F(f))     \\
                            & = (g\circ \b) \circ F(f)       \\
      \overset{\ref{comp2}} & {=} (\g\circ F(g)) \circ F(f)  \\
                            & = \g\circ (F(g) \circ F(f))    \\
                            & = \g\circ F(g \circ f)\qedhere
    \end{align*}
  \end{proof}
\end{lemma}

Using these two lemmas, we can now show that the $F$-algebras together with the morphisms form a category.

\begin{theorem}
  Given a category $\C$ and an endofunctor $F : \C \to \C$, the $F$-algebras together with the $F$-algebra morphisms form a category $F\text{-}\textbf{Alg}$.
  \begin{proof}It suffices to check the following. \hfill
    \begin{itemize}
      \item
            Each object $\tup{X, \a}$ has an identity morphism given by $\idc{X}$. By \cref{id_morph} this is indeed an $F$-morphism.
      \item
            For any two morphisms $\bigl(f: \tup{X,\a} \to \tup{Y, \b}\bigr)$ and $\bigl(g: \tup{Y,\b} \to \tup{Z, \g}\bigr)$, the composition is again a morphism by \cref{comp_morph}. \qedhere
    \end{itemize}
  \end{proof}
\end{theorem}

As we now have the category $F\text{-}\textbf{Alg}$, we can speak of the \textit{initial} object of this category.

\begin{definition}
  \label{initial_algebra_definition}
  An $F$-algebra $\tup{I,\g}$ is called an \textbf{initial} $F$-algebra if for any $F$-algebra $\tup{X,\a}$ there exists an unique $F$-morphism $\bigl(u_X: \tup{I,\g} \to \tup{X,\a}\bigr)$. We speak of \textbf{weakly initial} algebra if the uniqueness requirement is dropped.
\end{definition}

If it exists, this initial object is unique up to isomorphism. Therefore, we can speak of \textit{the} initial object \cite{Jacobs_Rutten_2011}. The uniqueness of the morphisms $u_X$ of \cref{initial_algebra_definition} can be stated in two equivalent ways as we show in \cref{uniqueness_rules}.

\begin{minipage}{.66\textwidth}
  \begin{proposition}
    \label{uniqueness_rules}
    Let $\tup{I, \gamma}$ be an $F$-algebra. The following are equivalent.
    \begin{enumerate}
      \item For all $F$-algebras $\tup{X,\alpha}$, if $\bigl(u_X, u_X': \tup{I, \gamma} \to \tup{X, \alpha}\bigr)$ then $u_X=u_X'$.
      \item For all $F$-algebras $\tup{X,\alpha}$ and $\tup{Y,\beta}$, if $\bigl(f: \tup{X, \alpha} \to \tup{Y, \beta}\bigr)$, $\bigl(u_X: \tup{I, \gamma} \to \tup{X, \alpha}\bigr)$ and $\bigl(u_Y: \tup{I, \gamma} \to \tup{Y, \alpha}\bigr)$, then $f \circ u_X = u_Y$.
    \end{enumerate}
    \begin{proof}{$(1 \implies 2)$}
      Assume we have a morphism $\bigl(f: \tup{X, \alpha} \to \tup{Y, \beta}\bigr)$. Then we also have that $\bigl(f \circ u_X : \tup{I, \gamma} \to \tup{Y, \beta}\bigr)$ is an $F$-morphism by \cref{comp_morph}. We also have that $u_Y$ is such a morphism. Thus by $1)$ we have $f \circ u_X = u_Y$.\\
      $(2 \implies 1)$ Take for $Y := X$. Let $\bigl(u_X, u_X': \tup{I, \gamma} \to \tup{X, \alpha}\bigr)$. Notice that $\bigl(\idc{X}: \tup{X, \alpha} \to \tup{X, \alpha}\bigr)$ is an $F$-morphism by \cref{id_morph}. Thus by $2)$ we have $\idc{X} \circ u_X = u_X'$. We conclude that $u_X = u_X'$.
    \end{proof}
  \end{proposition}
\end{minipage}
\begin{minipage}{.33\textwidth}
  \captionsetup{labelformat=empty}
  \centering
  \begin{tikzcd}
    I \arrow[d, "u_X"', bend right=49] \arrow[d, "u_X'", bend left=49] \arrow[d, "=", phantom] \\
    X
  \end{tikzcd}
  \\
  \begin{tikzcd}
    I \arrow[d, "u_X"'] \arrow[dd, "u_Y", bend left=49] \arrow[dd, "=", phantom, bend left] \\
    X \arrow[d, "f"']                                                                       \\
    Y
  \end{tikzcd}
\end{minipage}

\section{Coalgebra}
\label{coalgebra}
The dual notion of an algebra is a coalgebra. They can be used to categorically define coinductive data types such as streams and non-wellfounded trees \cite{Jacobs_Rutten_2011}.

\begin{definition}
  An $\textbf{F}$\textbf{-coalgebra} of an endofunctor $(F: \C \to \C)$ is a tuple $\tup{X, \alpha}$ where $X$, also called the \textbf{carrier}, is an object of $\C$ and $(\alpha: X \to F(X))$ is a $\C$-morphism.
\end{definition}

\begin{minipage}[t]{.66\textwidth}
  \begin{definition}
    Given $F$-coalgebras $\tup{X, \alpha}$ and $\tup{Y, \beta}$, we say that a $\C$-morphism $(f: X \to Y)$ is an $\textbf{F}$\textbf{-coalgebra morphism}, or $F$-morphism, if it makes the diagram on the right commute. In formulas, we require that $\b \circ f = F(f) \circ \a$. If $f$ is an $F$-coalgebra morphism, we denote this as $\bigl(f: \tup{X, \a} \to \tup{Y, \b}\bigr)$, implicitly stating that this commutativity requirement is satisfied.
  \end{definition}
\end{minipage}
\begin{minipage}[t]{.33\textwidth}
  \begin{center}
    \[\begin{tikzcd}
        X & {F(X)} \\
        Y & {F(Y)}
        \arrow["\alpha", from=1-1, to=1-2]
        \arrow["\beta"', from=2-1, to=2-2]
        \arrow["f"', from=1-1, to=2-1]
        \arrow["{F(f)}", from=1-2, to=2-2]
        \arrow["{=}"{description}, draw=none, from=1-1, to=2-2]
      \end{tikzcd}\]
  \end{center}
\end{minipage}

We have some similar results to the ones we saw for algebras. The proofs have been omitted as they are nearly identical to the ones for algebras.

\begin{lemma}
  \label{id_morph_co}
  For any $F$-coalgebra $\tup{X, \a}$, the identity morphism $(\idc{X}: X \to X)$ is an $F$-coalgebra morphism.
\end{lemma}

\begin{lemma}
  \label{comp_morph_co}
  The composition of two $F$-coalgebra morphisms is an $F$-coalgebra morphism.
\end{lemma}

These two lemmas allow us to define the category of $F$-coalgebras.

\begin{theorem}
  \label{coalg_category}
  Given a category $\C$ and an endofunctor $(F: \C \to \C)$, the $F$-coalgebras together with $F$-coalgebra morphisms form a category, denoted as $F$-\textbf{CoAlg}.
\end{theorem}

\begin{definition}
  \label{final_coalgebra_definition}
  An $F$-coalgebra $\tup{J, \g}$ is said to be \textbf{final} if for any $F$-coalgebra $\tup{X, \a}$, there exists a unique $F$-coalgebra morphism $\bigl(u_X: \tup{X, \a} \to \tup{J, \g}\bigr)$. We speak of \textbf{weakly final} coalgebra if the uniqueness requirement is dropped.
\end{definition}

If it exists, this final object is unique up to isomorphism. Therefore, we can speak of \textit{the} final object \cite{Jacobs_Rutten_2011} The uniqueness of the morphisms $u_X$ of \cref{final_coalgebra_definition} can be stated in two equivalent ways as we show in \cref{uniqueness_rules_co}.

\begin{minipage}{.66\textwidth}
  \begin{proposition}
    \label{uniqueness_rules_co}
    Let $\tup{J, \gamma}$ be an $F$-coalgebra. The following are equivalent.
    \begin{enumerate}
      \item For all $F$-algebras $\tup{X,\alpha}$, if $\bigl(u_X, u_X': \tup{X, \alpha} \to \tup{J, \gamma}\bigr)$ then $u_X=u_X'$.
      \item For all $F$-algebras $\tup{X,\alpha}$ and $\tup{Y,\beta}$, if $\bigl(f: \tup{X, \alpha} \to \tup{Y, \beta}\bigr)$, $\bigl(u_X: \tup{X, \alpha} \to \tup{I, \gamma}\bigr)$ and $\bigl(u_Y: \tup{Y, \alpha} \to \tup{J, \gamma}\bigr)$,\\
            then $u_X = u_Y \circ f$.
    \end{enumerate}
    \begin{proof}{$(1 \implies 2)$}
      Assume we have a morphism $\bigl(f: \tup{X, \alpha} \to \tup{Y, \beta}\bigr)$. Then we also have that $\bigl(u_Y \circ f : \tup{X, \gamma} \to \tup{J, \beta}\bigr)$ is an $F$-morphism by \cref{comp_morph_co}. We also have that $u_X$ is such a morphism. Thus by $1)$ we have $u_X = u_Y \circ f$.\\
      $(2 \implies 1)$ Take for $Y := X$. Let $\bigl(u_X, u_X': \tup{X, \alpha} \to \tup{J, \gamma}\bigr)$. Notice that $\bigl(\idc{X}: \tup{X, \alpha} \to \tup{X, \alpha}\bigr)$ is an $F$-morphism by \cref{id_morph_co}. Thus by $2)$ we have $u_X = u_X'\circ \idc{X}$, thus $u_X = u_X'$.
    \end{proof}
  \end{proposition}
\end{minipage}
\begin{minipage}{.33\textwidth}
  \captionsetup{labelformat=empty}
  \centering
  \begin{tikzcd}
    X \arrow[d, "u_X"', bend right=49] \arrow[d, "u_X'", bend left=49] \arrow[d, "=", phantom] \\
    J
  \end{tikzcd}
  \\
  \begin{tikzcd}
    X \arrow[d, "f"'] \arrow[dd, "u_X", bend left=49] \arrow[dd, "=", phantom, bend left] \\
    Y \arrow[d, "u_Y"']                                                                       \\
    J
  \end{tikzcd}
\end{minipage}

\newpage
\section{Type theory}

We assume some knowledge of type theory. We assume the reader is familiar with simply typed lambda calculus, the Curry-Howard correspondence, and the $\lambda$-cube (mainly $\lambda2$ and $\lambda$C) \cite{Girard_Taylor_Lafont_1993, Barendregt_1991, Nederpelt_Geuvers_2014}. We assume familiarity with typing judgments such as $(\Gamma \vdash a: A)$ and $\beta$-reduction. Whenever we have that $(\Gamma \vdash a: A)$, we say that $a$ is an inhabitant, element, or member of the type $A$. We often drop the context $\Gamma$ and write $(a: A)$. We furthermore expect some familiarity with $\Pi$-types, $\Sigma$-types, and equality types.

In this thesis, we make use of ideas from category theory to create inductive and coinductive data types. We keep these categorical statements separate from statements within the type system. We denote statements within the type system using \texttt{monospace font}. We believe that an explicit distinction is more accessible to the reader who is not deeply familiar with category theory.

Under the Curry-Howard correspondence, we can interpret types as propositions and terms as proofs. This means that we can use the system to reason about mathematical statements, and even use these proofs within our definitions. If a proposition holds, we can construct an inhabitant of the proposition, and vice versa. Whenever we prove a lemma and need the proof term, we name the proof term (the witness) next to the number of the lemma (see for example \cref{lem:LimList0,lem:LimListC,reclist_id}). We furthermore have that the arrow $\to$ and the implication symbol $\implies$ are actually one and the same. We still make use of both for emphasis.

We make use of three notions of equality: syntactical equality, $\beta$-equality, and propositional equality (sometimes called definitional equality). Within the type system, we can define terms and types. We say that two terms/types are syntactically equal if they are the same by definition, where we of course allow bounded variable renaming. We write $\texttt{a} :=\texttt{b}$ both when defining a term/type and to express that two terms/types are syntactically equal. We say that two terms are $\beta$-equal if they are equal up to $\beta$-reduction. We write $\texttt{a} \betar \texttt{b}$. We say that two terms are propositionally equal if there is a proof that the two terms are equal. We write $\texttt{a} = \texttt{b}$. This notation is also used to denote the corresponding \textit{equality type}. By the Curry-Howard correspondence, we thus have that two terms/types are propositionally equal if we have a proof term $\bigl(\Gamma \vdash \texttt{p} : (\texttt{a} = \texttt{b})\bigr)$. Often, we explicitly give the proof term, when showing that two terms are propositionally equal by writing $\texttt{a} \overset{\texttt{p}}{=} \texttt{b}$.

\begin{notation}
  We denote \textbf{function composition} as $f \circ g := \lambda x. f\ (g\ x)$.
\end{notation}

\chapter{The system}
\label{chap:system}

In this chapter, we outline the system in which we shall work. It was used by Awodey, Frey, and Speight in \cite{encodings} and detailed in the master thesis of Sam Speight \cite{speight}. This system shall be used throughout this thesis. The synopsis is that it is (intensional) Martin-Löf type theory, with one impredicative bottom universe $\U$, sigma-types ($\Sigma$-types), intensional identity types (equality types, or $=$-types), impredicative product types ($\Pi$-types) and function extensionality. Its inference rules are laid out in \cref{chap:complete_system}.

There are two ways to look at the system: as an extension of Girard's System F (also known as $\lambda$2) \cite{Girard_Taylor_Lafont_1993}, or as a subset of homotopy type theory (HoTT) with one impredicative bottom universe \cite{Hott}. The relation with System F is often emphasized by various authors \cite{encodings, speight,Xavier}. This relation boils down to the impredicativity that is often associated with System F and the style of creating data types. System F is very expressive in that it is possible to create both inductive and coinductive data types. Even though our system has humble beginnings in $\lambda$2, the various extensions make that we actually deal with a system similar in strength to that of the calculus of constructions $\lambda C$.

We explain our system using a constructive approach, starting with System F and working our way up. This means that we will not delve into the details of the syntax and the inference rules of the system here. Instead, we focus on the most important concepts. For a more formal definition we refer to \cref{chap:complete_system}, the HoTT book \cite{Hott} or the master thesis of Sam Speight \cite{speight}.

We want to shed some light on a foundational principle that underpins this thesis. That is the desire of \textit{defining} instead of \textit{postulating}. In the tradition of foundational mathematics and type theory specifically, there are many different systems with variable degrees of expressivity. These systems vary in the degree to which you can define certain structures or prove certain properties. We strive to create simple, yet powerful systems. There is thus a big difference between a system in which many concepts are postulated and a system in which many concepts can be defined. We shall make this distinction as clear as possible since it is the guiding principle that makes our current system so interesting.

In the next section, we start by showing the basic concepts of System F. After that, we extend these basic principles with the concepts of product-types, sigma-types, equality-types, and the axioms of function extensionality, uniqueness of identity proofs and existential identity.

\newpage
\section{System F}

System F, or polymorphic $\lambda$-calculus ($\lambda2$), was orignally introduced by Jean-Yves Girard in 1972 \cite{GIRARD197163} and John Reynolds in 1974 \cite{Reynolds74}. Its typing rules are simple, yet powerful and given in \cref{systemf}. We illustrate the power of System F by defining the natural numbers also known as the Church Numerals.

\define{\RecNatr}{RecNatr}{\texttt{rec}_{\N}^*}
\begin{example} We define the \textbf{natural numbers} in System F.
  \label{expl:nat}

  \begin{align*}
    \N^*         & := \prod (X: \U). X \to (X \to X) \to X                    \tag{formation}                                      \\
    \texttt{0}^* & := \lambda (X: \U). \lambda (z: X). \lambda (s: X \to X).\ z              \tag{introduction}                    \\
    \texttt{S}^* & := \lambda (n: \N^*) \lambda (X: \U). \lambda (z: X). \lambda (s: X \to X).\ s\ (n\ X\ z\ s) \tag{introduction}
  \end{align*}

  Naturally, we want to check that these definitions are well-typed. We check that the type of $\texttt{S}^*$ is indeed $\N^* \to \N^*$, so that the judgement $(\vdash \texttt{S}^* : \N^* \to \N^*)$ holds.

  \begin{gather*}
    \infer{\vdots}{
      \infer[\to\text{-elim}]{n:\N^*, X:\U, z: X, s: X \to X \vdash n\ X\ z : (X \to X) \to X}{
        \infer[\Pi\text{-elim}]{n:\N^*, X:\U \vdash n\ X : X \to (X \to X) \to X}{
          \infer[\text{id}]{n:\N^* \vdash n : \N^*}{}
          &
          \infer[\text{id}]{X : \U \vdash X : \U}{}
        }
        &
        \infer[\text{id}]{z : X \vdash z : X}{}
      }
      &
      \infer[\text{id}]{s : X \to X \vdash s : X \to X}{}
    }
    \\
    \infer[\text{def}]{\vdash \texttt{S}^* : \N^* \to \N^*}{%
      \infer[\to\text{-intro}]{\vdash \lambda n. \lambda X. \lambda z. \lambda s.\ s\ (n\ X\ z\ s) : \N^* \to \N^*}{
        \infer[\text{def}]{n:\N^* \vdash \lambda X. \lambda z. \lambda s.\ s\ (n\ X\ z\ s) : \N^*}{
          \infer[\Pi\text{-intro}]{n:\N^* \vdash \lambda X. \lambda z. \lambda s.\ s\ (n\ X\ z\ s) : \Pi (X: \U). X \to (X \to X) \to X}{
            \infer[\to\text{-intro}]{n:\N^*, X:\U \vdash \lambda z. \lambda s.\ s\ (n\ X\ z\ s) : X \to (X \to X) \to X}{
              \infer[\to\text{-intro}]{n:\N^*, X:\U, z: X \vdash \lambda s.\ s\ (n\ X\ z\ s) : (X \to X) \to X}{
                \infer[\to\text{-elim}]{n:\N^*, X:\U, z: X, s: X \to X \vdash s\ (n\ X\ z\ s) : X}{
                  \infer[\to\text{-elim}]{n:\N^*, X:\U, z: X, s: X \to X \vdash n\ X\ z\ s : X}{
                    \vdots
                  }
                  &
                  \infer[\text{id}]{s: X \to X \vdash s : X \to X}{}
                }
              }
            }
          }
        }
      }
    }
  \end{gather*}

  This example serves as an illustration of how typing judgments are made. This work is best left to a computer, but it is important to understand the principles behind it.

  We still need a way to define functions on the natural numbers. For this purpose, we need a recursor:

  \begin{align*}
    \RecNatr\ (X: \U)\ (z: X)\ (s: X \to X \to X)\ (n : \N^*) & := n\ X\ z\ s \tag{elimination}
  \end{align*}

\end{example}

Let's take a moment to consider one of the defining characteristics of System F: it is impredicative. Impredicativity refers to a style of definitions in which one quantifies over the whole universe. This allows for certain self-referential features that are not based on well-foundedness. In the case of the natural numbers, we see that we use $\prod (X : \U)$ to define them. This means that $X$ can be any type in the universe. Specifically, we can have $X := \N^*$. We thus define the natural numbers by quantifying over the whole universe, including the natural numbers that we just defined. A philosophical argument in favor of impredicativity could be to say that our definition is actually a description or schema to describe certain features. In this light, terms of the type $\N^*$ are those that, regardless of the choice of $X$, satisfy the description $X \to (X \to X) \to X$.

The fact that we can apply $\N^*$ to itself is not just a weird consequence, but a necessary feature. For example, if we want to define a function on the natural numbers, that doubles the result, we use the recursor that takes $\N^*$ as an argument:

\begin{align*}
  \texttt{doubleN}   & : \N^* \to \N^*                                                                        \\
  \texttt{doubleN}\  & := \RecNatr\ \N^*\ \texttt{0}^*\ (\lambda (n: \N^*).\ \texttt{S}^*\ (\texttt{S}^*\ n))
\end{align*}

In addition to the formation, introduction, and elimination rules, we also have that the expected computation rules, or $\beta$-rules, hold for the natural numbers. These rules show that the introduction rules and elimination rules interact in the expected way.

\begin{lemma} Let $(X: \U)$ be a type. Let $(z: X)$ be an element of $X$ and let $(s: X \to X)$ be some function on $X$. Furthermore, let $(n : \N^*)$ be a natural number. The following $\beta$-rules hold for $\N^*$.

  \begin{align*}
    \RecNatr\ X\ z\ s\ \texttt{0}^*      & = z \tag{computation}                         \\
    \RecNatr\ X\ z\ s\ (\texttt{S}^*\ n) & = s\ (\RecNatr\ X\ z\ s\ n) \tag{computation}
  \end{align*}

  \begin{proof}
    Let $(X: \U)$, $(x: X)$, $(s: X \to X)$ and $(n : \N^*)$ be as described. The $\beta$-rules hold by unfolding the definitions and applying $\beta$-reduction.

    \begin{align*}
      \RecNatr\ X\ z\ s\ \texttt{0}^*      & := \texttt{0}^*\ X\ z\ s                                                                                                    \\
                                           & := \Bigl(\lambda (X: \U). \lambda (z: X). \lambda (s: X \to X \to X).\ z\Bigr)\ X\ z\ s                                     \\
                                           & \betar z                                                                                                                    \\\\
      \RecNatr\ X\ z\ s\ (\texttt{S}^*\ n) & := (\texttt{S}^*\ n)\ X\ z\ s                                                                                               \\
                                           & := \Bigl(\lambda (n: \N^*). \lambda (X: \U). \lambda (z: X). \lambda (s: X \to X \to X).\ s\ (n\ X\ z\ s)\Bigr)\ n\ X\ z\ s \\
                                           & \betar s\ (n\ X\ z\ s)                                                                                                      \\
                                           & =: s\ (\RecNatr\ X\ z\ s\ n) \qedhere
    \end{align*}
  \end{proof}
\end{lemma}

To highlight the expressivity of System F, we give a few examples of types that you can define within the system. There are many more examples that we shall see throughout this thesis.

\begin{example} We use System F to define a unit, empty, product, and coproduct The details are worked out in \cref{datatypes:systemf}.
  \label{systemf-datatypes}

  \begin{align*}
    \textbf{1}^* & := \prod (X: \U). X \to X                       \\
    \textbf{0}^* & := \prod (X: \U). X                             \\
    A +^* B      & := \prod (X: \U). (A \to X) \to (B \to X) \to X \\
    A \times^* B & := \prod (X: \U). (A \to B \to X) \to X
  \end{align*}
\end{example}

\section{Extending System F}

In the previous section, we have seen that System F, or $\lambda 2$, is a very expressive system. It allows us to define many structures. In this section, we extend System F with the concepts of $\Pi$-types, $\Sigma$-types, $=$-types, the function extensionality axiom, the uniqueness of identity proofs axiom and a new existential identity axiom. For the complete set of inference rules, we refer to \cref{chap:complete_system}.

We postulate a cumulative hierarchy of universes $\U : \U_0 : \U_1 : \U_2 : \dots$, where the bottom universe $\U$ is impredicative\footnote{Note that in \cite{Hott}, $\U$ is used to denote $\U_i$ for some unspecified $i$.}. The cumulativity means that types present in the impredicative universe $\U$ are also present in $\U_0$ and types present in $U_i$ are also present in $U_{i+1}$.

We first extend System F with product types ($\Pi$-types). These product types are a generalization of the arrow types of System F that allow us to define dependent functions, where the \textit{type} of the output depends on the \textit{value} of the input. They are defined to allow for impredicative quantification over the universe $\U$.

\begin{definition} We define (impredicative) \textbf{product types}. We only give the formation and introduction rules. For the other rules, we refer to \cref{chap:complete_system}.

  \begin{gather*}
    \begin{aligned}
      \infer[\Pi\text{-form}_1]{\Gamma \vdash \prod (a: A). B : \U}{%
        \Gamma \vdash A : \U_i
       &
        \Gamma, a: A \vdash B : \U
      }
      \ \ \
       &
      \infer[\Pi\text{-form}_2]{\Gamma \vdash \prod (a: A). B : \U_i}{%
        \Gamma \vdash A : \U_i
       &
        \Gamma, a: A \vdash B : \U_i
      }
    \end{aligned}
    \\\\
    \infer[\Pi\text{-intro}]{\Gamma \vdash \lambda (a: A). b : \prod (a: A). B}{%
      \Gamma, a: A \vdash b : B
    }
  \end{gather*}

  The first formation rule allows for impredicative quantification since $A$ can be any type (since each type exists in some universe $U_i$). The second formation rule is the standard predicative formation rule for product types. Note that System F style arrow types can be defined in terms of product types. We shall use the notation $A \to B$ as shorthand for $\prod (a: A). B$. We finally assume the usual elimination and computation and uniqueness rules for $\Pi$-types as detailed in \cite{Hott,speight}.
\end{definition}

Next, we define sigma types ($\Sigma$-types). These types are a generalization of the product types definable in System F. They allow us to define dependent pairs, where the \textit{type} of the second component depends on the \textit{value} of the first component.

\begin{definition} We define strong \textbf{sigma types}. We only give the formation, introduction, and uniqueness rules. For the other rules, we refer to \cref{chap:complete_system}.

  \begin{gather*}
    \begin{aligned}
      \infer[\Sigma\text{-form}_1]{\Gamma \vdash \sum (a:A). B : \U}{%
        \Gamma \vdash A : \U
       &
        \Gamma, a: A \vdash B : \U
      }
      \ \ \
       &
      \infer[\Sigma\text{-form}_1]{\Gamma \vdash \sum (a:A). B : \U_i}{%
        \Gamma \vdash A : \U_i
       &
        \Gamma, a: A \vdash B : \U_i
      }
    \end{aligned}
    \\\\
    \infer[\Sigma\text{-intro}]{\Gamma \vdash \tup{a, b} : \sum (x: A). B[x]}{%
      \Gamma, x: A \vdash B[x] : \U_i
      &
      \Gamma \vdash a : A
      &
      \Gamma \vdash b : B[x := a]
    }
  \end{gather*}

  We shall use the notation $A \times B$ as shorthand for $\sum (x: A). B$. We furthermore have the two projection functions $(\pr1: \sum (x: A). B \to A)$ and $(\pr2: \sum (x: A). B \to B)$. We allow tuple destructuring and write $\lambda \tup{a, b}. M[a,b]$ as shorthand for $\lambda x. M[a := (\pr1\ x),\ b := (\pr2\ x)]$. Furthermore, the notation $\tup{a, b, c}$ is shorthand for $\tup{a, \tup{b, c}}$, and we syntactically define $\pr3$. We finally assume the usual elimination and computation and uniqueness rules for $\Sigma$-types as detailed in \cite{Hott,speight} and \cref{chap:complete_system}.

  \subsection*{Embeddings}

  We can use $\Sigma$-types to create embeddings or subtypes. Embeddings are the type theoretic analog of subsets in set theory. Given a predicate $(\texttt{IsEven}: \N \to \U)$, we can create a subtype $\texttt{even} := \sum (n: \N). \texttt{IsEven}\ n$. Elements of this type are the even natural numbers. More precisely, they are tuples $\tup{n, p}$ where $n$ is a natural number, and $p$ is a proof of the $\texttt{IsEven}$ predicate that states that $n$ is indeed even. In summary, we call a $\Sigma$-type an embedding if the second component is a proposition (is of type $\U$).

  As a consequence of the $\eta$-rule for $\Sigma$-types, we get the sigma injection principle $(\hookrightarrow)$. This principle states that, if the sigma type is an embedding, then the equality of elements can be shown by proving the equality of the first component. Compare this to the concept of subsets: if we take a set $X$ and some subset $Y := \{ x \in X \mid P(x) \} \of X$, the subset $Y$ inherits the equality relation of $X$. In other words, if $x = x'$ in X, then $x = x'$ obviously still holds for $(x, x' \in Y)$.

  This principle is formalized below in the following lemma.

  \begin{lemma} We have the following sigma injection principle.

    \begin{align*}
      (\hookrightarrow) : \prod (X:\ \U). \prod (P: X \to \U). \prod \bigl(y, y': \sum (x: X). P\ x\bigr). y = y' \iff \pr1\ y = \pr1\ y'
    \end{align*}

    \begin{proof}
      We give the proof in \cref{chap:complete_system}.
    \end{proof}
  \end{lemma}

\end{definition}

Next, we define intentional equality types ($=$-types), originally called identity types. These types allow us to define the notion of propositional equality.

\begin{definition} We define \textbf{equality types}. We give the formation rules and the introduction rule. For the other rules, we refer to \cref{chap:complete_system}.

  \begin{gather*}
    \begin{aligned}
      \infer[\text{=-form}_1]{\Gamma \vdash (a =_A b) : \U}{%
        \Gamma \vdash A : \U
       &
        \Gamma \vdash a : A
       &
        \Gamma \vdash b : A
      }
       &
      \ \ \
      \infer[\text{=-form}_1]{\Gamma \vdash (a =_A b) : \U_i}{%
        \Gamma \vdash A : \U_i
       &
        \Gamma \vdash a : A
       &
        \Gamma \vdash b : A
      }
    \end{aligned}
    \\
    \\
    \infer[\text{=-intro}]{\Gamma \vdash \texttt{refl} : a =_A a}{%
      \Gamma \vdash A : \U_i
      &
      \Gamma \vdash a : A
    }
  \end{gather*}

  In the future, we drop the subscript $A$ and write $a = b$. When the proof term $(p : a = b)$ is present, we often make its use explicit by writing $a \overset{p}{=} b$.

\end{definition}

Next, we add three axioms to the system. The first axiom is the axiom of function extensionality. This axiom states that any two functions that are equal for all values are equal. This axiom is essential in proving many theorems and indispensable when defining W-types.

\define{\funext}{funext}{\texttt{FunExt}}
\begin{axiom} Let $X, Y$ be types in some universe. The \textbf{function extensionality} axiom states that any two functions $\bigl(f, g : \prod (x: X). Y\bigr)$ are equal if they are equal for all values $(x : X)$. We formally write this as follows:

  \begin{align*}
    \funext : \Bigl( \prod(x: X). f\ x = g\ x \Bigr) \implies f = g
  \end{align*}

\end{axiom}

We add the axiom of uniqueness of identity proofs (UIP). This axiom states that if we have two proofs of a statement, then these proofs are propositionally equal. Strictly speaking, we could do without this axiom by using the notion of $0$-types by defining $\texttt{Set}$ as those types for which UIP holds. This approach was taken in \cite[Introduction]{encodings} and opens the door for higher inductive types such as $\mathcal{S}_1$. An interesting avenue for future work is to remove this axiom and adapt the relevant defintions.

\newpage
\define{\UIP}{UIP}{\texttt{UIP}}
\begin{axiom} The \textbf{uniqueness of identity proofs} axiom states that any two proofs of equality between two elements are equal. We formally write this as follows:

  \begin{align*}
    \UIP : \prod (X: \U) \prod (x, y: X) \prod (p, q: x = y). p = q
  \end{align*}

\end{axiom}

In \cref{existential-types}, we add another axiom to the system (\cref{axiom:existsid}). This axiom is best explained in the context of that section. It states that we have a certain identity on existential types.

Now that we have defined our system, we can properly begin our journey. In the next chapter, we introduce the technique of \cite{encodings} by creating an impredicative encoding of a list type.

\chapter{Lists}
\label{chap:list}

In this chapter, we apply the encoding technique of \cite{encodings} to create a finite list data type that has an induction principle. Our goal is to introduce the reader to the principle idea of encoding the uniqueness principle of an inductive type within its definition.

We begin by giving the common impredicative definition of impredicative lists and show that the $\beta$-rules, or computation rules, are satisfied for this definition. Next, we take a look at the categorical definition of inductive lists using a suitable initial algebra. We then combine these definitions into an impredicative encoding of a list type that also satisfies the $\eta$-rule. We finish by proving the induction principle for finite lists.

These lists shall range over some element type $(E : \U)$, so we speak of lists \textit{over} $E$. We globally assume that $\bigl(\Gamma \vdash E : \U\bigr)$. The following definitions can be straightforwardly generalized to any element type.

\section{Impredicative list type}

Let us begin by giving an impredicative finite list type \cite[11.5.2 Lists]{Girard_Taylor_Lafont_1993}.

\define{\Listr}{Listr}{\texttt{List}^*}
\newcommand{\Listrr}[1]{\hyperref[def:Listrr]{\texttt{\texttt{List}}^*_#1}}
\label{def:Listrr}
\define{\nilr}{nilr}{\texttt{nil}^*}
\define{\consr}{consr}{\texttt{cons}^*}

\begin{definition}\label{def:impr_list}
  We define an \textbf{impredicative list} data type $\Listr$ over some element type $(E : \U)$. We also define two constructors: $\nilr$ and $\consr$.

  \begin{align*}
    \Listr                     & := &  & \prod (X:\U)\.X\to(E\to X\to X)\to X \tag{formation}                          \\
    \nilr                      & := &  & \lambda (X:\U). (x:X). (f: E\to X\to X). x \tag{introduction}                 \\
    \consr\ (e:E)\ (l: \Listr) & := &  & \lambda (X:\U). (x:X). (f: E\to X\to X). f\ e\ (l\ X\ x\ f)\tag{introduction}
  \end{align*}
\end{definition}

\begin{example}
  Using the element type $E := \N$, we can define the list $[1,2,3]$ or $1 : 2 : 3 : []$ as follows:

  \begin{align*}
    [\ 1,\ 2,\ 3\ ] & := &  & \consr\ 1\ \bigl(\consr\ 2\ (\consr\ 3\ \nilr)\bigr) : \Listrr{\N}
  \end{align*}

  It is clear that $\nilr$ and $\consr$ are of the appropriate type. For an example of a full derivation, we refer to \cref{expl:nat}.

  \begin{align*}
    \Gamma                    & \vdash \nilr        & : &  & \Listr \hspace{20em} \\
    \Gamma, (e:E), (l:\Listr) & \vdash \consr\ e\ l & : &  & \Listr \hspace{20em}
  \end{align*}

  Note that we can also create lists of lists. For example $(\Gamma \vdash \consr\ \nilr\ \nilr : \Listr_{\Listrr{\N}})$, which is a list with one element: an empty list.
\end{example}

Datatypes are not of much use if we can only construct data: we also want to \textit{do} things with the data. Suppose we want to perform some operation on these lists, for example, mapping the list, or taking the $\texttt{head}$ or $\texttt{tail}$. To this end, we need an elimination principle or \textit{recursor}.

This recursor takes four arguments. The first argument $(X: \U)$ specifies the output type. If we want to create a function from lists to natural numbers, we would set $X := \N$. If we output a new list, we would set $X := \Listr$. The second argument $(x: X)$ specifies what we map the empty list to. The third argument $(g: E \to X \to X)$ specifies the recursive call. The fourth argument $(l: \Listr)$ is the list to operate on.

\define{\RecListr}{RecListr}{\texttt{rec}^*_{\texttt{l}}}
\begin{definition}We define the \textbf{recursor} for the $\Listr$ type.

  \begin{align*}
    \RecListr             & : \prod (X:\U). X \to (E \to X \to X) \to \Listr \to X \tag{elimination} \\
    \RecListr\ X\ x\ g\ l & := l\;X\;x\;g
  \end{align*}
\end{definition}

\begin{example}
  We use the list recursor to define a function that sums all elements in a list. We set $X := \N$, $x := 0$ and $g := \lambda e. \lambda n. e + n$.

  \begin{align*}
    \texttt{sum} & : \Listrr{\N} \to \N                               \\
    \texttt{sum} & := \RecListr\ \N\ 0\ (\lambda e. \lambda n. e + n)
  \end{align*}
\end{example}

\begin{example}
  We use the list recursor to define a function that maps all elements in a list to their square. We set $X := \Listrr{\N}$, $x := \nilr$ and $g := \lambda e. \lambda l. \consr\ (e * e)\ l$.

  \begin{align*}
    \texttt{square} & : \Listrr{\N} \to \Listrr{\N}                                                \\
    \texttt{square} & := \RecListr\ \Listrr{\N}\ \nilr\ (\lambda e. \lambda l. \consr\ (e * e)\ l)
  \end{align*}
\end{example}

This recursor then needs to $\beta$-reduce in an expected way. We call these expected reductions the $\beta$-rules or computation rules. We show that this recursor indeed satisfies the expected $\beta$-rules.

\begin{proposition}
  The list recursor $\RecListr$ satisfies the following $\beta$-rules.
  Let $(X : \U)$, $(x: X)$ and $(g: E \to X \to X)$.
  \label{list_beta}

  \begin{align*}
    \RecListr\ X\ x\ g\ \nilr          & \betar x \tag{computation}                             \\
    \RecListr\ X\ x\ g\ (\consr\ e\ l) & \betar g\ e\ (\RecListr\ X\ x\ g\ l) \tag{computation}
  \end{align*}

  \begin{proof}
    Let $(X : \U)$, $(x: X)$ and $(g: E \to X \to X)$. We derrive the following:

    \begin{align*}
      \RecListr\ X\ x\ g\ \nilr          & := \nilr\ X\ x\ g          \\ &\betar x\\
      \RecListr\ X\ x\ g\ (\consr\ e\ l) & := (\consr\ e\ l)\ X\ x\ g \\
                                         & \betar g\ e\ (l\ X\ x\ g)  \\ &=: g\ e\ (\RecListr\ X\ x\ g\ l) \qedhere
    \end{align*}
  \end{proof}
\end{proposition}

To summarize, we have defined a finite list type $\Listr$. We have two constructors or introduction rules: $\nilr$ and $\consr$. We have a recursion principle $\RecListr$ that satisfies the expected $\beta$-rules. Even though this definition seems to behave quite well out of the box we are unable to show the induction principle \cite{NoInduction}. This is because, even though the recursor satisfies the $\beta$-rules, it fails to satisfy the $\eta$-rule. This rule states that the recursor $\RecListr\ X\ x\ g$ is unique in an appropriate sense. We would expect, for example that $\RecListr\ \Listr\ \nilr\ \consr = \id{\Listr}$. Using this $\eta$-rule, we can prove the induction principle.

In the next section, we jump to the categorical definition of lists using initial algebras. We then combine these definitions to create an impredicative encoding of a list type that satisfies the $\eta$-rule. We finish by proving the list induction principle.

\section{Categorical lists}

In this section, we use the theory of algebras of \cref{sec:algebras} using a suitable functor $\mathcal{L}$. From the theory of algebra, we know that the initial algebra of $\mathcal{L}$-\textbf{Alg} is a good candidate for the definition of an inductive list type. We shall take inspiration from the uniqueness property of this initial algebra to define an impredicative list encoding that satisfies the $\eta$-rule.

\begin{definition} We define the endofunctor $(\mathcal{L}: \C \to C)$. We leave the category $\C$ unspecified, assuming it is a category with finite products and coproducts and a unit object $\1$, for example the category $\textbf{Set}$. We also assume that $\C$ has some object $E$. We define $\mathcal{L}$ as follows:

  \begin{align*}
    \mathcal{L}(X)          & := \1 + (E\times X)                             \\
    \mathcal{L}(f: X \to Y) & := [\inlc, \tup{e,x} \mapsto \inrc\tup{e,f(x)}]
  \end{align*}

\end{definition}

An $\mathcal{L}$-algebra is a tuple $\tup{X, \alpha}$, where $\mathcal{L}(X) \overset{\alpha}{\to} X$. If we unfold the definition of $\mathcal{L}$, we get the following: $\1 + (E\times X) \overset{\alpha}{\to} X$. We thus see that $\alpha$ consists of two parts: a function $\1 \to X$ and a function $(E\times X) \to X$. We write $\alpha := [x,g]$, where $(x: \star \to X)$ and $(g: (E\times X) \to X)$.

We instantiate the definition of $\mathcal{L}$-morphism given in \cref{def:alg-morphism}, where we have that \\$(y: \star \to Y)$ and $(h: (E\times Y) \to Y)$.

  \begin{figure}[H]
    \centering
    \begin{tikzcd}
      {\mathbf{1}+(E\times X)} & X \\
      {\mathbf{1}+(E\times Y)} & Y
      \arrow["{{[x,g]}}", from=1-1, to=1-2]
      \arrow["{{[\inlc, \tup{e,x} \mapsto \inrc \tup{e,f(x)}]}}"', from=1-1, to=2-1]
      \arrow["{=}"{description}, draw=none, from=1-1, to=2-2]
      \arrow["f", from=1-2, to=2-2]
      \arrow["{{[y,h]}}"', from=2-1, to=2-2]
    \end{tikzcd}

    \caption{Commutative diagram of a $\mathcal{L}$-morphism.}
  \end{figure}

  If we write out this diagram in formulas we derive the following commutativity requirement:

  \begin{align}
    f \circ [x, g]         & = [y, h] \circ [\inlc, \tup{e,x} \mapsto \inrc\tup{e,f(x)}]                           & \iff \nonumber \\
    [f \circ x, f \circ g] & = [y, h \circ (\tup{e,x} \mapsto \tup{e,f(x)})]                                       & \iff \nonumber \\
    f \circ x = y          & \land f \circ g = h \circ (\tup{e,x} \mapsto \tup{e,f(x)}) \label{commutativity_list}
  \end{align}

  Recall that the initial object $\tup{I, \gamma}$ of the category of $\mathcal{L}$-algebras $\mathcal{L}\text{-}\textbf{Alg}$ has exactly one morphism $\bigl(u_X: \tup{I, \gamma} \to \tup{X, \alpha}\bigr)$ for each object $\tup{X, \alpha}$ in the category $L\text{-}\textbf{Alg}$:

  \begin{figure}[H]
    \centering
    \begin{tikzcd}
      1 + (E\times I) \arrow[r, "\gamma"] \arrow[d, "\mathcal{L}(u_X)"', dashed] & I \arrow[d, "\exists!u_X", dashed] \\
      \arrow[ur, phantom, "=" description]
      \mathbf{1}+(E\times X) \arrow[r, "\alpha"]                         & X
    \end{tikzcd}
    \caption{Commutative diagram of the initial $\mathcal{L}$-algebra.}
  \end{figure}

  These $\mathcal{L}$-morphisms $u_X$ should of course satisfy the commutativity requirement of \cref{commutativity_list}. Writing down the uniqueness requirement is a bit more tricky. We use requirement 2 from \cref{uniqueness_rules}. This states that, given a morphism $\bigl(f: \tup{X, \alpha} \to \tup{Y, \beta}\bigr)$, and morphisms $(u_X : I \to X)$ and $(u_Y: I \to Y)$ (which all satisfy \cref{commutativity_list}), we have that $f \circ u_X = u_Y$. This construction is shown \cref{fig:uniqueness_list}. Compare the right-hand side of this diagram with requirement 2 from \cref{uniqueness_rules}.

  \begin{figure}[H]
    \centering
    \begin{tikzcd}
      {\mathbf{1}+(E\times I)} & I \\
      {\mathbf{1}+(E\times X)} & X \\
      {\mathbf{1}+(E\times Y)} & Y
      \arrow["\gamma", from=1-1, to=1-2]
      \arrow["{\mathcal{L}(u_X)}"', from=1-1, to=2-1]
      \arrow["{=}"{description, pos=0.4}, draw=none, from=1-1, to=2-2]
      \arrow["{u_X}"', from=1-2, to=2-2]
      \arrow[""{name=0, anchor=center, inner sep=0}, "{u_Y}", bend left=49, from=1-2, to=3-2]
      \arrow["{[x, g]}", from=2-1, to=2-2]
      \arrow["{\mathcal{L}(f)}"', from=2-1, to=3-1]
      \arrow["{=}"{description, pos=0.4}, draw=none, from=2-1, to=3-2]
      \arrow["f"'{pos=0.3}, from=2-2, to=3-2]
      \arrow["{[y,h]}", from=3-1, to=3-2]
      \arrow["="{marking, allow upside down, pos=0.4}, draw=none, from=2-2, to=0]
    \end{tikzcd}
    \caption{Diagram of the uniqueness requirement of an initial $\mathcal{L}$-algebra.}
    \label{fig:uniqueness_list}
  \end{figure}

  \section{Impredicative encoding of lists}

  We are now ready to translate the categorical requirement of the initial $\mathcal{L}$-algebra back to type theory. We first need to create an analog of $\mathcal{L}$-algebras within type theory. We use triples $\tup{X, x, g}$ where $(X: \U)$, $(x: X)$ and $(g: E \to X \to X)$. These triples can be thought of as the translation of categorical $\mathcal{L}$-algebras $\tup{X, [x, g]}$. We start by giving a type theoretical predicate that states that a morphism $(f: X \to Y)$ is an $\mathcal{L}$-algebra morphism.

  \define{\MorphList}{MorphList}{\texttt{MorphList}}
  \begin{definition}
    Let $(X, Y: \U), (x: X), (y: Y), (g: E \to X \to X)$ and $(h: E \to Y \to Y)$. We define a predicate $\MorphList$ that states that $(f: X \to Y)$ forms an $\mathcal{L}$-algebra morphism. Compare this predicate with \cref{commutativity_list}.

    \begin{align*}
      \MorphList\ X\ x\ g\quad Y\ y\ h\quad (f: X\to Y) := & \; (f\ x) = y \land \prod (e:\ E). f \circ (g\;e) = (h\;e) \circ f
    \end{align*}

  \end{definition}

  We can now define a subtype of $\Listr$ that encodes the uniqueness requirement of the initial $\mathcal{L}$-algebra. We do this by essentially mimicking the uniqueness rule outlined in \cref{fig:uniqueness_list} (where $u_X := \RecListr\ X\ x\ g$ and $u_Y := \RecListr\ Y\ y\ h$). We require that $f\ (\RecListr\ X\ x\ g) = \RecListr\ Y\ y\ h$ for all $\mathcal{L}$-morphisms $\bigl(f: \tup{X, [x, g]} \to \tup{Y, [y, h]}\bigr)$.

  \newpage
  \define{\List}{List}{\texttt{List}}
  \define{\LimList}{LimList}{\texttt{LimList}}
  \begin{definition}
    We define an \textbf{inductive list} type.

    \begin{align*}
      \List :=                 & \sum (l:\Listr)\.\texttt{LimList}\ l \tag{formation}                                            \\
      \LimList\ (l: \Listr) := & \prod_{X, Y:\U}\prod_{\substack{x: X, y: Y}}\prod_{f: X \to Y}\prod_{\substack{g: E \to X \to X \\ h: E \to Y \to Y}} \\
                               & (\MorphList\ X\ x\ g\quad Y\ y\ h\quad f) \implies                                              \\
                               & f (\RecListr\ X\ x\ g\ l) = \RecListr\ Y\ y\ h\ l
    \end{align*}
  \end{definition}

  Let's take a moment to understand this definition. The $\List$ type is defined using a sigma-type. In other words, its inhabitants are pairs $\tup{l, p}$, where $(l: \Listr)$ is the previously defined impredicative list (\cref{def:impr_list}). The second component $p$ is a proof term that $(\LimList\ l)$ holds. This proof states that for all $\mathcal{L}$-algebras $\tup{X, x, g}$ and $\tup{Y, y, h}$, and all $\mathcal{L}$-algebra morphisms $f$, we have that $f\ (\RecListr\ X\ x\ g\ l) = \RecListr\ Y\ y\ h\ l$.

  \define{\nil}{nilr}{\texttt{nil}}
  \define{\cons}{cons}{\texttt{cons}}
  \define{\RecList}{RecList}{\texttt{rec}_{\texttt{l}}}

  This new $\List$ type could very well be the empty type. We could have that no lists comply with the $\LimList$ predicate. Fortunately, we both have that $\nilr$ and $\consr$ satisfy the $\LimList$ predicate. We can thus define $(\nil: \List)$ and $(\cons: E \to \List \to \List)$ in a straightforward manner.

  \define{\LimListZ}{LimListZ}{\texttt{LimList0}}

  \begin{lemma}[$\LimListZ$] \label{lem:LimList0} The proposition $(\LimList\ \nilr)$ holds.
    \begin{proof}
      Let $(X, Y:\U)$, $(x: X)$, $(y: Y)$, $(f: X \to Y)$, $(g: E \to X \to X)$ and $(h: E \to Y \to Y)$. We assume that we have a proof term of $(\MorphList\ X\ x\ g\quad Y\ y\ h\quad f)$. Therefore we obtain that $f(x) = y$. By the $\beta$-rules of $\RecListr$ (\cref{list_beta}), we have that $\RecListr\ X\ x\ g\ \nilr \betar x$ and $\RecListr\ Y\ y\ h\ \nilr \betar y$. We conclude that $f(\RecListr\ X\ x\ g\ \nilr) \betar f(x) = y \betar \RecListr\ Y\ y\ h\ \nilr$.
    \end{proof}
  \end{lemma}

  \define{\LimListC}{LimListC}{\texttt{LimListC}}

  \begin{lemma}[$\LimListC$] \label{lem:LimListC} For $(e: E)$ and $(l: \List)$ we have that $(\LimList\ (\consr\ e\ (\pr1\ l)))$ holds.
    \begin{proof}
      Let $(X, Y:\U)$, $(x: X)$, $(y: Y)$, $(f: X \to Y)$, $(g: E \to X \to X)$ and $(h: E \to Y \to Y)$. We assume that we have a proof term of $(\MorphList\ X\ x\ g\quad Y\ y\ h\quad f)$. Therefore we have $(\texttt{morph}: f \circ (g\;e) = (h\;e) \circ f)$. Let $(e: E)$ and $(l: \List)$.
      We have that $(\LimList\ (\pr1\ l))$ holds. Note that $(\pr2\ l)$ has type $(\LimList\ (\pr1\ l))$. We can apply $(\LimList\ (\pr1\ l))$ with the appropriate variables $(X, Y, x, y, f, g, h)$. We obtain $\bigl(\texttt{comm}: f (\RecListr\ X\ x\ g\ (\pr1\ l)) = \RecListr\ Y\ y\ h\ (\pr1\ l)\bigr)$. We derive the following:

      \begin{align*}
        f(\RecListr\ X\ x\ g\ (\consr\ e\ (\pr1\ l))) & \betar f(g\ e\ (\RecListr\ X\ x\ g\ (\pr1\ l)))            \\
        \overset{\texttt{morph}}                      & {=} h\ e\ (f (\RecListr\ X\ x\ g\ (\pr1\ l)))              \\
        \overset{\texttt{comm}}                       & {=} h\ e\ (\RecListr\ Y\ y\ h\ (\pr1\ l))                  \\
                                                      & \betar \RecListr\ Y\ y\ h\ (\consr\ e\ (\pr1\ l)) \qedhere
      \end{align*}
    \end{proof}
  \end{lemma}

  Now that we have proven that $\nil$ and $\cons$ satisfy the $\LimList$ predicate, we can create new introduction rules for the inductive list type $\List$. Note that we actually embed the proofs of the $\LimList$ predicate within the definition!

  \newpage
  \begin{definition}
    We define $\nil$, $\cons$ and $\RecList$ for the $\List$ type using $\LimListZ$ and $\LimListC$.

    \begin{align*}
      \nil :=                      & \; \tup{\nilr, \LimListZ} \tag{introduction}                            \\
      \cons\ (e: E)\ (l: \List) := & \; \tup{\consr\ e\;(\pr1\ l), \LimListC\ e\ l} \tag{introduction}       \\
      \intertext{We can also define a new recursor that simply uses the old recursor.}                       \\
      \RecList                     & : \prod (X:\U). X \to (E \to X \to X) \to \List \to X \tag{elimination} \\
      \RecList\ X\ x\ c\ l         & := \RecListr\ X\ x\ c\ (\pr1\ l)
    \end{align*}
  \end{definition}

  Given these new introduction and elimination rules, we again need to check that the computation rules ($\beta$-rules) are still satisfied.

  \begin{proposition}
    The list recursor $\RecList$ satisfies the following $\beta$-rules.
    Let $(X : \U),\ (x: X)$ and $(f: E \to X \to X)$.

    \begin{align*}
      \RecList\ X\ x\ f\ \nil          & \betar x \tag{computation}                            \\
      \RecList\ X\ x\ f\ (\cons\ e\ l) & \betar f\ e\ (\RecList\ X\ x\ f\ l) \tag{computation}
    \end{align*}

    \begin{proof}
      By $\beta$-reduction.
    \end{proof}
  \end{proposition}

  With this recursor, we have that the triple $\tup{\List, \nil, \cons}$ acts as the initial $\mathcal{L}$-algebra. This means that for all triples $\tup{X, x, g}$ that act like $\mathcal{L}$-algebra, we have that $(\RecList\ X\ x\ g : \List \to X)$ acts as the unique morphism $u_X$. We ensure that $(\RecList\ X\ x\ g)$ is an $\mathcal{L}$-algebra morphism by proving the following lemma.

  \begin{minipage}[t]{.66\textwidth}
    \begin{lemma}
      \label{reclist_morphlist}
      For all $(X: \U)$, $(x: X)$ and $(g: E \to X \to X)$ we have that $(\RecList\ X\ x\ g)$ is an $\mathcal{L}$-algebra morphism. We thus have a proof of the following proposition:

      \begin{gather*}
        \bigl(\MorphList\ \List\ \nil\ \cons \quad X\ x\ g\quad (\RecList\ X\ x\ g) \bigr)
      \end{gather*}
    \end{lemma}
  \end{minipage}
  \begin{minipage}[t]{.33\textwidth}
    \[\begin{tikzcd}
        {1 + (E \times \List)} & \List \\
        {1+(E \times X)} & X
        \arrow["{[\nil, \cons]}", from=1-1, to=1-2]
        \arrow["{\mathcal{L}(\RecList\ X\ x\ g)}"', from=1-1, to=2-1]
        \arrow["{=}", shift right=3, draw=none, from=1-1, to=2-2]
        \arrow["{\RecList\ X\ x\ g}", from=1-2, to=2-2]
        \arrow["{[x, g]}"', from=2-1, to=2-2]
      \end{tikzcd}\]
  \end{minipage}
  \begin{proof}
    Let $(X: \U)$, $(x: X)$ and $(g: E \to X \to X)$.
    For the left conjunct, we have $(\RecList\ X\ x\ g)\ \nil \betar x$ by the first $\beta$-rule.
    For the right conjunct, we have the following reduction using beta equality and function extensionality.

    \begin{align*}
                               & (\RecList\ X\ x\ g) \circ (\cons\;e) = (g\;e) \circ (\RecList\ X\ x\ g) \overset{\funext} & {\iff} \\
      \forall (l: \List).\quad & (\RecList\ X\ x\ g) (\cons\ e\ l) = g\ e\ (\RecList\ X\ x\ g\ l) \overset{\beta}          & {\iff} \\
      \forall (l: \List).\quad & g\ e\ (\RecList\ X\ x\ g\ l) = g\ e\ (\RecList\ X\ x\ g\ l)
    \end{align*}
    Thus we conclude that $(\MorphList\ \List\ \nil\ \cons\quad X\ x\ g \quad (\RecList\ X\ x\ g))$ holds
  \end{proof}

  Before we can prove the $\eta$-rule for the recursor, we need the following helper lemma.

  \define{\RecId}{RecId}{\texttt{RecId}}

  \begin{lemma}[$\RecId$]
    $\RecList\ \List\ \nil\ \cons = \id{\List}$

    \[\begin{tikzcd}
        {\mathbf{1}+(E\times \List)} & \List \\
        {\mathbf{1}+(E\times \List)} & \List
        \arrow["{[\nil,\cons]}", from=1-1, to=1-2]
        \arrow[from=1-1, to=2-1]
        \arrow["{=}"{description}, draw=none, from=1-1, to=2-2]
        \arrow[""{name=0, anchor=center, inner sep=0}, "\id{\List}"', from=1-2, to=2-2]
        \arrow[""{name=1, anchor=center, inner sep=0}, "{\RecList\ \List\ \nil\ \cons}"{pos=0.54}, bend left=49, from=1-2, to=2-2]
        \arrow["{[\nil,\cons]}"', from=2-1, to=2-2]
        \arrow["{=}"{marking}, draw=none, from=0, to=1]
      \end{tikzcd}\]

    \label{reclist_id}
    \begin{proof}
      We first reduce the statement using function extensionality and $\beta$-reduction. In the $(\hookrightarrow)$ step, we use the fact that $\List^* \hookrightarrow \List$ is an embedding.

      \begin{align}
                                    & \RecList\ \List\ \nil\ \cons = \id{\List} \overset{\funext}           & {\iff} \nonumber \\
        \forall\ (l: \List). \qquad & \RecList\ \List\ \nil\ \cons\ l  = l \overset{\hookrightarrow}        & {\iff} \nonumber \\
        \forall\ (l: \List). \qquad & \ \pr1\ (\RecList\ \List\ \nil\ \cons\ l) = \pr1\ l \overset{\funext} & {\iff} \nonumber \\
        \forall\ \substack{(l: \List), (X:\U),                                                                                 \\(x: X), (g: E \to X\to X)}. \qquad &\ \bigl(\pr1\ (\RecList\ \List\ \nil\ \cons\ l)\bigr)\ X\ x\ g = (\pr1\ l)\ X\ x\ g \label{RecIdReduc}
      \end{align}

      By \cref{reclist_morphlist} we known that $((\RecList\ X\ x\ g): \List \to X)$ is an $\mathcal{L}$-morphism. By looking at $(p:= \pr2\ l)$, which has type $\LimList(\pr1\ l)$, we obtain the following equation:

      \begin{align}
        (\RecList\ X\ x\ g) (\RecListr\ \List\ \nil\ \cons (\pr1\ l)) = (\RecListr\ X\ x\ g\ (\pr1\ l)) \label{LimListPr}
      \end{align}

      We conclude the proof by showing \cref{RecIdReduc}. Let $(l: \List), (X:\U), (x: X)$, and $(g: E \to X\to X)$.

      \begin{align*}
        \bigl(\pr1\ (\RecList\ \List\ \nil\ \cons\ l)\bigr)\ X\ x\ g & \betar \RecListr\ X\ x\ g\ \bigl(\pr1\ (\RecList\ \List\ \nil\ \cons\ l)\bigr) \\
                                                                     & \betar \RecList\ X\ x\ g\ (\RecList\ \List\ \nil\ \cons\ l)                    \\
                                                                     & = (\RecList\ X\ x\ g)\ (\RecList\ \List\ \nil\ \cons\ l)                       \\
                                                                     & := (\RecList\ X\ x\ g)\ (\RecListr\ \List\ \nil\ \cons\ (\pr1\ l))             \\
                                                                     & \overset{\ref{LimListPr}}{=} (\RecListr\ X\ x\ g (\pr1\ l))                    \\
                                                                     & := (\pr1\ l)\ X\ x\ g \qedhere
      \end{align*}
    \end{proof}
  \end{lemma}

  \begin{theorem}
    \label{thm:eta_list}
    The encoding $\List$ satisfies the $\eta$-rule which states that for all $(X: \U),\ (x: X),\ (g: E \to X \to X)$ and $(f: \List \to X)$ we have:

    \begin{gather}
      \label{eta_rule_list_p}
      (\MorphList\ \ \List\ \nil\ \cons \quad X\ x\ g \quad f) \implies f = \RecList\ X\ x\ g \tag{uniqueness}
    \end{gather}

    \[\begin{tikzcd}
        {\mathbf{1}+(E\times \List)} & \List \\
        {1+(E\times X)} & X
        \arrow["{[\nil,\cons]}", from=1-1, to=1-2]
        \arrow[from=1-1, to=2-1]
        \arrow["{=}"{description}, draw=none, from=1-1, to=2-2]
        \arrow[""{name=0, anchor=center, inner sep=0}, "f"', from=1-2, to=2-2]
        \arrow[""{name=1, anchor=center, inner sep=0}, "{\RecList\ X\ x\ g}"{pos=0.54}, bend left=49, from=1-2, to=2-2]
        \arrow["{[x, g]}"', from=2-1, to=2-2]
        \arrow["{=}"{marking}, draw=none, from=0, to=1]
      \end{tikzcd}\]

    \begin{proof}
      Let $(X: \U), (x: X), (g: E \to X \to X), (f: \List \to X)$ and $(l : \List)$. \\
      Assume we have a proof term of $(\MorphList\ \ \List\ \nil\ \cons \quad X\ x\ g \quad f)$.

      By looking at $(p:= \pr2\ l)$, which has type $\LimList(\pr1\ l)$, we obtain the following equation:

      \begin{align}
        f\ (\RecListr\ \List\ \cons\ \nil\ (\pr1\ l)) = \RecListr\ X\ g\ (\pr1\ l) \label{LimListPr2}
      \end{align}

      We derive the following:

      \begin{align*}
        f\ l \overset{\RecId} & {=} f\ (\RecList\ \List\ \nil\ \cons\ l) \hspace{4em} \\
                              & := f\ (\RecListr\ \List\ \nil\ \cons\ (\pr1\ l))      \\
        \overset{p}           & {=} \RecListr\ X\ x\ g\ (\pr1\ l)                     \\
                              & =: \RecList\ X\ x\ g\ l
      \end{align*}
      By functional extensionality, we conclude that $f = \RecList\ X\ x\ g$.
    \end{proof}
  \end{theorem}

  To summerize this section: we employed the technique of Awodey, Frey and Speight to encode the impredicative list type of the previous section togeter with a proof that the recursor is unique. This created an embedding $\Listr \hookrightarrow \List$. We showed that this definition was well defined by updating the $\nil$ and $\cons$ constructors. We showed that the $\beta$-rules and the $\eta$-rule are satisfied by this new definition.

  \section{List induction principle}

  \define{\IndList}{IndList}{\texttt{IndList}}

  We are now ready to show that we have an induction principle for the new $\List$ type. Just like induction for the natural numbers, the list induction principle is used to prove facts about \textit{all} lists. If some predicate $(P: \List \to \U)$ holds for $\nil$ and it is preserved under $\cons$, then we expect the predicate to hold for all lists.

  \begin{definition} We define the \textbf{list induction principle}. Let $(P: \List \to \U)$ be a predicate on lists. If $P$ holds for $\nil$ and is preserved under $\cons$, then $P$ holds for all lists. Formally, we have:
    \label{def:indlist}

    \begin{align*}
      \IndList\ P := (P\;\nil) \to \Bigl(\prod_{l: \List}\prod_{e: E}\ (P\ l) \to (P\ (\cons\ e\ l))\Bigr)\to \prod_{l: \List} P\ l
    \end{align*}

  \end{definition}

  To prove the list induction principle, we make use of a helper type $T := (\sum (l: \List). P\ l)$. This type is a pair of a list $l$ and a proof that $P$ holds for $l$. We show the induction principle in the following steps.

  \newcommand{\p}[1]{\texttt{p#1}}
  \newcommand{\h}[1]{\texttt{h#1}}

  \begin{minipage}[t]{0.66\textwidth}
    \begin{enumerate}
      \item We show that $T := (\sum (l: \List). P\ l)$ forms an $\mathcal{L}$-algebra.
      \item We have that $\pr1$ is an $\mathcal{L}$-morpism $(\pr1\ : T \to \List)$.
      \item By initiality of $\List$ we have an $\mathcal{L}$-morpism $(\texttt{u} : \List \to T)$.
      \item We have that $(\pr1\ \circ \texttt{u})$ is a $\mathcal{L}$-morphism from $\List$ to $\List$.
      \item We have that $\id{\List}$ is also an $\mathcal{L}$-morphism from $\List$ to $\List$.
      \item By the of $\List$ we conclude that $\pr1\ \circ u = \id{{}\List}$.
      \item For all $l$ we have that $\pr1\ (u\ l) = l$ and thus conclude that $(\pr2\ (u\ l): P\ l)$.
    \end{enumerate}
  \end{minipage}
  \begin{minipage}[t]{0.33\textwidth}
    \[\begin{tikzcd}
        \L(\List) \arrow[d, "\L(\texttt{u})"', dashed] \arrow[r, "{[\nil, \cons]}"] & \List \arrow[d, "\texttt{u}", dashed] \arrow[dd, "\id{{}\List}", dashed, bend left=49] \\
        \arrow[ur, phantom, "=" description]
        \L(T) \arrow[r, "{[\h1, \h2]}"] \arrow[d, "\L(\pr1)"']                  & T \arrow[d, "\pr1"]                                                                   \\
        \arrow[ur, phantom, "=" description]
        \L(\List) \arrow[r, "{[\nil, \cons]}"]                              & \List
      \end{tikzcd}\]
  \end{minipage}

  \begin{lemma} Let us assume a predicate $(P: \List \to \U)$. Furthermore, we assume the following:
    \label{lemma:algebra}

    \begin{align*}
       & \p0: P\ \nil                                                        \\
       & \p{n} : \prod_{l: \List}\prod_{e: E}\ (P\ l) \to (P\ (\cons\ e\ l))
    \end{align*}

    Let $(T := \sum (l: \List). P\ l)$. We have that $\tup{T, \tup{\h1, \h2}}$ forms an $\mathcal{L}$-algebra, where $\h1$ and $\h2$ are defined as follows:

    \begin{align*}
      \h1 & := \tup{\nil, \p0}                                                                  \\
      \h2 & := \lambda (e: E) (t: T). \tup{\cons\ e\ (\pr1\ t), \p{n}\ (\pr1\ t)\ e\ (\pr2\ t)}
    \end{align*}
    \begin{proof}
      We have to show that the following two equations hold:

      \begin{align}
        \pr1\ \h1            & = \nil \label{eq:pr1_h1}                   \\
        \pr1\ \circ (\h2\ e) & = (\cons\ e) \circ \pr1\ \label{eq:pr1_h2}
      \end{align}

      Both equations hold by simple $\beta$-reduction. For \cref{eq:pr1_h1} we have:

      \begin{align*}
        \pr1\ \h1 & := \pr1\ \tup{\nil, \p0} \betar \nil
      \end{align*}

      For \cref{eq:pr1_h2} we have:

      \begin{align*}
        \pr1\ \circ (\h2\ e) & := \pr1\ \tup{\cons\ e\ (\pr1\ t), \p{n}\ (\pr1\ t)\ e\ (\pr2\ t)} \\
                             & \betar \cons\ e\ (\pr1\ t) \qedhere
      \end{align*}
    \end{proof}
  \end{lemma}

  \begin{lemma} We have that $(\pr1\ : T \to \List)$ is an $\mathcal{L}$-morphism. Formally, we have that the following holds:
    \label{lemma:pr1_listmorphism}

    \begin{align}
      \bigl(\MorphList\ T\ \h1\ \h2 \quad \List\ \nil\ \cons \quad \pr1\bigr)
    \end{align}

    \begin{proof}
      We have to show that the following two equations hold:

      \begin{align}
        \pr1\ \h1            & = \nil \hspace{5em}  \label{morph_pr_1} \\
        \pr1\ \circ (\h2\ e) & = (\cons\ e) \circ f \label{morph_pr_2}
      \end{align}

      Both hold by simple $\beta$-reduction, where you need to instantiate \cref{morph_pr_2} with $(t : T)$.
    \end{proof}
  \end{lemma}

  We can now combine the previous two lemmas to show that the induction principle holds for $\List$.

  \begin{theorem}[$\IndList$] The list induction principle $\IndList$ described in \cref{def:indlist} holds for the $\List$ type. \label{thm:indlist}
    \begin{proof}
      Let us assume a predicate $(P: \List \to \U)$. Furthermore, we assume the following:

      \begin{align*}
         & \p0: P\ \nil                                                        \\
         & \p{n} : \prod_{l: \List}\prod_{e: E}\ (P\ l) \to (P\ (\cons\ e\ l))
      \end{align*}

      We follow the steps described before.

      \begin{enumerate}
        \item By \cref{lemma:algebra} we have that $T := (\sum (l: \List). P\ l)$ forms an $\mathcal{L}$-algebra.

        \item By \cref{lemma:pr1_listmorphism} we have that $(\pr1\ : T \to \List)$ is an $\mathcal{L}$-morphism.

        \item By initiality of $\List$, we have the unique morphism $(u : \List \to T)$ given by $u := (\RecList\ T\ \h1\ \h2)$.

        \item By a straightforward adaptation of \cref{comp_morph} we have that $(\pr1\ \circ \texttt{u})$ is an $\mathcal{L}$-morphism.

        \item By a straightforward adaptation of 0\cref{id_morph} we have that $\id{\List}$ is an $\mathcal{L}$-morphism.

        \item By \cref{thm:eta_list} we have that $\pr1\ \circ\ \texttt{u} = (\RecList\ \List\ \nil\ \cons)$. By \cref{reclist_id} we have that $(\RecList\ \List\ \nil\ \cons) = \id{\List}$. We combine these to obtain that $\pr1\ \circ\ \texttt{u} = \id{\List}$.

        \item Let $(l : \List)$ be any list. We want to show that $(P\ l)$ holds. We just saw that $\pr1\ (\texttt{u}\ l) = l$. We can thus obtain a proof of $(P\ l)$ by looking at the second projection $\bigl(\pr2\ (\texttt{u}\ l): P\ l\bigr)$.

      \end{enumerate}

    \end{proof}
  \end{theorem}

  \chapter{Quotients}
  \label{chap:quotients}

  In this chapter, we discuss the concept of quotient types. Quotient types are a way to define types where some values of a type $(D: \U)$ are identified by a relation $(R: D\to D\to \U)$. You can compare this to the concept of quotients in set theory. It is possible to define quotient types in impredivative type theory. Similarly to lists, this definition does satisfy the expected $\beta$-rule, but fails to satisfy the $\eta$-rule. We show that it is possible to extend the technique of \cite{encodings} to define quotients with an $\eta$-rule. We follow the same basic approach as for lists. We use this improved quotient type to show that the class-function, that lifts an element of the base type to an element of the quotient type, is indeed a surjective function. In \cref{chap:stream} we use the quotients as defined here to create stream types.

  \section{Impredicative quotient type}

  Quotient types can best be understood through an example.

  \define{\even}{even}{\texttt{even}}
  \define{\clsr}{cls}{\texttt{cls}^*}
  \begin{example}
    Let us create a quotient type that identifies all even numbers. In set theory, we would write this as $\N / \texttt{even} = \{[0], [1]\}$. We define the relation \texttt{even} as follows:

    \begin{align*}
      \even       & : \N \to \N \to \U                            \\
      \even\ n\ m & := \texttt{isEven}\ n \iff \texttt{isEven}\ m
    \end{align*}

    The quotient type $\texttt{quot}^*\ N\ \even$ should be similar to the set-theoretic quotient $\N / \even$. We want a function $(\clsr: \N \to \texttt{quot}\ \N\ \even)$ that maps a natural number to its `equivalence class' such that:

    \begin{align*}
      \forall (n, m: \N)\qquad \even\ n\ m \to \clsr\ n = \clsr\ m
    \end{align*}

  \end{example}

  Let us now give a more formal definition of quotient types. But first, we define a predicate that states that a function behaves the same on all equivalence classes of a relation $R$.

  \define{\EqCls}{EqCls}{\texttt{EqCls}}
  \begin{definition}
    A function $f$ behaves the same on all equivalence classes of $R$ if the following predicate holds:

    \begin{align*}
      \EqCls\ (f: D\to E)\ (R: D \to D \to \U) & := \prod (x, y: D). R\ x\ y \implies f\ x = f\ y
    \end{align*}
  \end{definition}

  \define{\quotr}{quotr}{\texttt{quot}^*}
  \begin{definition} We define \textbf{quotient types}. Let $(D: \U)$ and $(R: D \to D \to \U)$.

    \begin{align*}
      \quotr\ D\ R := \prod (C: \U). \prod (f: D \to C). \EqCls\ f\ R \to C \tag{formation}
    \end{align*}

    We create the class function as follows:

    \begin{align*}
      \clsr         & : D \to \quotr\ D\ R \tag{introduction}  \\
      \clsr\ (d: D) & := \lambda C. \lambda f. \lambda H. f\ d
    \end{align*}
  \end{definition}

  We are able to lift any function $(f: D \to E)$ that behaves the same on all equivalence classes to a function $(\overline{f}: \quotr\ D\ R \to E)$. This lifting is done by the recursor for quotient types.

  \define{\RecQuotr}{RecQuotr}{\texttt{rec}^*_{\texttt{q}}}
  \begin{definition}
    We define the \textbf{recursor for quotient types} as the function that lifts functions that behave the same on all equivalence classes.

    \begin{align*}
      \RecQuotr :           & \prod (C: \U). \prod (f: D\to C).         \tag{elimination} \\
                            & \prod (H: \EqCls\ f\ R). \prod (q: \quotr\ D\ R).\ C        \\
      \RecQuotr\ C\ f\ H\ q & := q\ C\ f\ H
    \end{align*}

  \end{definition}

  \begin{notation}
    We often write $\overline{f} := (\lambda q.\ \RecQuotr\ C\ f\ H\ q)$ if $C$ and $H$ are clear from the context.
  \end{notation}

  We expect that the lifted function satisfies $\overline{f} \circ \clsr = f$. This gives us the $\beta$-rule for quotient types as shown in \cref{beta-quot-cat}.

  \begin{figure}[H]
    \centering
    \begin{tikzcd}
      D \arrow[d, "f"'] \arrow[r, "\clsr"] & D / R \arrow[ld, "\overline{f}"] \arrow[ld, "=", phantom, shift right=4] \\
      C                                   &
    \end{tikzcd}
    \caption{Commutative diagram of the $\beta$-rule for quotient types.}
    \label{beta-quot-cat}
  \end{figure}

  \begin{lemma}
    The $\beta$-rule for quotient types is satisfied. For $(D: \U), (R: D \to D \to \U), (f: D \to C)$ and $(H: (\EqCls\ f\ R))$ we have that $\overline{f} \circ \clsr = f$. In other words:

    \begin{align*}
      (\RecQuotr\ C\ f\ H) \circ \clsr = f \tag{computation}
    \end{align*}

    \begin{proof}
      Let $D, R, f$ and $H$ be as described above. Let $(x: D)$.

      \begin{align*}
        \overline{f}\ (\clsr\ x) & := \RecQuotr\ C\ f\ H\ (\clsr\ x) \\
                                 & \betar (\clsr\ x)\ C\ f\ H        \\
                                 & \betar f\ x \qedhere
      \end{align*}
    \end{proof}
  \end{lemma}

  \define{\lemquot}{lemquot}{\texttt{EqCls1}}
  \begin{lemma}[\lemquot]
    Let $(D: \U)$ and $(R: D \to D \to \U)$. For $(x, y: D)$ we have:
    \label{quotient_classes}

    \begin{align*}
      R\ x\ y \implies \clsr\ x = \clsr\ y
    \end{align*}

    \begin{proof}
      Let $D$ and $R$ be as described above. Let $(x, y: D)$ and assume $(R\ x\ y)$. We initialize $(\clsr\ x)$ using $(C: \U)$, $(f: D \to C)$ and $(H: (\EqCls\ f\ R))$. The statement follows by function extensionality.

      \begin{align*}
        (\clsr\ x)\ C\ f\ H & \betar f\ x \overset{H}{=} f\ y \\
                            & \betar (\clsr\ y)\ C\ f\ H      \\
                            & \betar f\ y \qedhere
      \end{align*}
    \end{proof}
  \end{lemma}

  \begin{note}
    \label{note:equivalence}
    Before we continue, here is a quick note about the relation $R$. From set theory, we expect that $R$ needs to be an equivalence relation. We expect it to be \textit{reflexive}, \textit{symmetric}, and \textit{transitive}. However, none of the definitions given in this section enforce this expectation. Nevertheless, by definition, any function $f$ that acts on elements of $\quotr D\ R$ needs to satisfy $\EqCls\ f\ R$. If $R$ is not reflexive, so that we don't have $R\ x\ x$ for some $(x: D)$, we obviously still have that $f\ x = f\ x$. The same goes for symmetry: if $R\ x\ y$ holds, but, $R\ y\ x$ does not hold for some $(x, y : D)$, we still have that $R\ x\ y \implies f\ x = f\ y$ by $\EqCls\ f\ R$. Therefore, we also have $f\ y = f\ x$, by the reflexivity of $=$. To complete this reasoning, if we have $R\ x\ y$ and $R\ y\ z$, but don't have $R\ x\ z$ for some $(x, y, z: D)$, we still have that $f\ x = f\ y$ and $f\ y = f\ z$. By transitivity of $=$ we thus have that $f\ x = f\ z$. The behavior of $\quotr$ can only be observed as if $R$ were an equivalence relation. This definition of quotients is thus built on the universal property in a categorical setting, rather than the set-theoretic equivalence class definition. Taking the reflexive, symmetric, and transitive closure of $R$ would thus not matter (although we shall not explore this avenue further).
  \end{note}

  We would hope to have that the function $\overline{f}$ is indeed \textit{the} lifted function and that any other function $h$ that satisfies the $\beta$-rule is equal to this $f$. This uniqueness requirement or $\eta$-rule is unfortunately not satisfied. In the next section, we take a look at a categorical definition of quotients. Using this definition and its uniqueness requirement, we can apply the same encoding technique as before to create a quotient type that satisfies this uniqueness property.

  \section{Categorical quotients}

  We take a quick look at the categorical requirements for a quotient type that satisfies the $\eta$-rule. Just like we have seen for lists, we want quotients to satisfy the following uniqueness rule:

  \begin{figure}[H]
    \centering
    \begin{tikzcd}
      D \arrow[d, "f"'] \arrow[r, "\clsr"] & D / R \arrow[ld, "\exists! \overline{f}", dashed] \arrow[ld, "=", phantom, shift right=4] \\
      C                                   &
    \end{tikzcd}
    \caption{Uniqueness rule for quotients.}
    \label{quot_uniq}
  \end{figure}

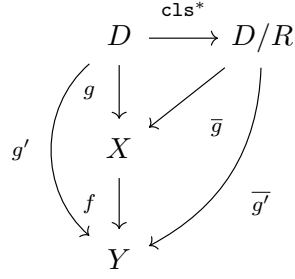
\begin{figure}[H]
    \centering
    \begin{tikzcd}
      D \arrow[d, "g"'] \arrow[r, "\clsr"] \arrow[dd, "g'"', bend right=49] & D/R \arrow[ld, "\overline{g}"] \arrow[ldd, "\overline{g'}", bend left] \\
      X \arrow[d, "f"']                                                    &                                                                        \\
      Y                                                                    &
    \end{tikzcd}
    \caption{Commutative diagram of the $\eta$-rule for quotient types.}
    \label{quot_eta}
  \end{figure}

  Just as seen before, we can give an alternative definition. Suppose we have morphisms $(g: D\to X)$ and $(g': D\to Y)$ that both satisfy the additional requirement that $(\EqCls\ g\ R)$ and $(\EqCls\ g'\ R)$. By the $\beta$-rule, we know that $\overline{g}\circ \clsr = g$ and $\overline{g'}\circ \clsr = g'$. If we have a morphism $(f: X \to Y)$, such that $g' = f \circ g$, we want to have that $\overline{g'} = f \circ \overline{g}$ (the bottom right triangle in \cref{quot_eta}). This is the $\eta$-rule for quotient types.

  \section{Impredicative encoding of quotients}
  \label{encoding-quot}

  We are now ready to define an impredicative encoding that satisfies the $\eta$-rule. Note that we follow the same approach as we did for encoding $\List$ in the previous chapter. We encode the uniqueness rule of \cref{quot_eta}. As an additional requirement, we need to have that the morphisms satisfy the $\EqCls$ predicate.

  \define{\quot}{quot}{\texttt{quot}}
  \define{\LimQuot}{LimQuot}{\texttt{LimQuot}}
  \define{\MorphQuot}{MorphQuot}{\texttt{MorphQuot}}
  \define{\RecQuot}{RecQuot}{\texttt{rec}_{\texttt{q}}}

  \begin{definition} We define an \textbf{inductive quotient type} $\quot$.
    \hspace{2em}\\
    \begin{align*}
      \quot\ (D: \U)\ (R: D \to D \to \U) & := \sum (q: \quotr\ D\ R). \LimQuot\ q        \tag{formation}        \\
      \LimQuot\ (q: \quotr\ D\ R)         & := \prod (X, Y: \U). \prod \big(\substack{g : D\to X                 \\ g': D \to Y}\big). \prod (f: X \to Y). \\
                                          & (H: (\EqCls\ g\ R)) \land (H': (\EqCls\ g'\ R)) \land f \circ g = g' \\
                                          & \hspace{9em}\implies                                                 \\
                                          & f\ (\RecQuotr\ E\ g\ H\ q) = \RecQuotr\ E\ g'\ H'\ q
    \end{align*}

    We define the quotient recursor as follows:

    \begin{align*}
      \RecQuot\ C\ f\ H\ q & := \RecQuotr\ C\ f\ H\ (\pr1\ q)
    \end{align*}
  \end{definition}

  \define{\cls}{cls}{\texttt{cls}}

  Note the similarity to the definition of $\List$. To define the new $\cls$ function, we need to show that $(\LimQuot\ (\cls\ d))$ holds for all $(d: D)$.

  \define{\LimQuotCls}{LimQuotCls}{\texttt{LimQuotCls}}

  \begin{lemma}[\LimQuotCls]
    For all $(D: \U)$, we have that $(\LimQuot\ (\cls\ d))$ holds.
    \begin{proof}
      Let $(d: D)$. Let $(X,Y: \U)$, $(g: D\to X)$, $(g': D\to Y)$ and $(f: X\to Y)$. Assume $(H: (\EqCls\ g\ R))$, $(H': (\EqCls\ g'\ R))$ and $(*):= f\circ g = g'$. We need to show that $f\ (\RecQuotr\ E\ g\ H\ (\clsr\ d)) = \RecQuotr\ E\ g'\ H'\ (\clsr\ d)$.

      \begin{align*}
        f\ (\RecQuotr\ E\ g\ H\ (\clsr\ d)) & \betar f\ ((\clsr\ d)\ E\ g\ H)                  \\
                                            & \betar f\ (g\ d)                                 \\
                                            & \overset{(*)}{=} g'\ d                           \\
                                            & \betar (\clsr\ d)\ E\ g'\ H'                     \\
                                            & \betar \RecQuotr\ E\ g'\ H'\ (\clsr\ d) \qedhere
      \end{align*}
    \end{proof}
  \end{lemma}

  \begin{definition}
    Let $(D: \U)$ and $(R: D \to D \to \U)$. We define the \textbf{class function} $\cls$ for $\quot\ D\ R$ as follows:

    \begin{align*}
      \cls\ (d: D) & := \tup{\clsr\ d,\ \LimQuotCls\ d} \tag{introduction}
    \end{align*}
  \end{definition}

  We also still have that $\cls\ x = \cls\ y$ for all $(x, y: D)$ such that $R\ x\ y$.

  \define{\EqClsCls}{EqClsCls}{\texttt{EqCls2}}
  \begin{lemma}[\EqClsCls]
    Let $(D: \U)$ and $(R: D \to D \to \U)$. We have that $(\EqCls\ \cls\ R)$ holds. In other words, for $(x, y: D)$ we have:
    \label{quotient_classes2}

    \begin{align*}
      R\ x\ y \implies \cls\ x = \cls\ y
    \end{align*}

    \begin{proof}
      Assume $(R\ x\ y)$. By \cref{quotient_classes} we have that $\clsr\ x = \clsr\ y$. We also have that $\LimQuot(\clsr\ x) = \LimQuot(\clsr\ y)$ by $\UIP$. \\
      We conclude that $\cls\ x = \tup{\clsr\ x, \LimQuot(\clsr\ x)} = \tup{\clsr\ y, \LimQuot(\clsr\ y)} = \cls\ y$.
    \end{proof}
  \end{lemma}

  We are now ready to show that the $\eta$-rule is satisfied by the new quotient type. We first need a few lemmas. We show that the $\beta$-rule remains satisfied for $\quot$. We then show that the identity case of the $\eta$-rule holds: $\overline{\cls} = \id{\quot\ D\ R}$. Finally, we show the complete $\eta$-rule.

  \begin{lemma}
    \label{beta-quot}
    The $\beta$-rule for $\quot$ is satisfied.
    Let $(D: \U)$ and $(R: D \to D \to \U)$ form a quotient $\quot\ D\ R$ and let $(C:\U)$, $(g: D \to C)$ and $(H: \EqCls(g, R))$ we have that $\overline{g} \circ \cls = g$. In other words, we have:

    \begin{align*}
      (\RecQuot\ C\ g\ H) \circ \cls = g \tag{computation}
    \end{align*}

    \begin{proof}
      Let $D, R, g$ and $H$ be as described above. Let $(d: D)$.

      \begin{align*}
        (\overline{g} \circ \cls)\ d & := \RecQuot\ C\ g\ H\ (\cls\ d)                                       \\
                                     & := \RecQuotr\ C\ g\ H\ (\pr1\ (\cls\ d))                              \\
                                     & \betar \RecQuotr\ C\ g\ H\ (\pr1\ \tup{\clsr\ d, \LimQuot(\clsr\ d)}) \\
                                     & \betar \RecQuotr\ C\ g\ H\ (\clsr\ d)                                 \\
                                     & \betar g\ d \qedhere
      \end{align*}

    \end{proof}
  \end{lemma}

  \define{\IdLift}{IdLift}{\texttt{IdLift}}

  \begin{lemma}[\IdLift]
    \label{idLift}
    Let $(D: \U)$ with $(R: D \to D \to \U)$ be a quotient $(D / R := \quot\ D\ R)$.
    We have $\overline{\cls} = \id{D/R}$.

    \begin{minipage}{.66\textwidth}
      \begin{align*}
        \RecQuot\ D/R\ \cls\ \EqClsCls = \id{D/R}
      \end{align*}
    \end{minipage}
    \begin{minipage}{.33\textwidth}
      \begin{tikzcd}
        D & {D/R} \\
        {D/R}
        \arrow["\cls", from=1-1, to=1-2]
        \arrow["\cls"', from=1-1, to=2-1]
        \arrow[""{name=0, anchor=center, inner sep=0}, "{{\id{D/R}}}", from=1-2, to=2-1]
        \arrow["{=}"{description, pos=0.7}, draw=none, from=1-1, to=0]
      \end{tikzcd}
    \end{minipage}

    \begin{proof}
      We first reduce the statement using function extensionality and $\beta$-reduction. In the $(\hookrightarrow)$ step, we use the fact that $\quotr\ D\ R \hookrightarrow \quot\ D\ R$ is an embedding.

      \begin{align}
                                  & \RecQuot\ D/R\ \cls\ \EqClsCls = \id{D/R} \overset{\funext}             & {\iff} \nonumber \\
        \forall\ (q: D/R). \qquad & \RecQuot\ D/R\ \cls\ \EqClsCls\ q  = q \overset{\hookrightarrow}        & {\iff} \nonumber \\
        \forall\ (q: D/R). \qquad & \ \pr1\ (\RecQuot\ D/R\ \cls\ \EqClsCls\ q) = \pr1\ q \overset{\funext} & {\iff} \nonumber \\
        \forall\ \substack{(q: D/R), (C:\U),                                                                                   \\(g: D \to C), (H: \EqCls\ g\ R)}. \qquad &\ \bigl(\pr1\ (\RecQuot\ D/R\ \cls\ \EqClsCls\ q)\bigr)\ C\ f\ H = (\pr1\ q)\ C\ f\ H \label{QuotIdReduc}
      \end{align}

      We want to apply the $(\LimQuot\ (\pr1\ q))$ proposition.  The function $((\RecList\ X\ x\ g): \List \to X)$ will play the role of $f$.

      \begin{enumerate}
        \item We have $(\EqCls\ g\ R)$ by $H$.
        \item We have $(\EqCls\ \cls\ R)$ by \cref{quotient_classes2}.
        \item We have $(\RecQuot\ C\ g\ H) \circ \cls = g$ by the $\beta$-rule of \cref{beta-quot}.
      \end{enumerate}

      We can thus use the $(\LimQuot\ (\pr1\ q))$ proposition to show that the following equation holds:

      \begin{align}
        (\RecQuot\ C\ g\ H)\ (\RecQuotr\ D/R\ \cls\ \EqClsCls\ (\pr1\ q)) = (\RecQuotr\ C\ g\ H\ (\pr1\ q)) \label{PrQuot1}
      \end{align}

      We conclude the proof by showing \cref{QuotIdReduc}. Let $(q: D/R), (C:\U), (f: D \to C)$, and $(H: \EqCls\ f\ R)$.

      \begin{align*}
        \bigl(\pr1\ (\RecQuot\ D/R\ \cls\ \EqClsCls\ q)\bigr)\ C\ g\ H & \betar \RecQuotr\ C\ g\ H\ \bigl(\pr1\ (\RecQuot\ D/R\ \cls\ \EqClsCls\ q)\bigr) \\
                                                                       & \betar \RecQuot\ C\ g\ H\ (\RecQuot\ D/R\ \cls\ \EqClsCls\ q)                    \\
                                                                       & = (\RecQuot\ C\ g\ H)\ (\RecQuot\ D/R\ \cls\ \EqClsCls\ q)                       \\
                                                                       & := (\RecQuot\ C\ g\ H)\ (\RecQuotr\ D/R\ \cls\ \EqClsCls\ (\pr1\ q))             \\
                                                                       & \overset{\ref{PrQuot1}}{=} (\RecQuotr\ C\ g\ H\ (\pr1\ q))                       \\
                                                                       & := (\pr1\ q)\ C\ g\ H\ \qedhere
      \end{align*}
    \end{proof}

  \end{lemma}

  Using the previous lemma, we show that the general $\eta$-rule is satisfied.

  \begin{minipage}[t]{.66\textwidth}
    \begin{theorem} The $\eta$-rule holds for $\quot$.
      \label{eta-quot}
      Let $(D: \U)$ with $(R: D \to D \to \U)$ be a quotient $D / R := \quot\ D\ R$.
      Suppose $(g: D \to C)$ with $(H: (\EqCls\ g\ R))$ and $(C: \U)$. Furthermore, suppose we have a function $(f: D/R \to C)$ with $f \circ \cls = g$. Then we have that $f = \overline{g}$. In formulas:

      \begin{align*}
        \prod (H: \EqCls(g, R)). (f \circ \cls = g) \implies f = \RecQuot\ C\ g\ H \tag{uniqueness}
      \end{align*}

    \end{theorem}
  \end{minipage}
  \begin{minipage}[t]{.33\textwidth}
    \[\begin{tikzcd}
        D & {D/R} \\
        C
        \arrow["\cls", from=1-1, to=1-2]
        \arrow["f"', from=1-1, to=2-1]
        \arrow[""{name=0, anchor=center, inner sep=0}, "{\exists! \overline{f}}", dashed, from=1-2, to=2-1]
        \arrow["{=}"{description, pos=0.1}, draw=none, from=0, to=1-1]
      \end{tikzcd}\]
  \end{minipage}
  \begin{proof}
    Assume we have $D, R, C, g, H$ and $f$ as described. We want to show that $f = \RecQuot\ C\ g\ H$.

    Let $(\tup{q, p}: D/R)$ where $(q: \quotr D\ R)$ and $(p: \LimQuot(q))$. We have that $f \circ \cls = g$ by definition. Obviously we have $(H': (\EqCls\ f \circ \cls\ R))$. We thus have that all requirements of $p$ are satisfied. The statement follows by function extensionality.

    \begin{align*}
      f\ \tup{q, p} \overset{\IdLift} & {=} f\ (\RecQuot\ D/R\ \cls\ \EqClsCls\ \tup{q, p}) \\
      \overset{p}                     & {=}\ \RecQuot\ C\ g\ H\ \tup{q, p} \qedhere
    \end{align*}

  \end{proof}
  \section{Quotient induction priciple}

  Whenever we can show that some property $P$ holds for all $((\cls\ d): D)$ we of course want to obtain that $P$ holds for all $(q: D/R)$. This fact does not hold for the $\quotr$ type. The improved $\quot$ type does satisfy this property since it follows from the $\eta$-rule. We show this in the following theorem.

  \begin{theorem}
    \label{quot_induction}
    Let $(D: \U)$ with $(R: D \to D \to \U)$ be a quotient $D / R := \quot\ D\ R$. Let $(P: D/R \to \U)$ be some property. We then have the that:

    \begin{align*}
      \prod (x: D).\ P\ (\cls\ x) \implies \prod (q: D/R).\ P\ q
    \end{align*}

    \begin{proof}
      Assume we have $(A: \prod (x: D).\ P\ (\cls\ x))$.
      Define $(\sum (q: D/R). P\ y)$ and $g(x) := \tup{\cls\ x, A\ (\cls\ x)}$.

      In this proof, we shall make the following diagram commute in all configurations:

      \begin{figure}[H]
        \centering
        \begin{tikzcd}
          D \arrow[dd, "g"] \arrow[rr, "\cls"] \arrow[dddd, "\cls"', bend right=60] &    & D/R \arrow[lldd, "\bar{g}"] \arrow[lldddd, "\overline{\cls}", bend left] \\
          & {} &                                                                                                       \\
          \sum (q: D/R). P\ y \arrow[dd, "\pr1"]                                                                                                   &    &                                                                                                       \\
          &    &                                                                                                       \\
          D/R                                                                                                                                        &    &
        \end{tikzcd}
      \end{figure}

      We have that $R\ x\ y \implies g\ x = g\ y$ by \cref{quotient_classes} and because $A\ x$ is a proof of $P\ x$. We thus have that $\overline{g}$ is well-defined and we have that $\overline{g} \circ \cls = g$. Furthermore, we notice that $(\cls: D \to D/R)$ can be lifed to $(\overline{\cls}: D/R \to D/R)$ such that $\overline{\cls} \circ \cls = \cls$ which is well defined by \cref{quotient_classes}. Finally, we notice that $\cls = \pr1 \circ g$. By \cref{idLift} we have that $\overline{\cls} = \id{D/R}$. By \cref{eta-quot} we have that $\pr1 \circ \overline{g} = \id{D/R}$. We fix a quotient $(q: D/R)$. We then have that $\pr{1} (\overline{g}\ q) = q$. But then we find that $\pr2\ (\overline{g}\ q)$ is of type $P\ q$. We conclude that we have a proof of type $P\ q$ for any $(q: D/R)$.
    \end{proof}
  \end{theorem}

  In the next chapter, we take a look at the coinductive stream type. Like before, we encode the uniqueness property of streams within their definition. In our continuous quest to satisfy the $\eta$-rules, we shall make use of the improved quotient types to show that our stream type satisfies the $\eta$-rule of streams.

  \chapter{Streams}
  \label{chap:stream}

  In this chapter, we develop an impredicative encoding of a stream type that satisfies the expected $\beta$-rules and the $\eta$-rule. These streams are an example of a coinductive type. Coinductive types are the dual notion of inductive types and describe (possibly) infinite data structures. These streams, or infinite lists, shall range over some element type $(E : \U)$. As before, we assume that $(\Gamma \vdash E : \U)$ holds globally. The definitions in this chapter can be straightforwardly generalized to any element type.

  To create streams, we need the concept of existential types. They are the type theoretical analog of existential quantifiers in set theory. We use them to hide the implementation of streams. We continue by giving a common impredicative definition of streams and show that the $\beta$-rules are satisfied for this definition. Next, we take a look at the categorical definition of coinductive streams using a suitable final coalgebra. We combine these definitions into an impredicative encoding of a stream type that also satisfies the $\eta$-rule and prove the coinduction principle for streams.

  \section{Existential types}
  \label{existential-types}

  In System F, it is possible to create existential types \cite{ChurchScottDataTypes}. They can be used to \textit{hide} the implementation of a type and only give access to an interface. An existential type $\exists (X: \U). P(X)$ is constructed using some type $(X: \U)$ and a type family $(P : \U \to \U)$. Just like an existential quantifier, we can create an instance of the existential type by giving a sample $(X: \U)$ and a witness $(p : P(X))$. Once $X$ has been packed, you cannot retrieve it anymore. In this way, the implementation (the concrete value of $X$) is hidden from the outside world and only the interface (the type family $P$) is accessible. This means we can operate on $P$, but not retrieve information about $X$.

  \define{\pack}{pack}{\texttt{pack}}
  \define{\RecExists}{RecExists}{\texttt{rec}_\exists}
  \begin{definition} We define the \textbf{existential type}. Let $(P : \U \to \U)$.

    \begin{align*}
      \exists (X: \U). P(X) & := \prod (Y: \U). \bigl(\prod (X: \U). P(X) \to Y\bigr) \to Y \tag{formation}
    \end{align*}

    We denote these existential types using the shorthand notation $\exists X.P := \exists (X: \U). P(X)$. Inhabitants of these existential types are created using the $\pack$ function.

    \begin{align*}
      \pack\      & : \prod (X: \U). \prod (p: P(X)).\ \exists X.A \tag{introduction}             \\
      \pack\ X\ p & := \lambda (Y: \U). \lambda \bigl(k: \prod (X: \U). P(X) \to Y\bigr). k\ X\ p
    \end{align*}

    We can operate on the existential type using the following recursor. Note that $(\widetilde{X}: \U)$, has no connection to the original value of $(X: \U)$ that was used to $\pack$ the existential type. Therefore, the output type $(Y : \U)$ cannot depend on the concrete implementation of $X$.

    \begin{align*}
      \RecExists          & := \prod (Y: \U). \bigl(\prod (\widetilde{X}: \U). P(\widetilde{X}) \to Y\bigr). \exists X.P \to Y \tag{elimination} \\
      \RecExists\ Y\ f\ e & := e\ Y\ f
    \end{align*}

    We can adapt this definition for bounded existential types like $\exists (a: A). P(a)$ by replacing $(X: \U)$ with $(A: a)$ and taking the type $(P : A \to \U)$.
  \end{definition}

  This definition is best understood through an example. We sketch why we cannot retrieve the value of $X$ once it has been packed.

  \begin{example}
    Define $\exists X. (X \times X)$. We can create a pair of natural numbers and encapsulate it in the existential type. We thus have $(\pack\ \N\ \tup{3,3} : \exists X. (X \times X))$. When we try to get the value $3$ out of the existential type, we quickly see that this is not possible. The recursor only allows us to operate on the value for \textit{some} type $X$. We have forgotten that this type was $\N$. Below is a failed attempt to extract $3$ from the existential type. This will not type-check because $x$ is not of type $\N$.

    \setulcolor{red}
    \begin{align*}
      \RecExists\ \N\ (\lambda X. \lambda (\tup{x, y}: X \times X). \quad \fbox{\textcolor{red}{$x$}} \quad)\ (\pack\ \N\ \tup{3,3})
    \end{align*}
  \end{example}

  \begin{example} We define the existential type $\exists X. X \times (X \to X)$ that encapsulates the behaviour of the natural numbers. This definition hides the concrete implementation from the interface. We have for example $N := \pack\ \N^*\ \tup{\texttt{0}^*, \texttt{S}^*}$. When we operate on $N$, there is no way to see that we used $\N$. If we somehow define another implementation of the natural numbers, for example using binary lists, we can use $\pack$ to hide this implementation.
  \end{example}

  Before we continue, we first have to add an axiom about equality on existential types. When we \textit{unpack} an existential type $(t: \exists X. P)$ using the recursor $\RecExists$, we obtain values $(X: \U)$ and $(p: P(X))$. These values are unrelated to the values we used to create the member $t$. This is by design. There is however a problem. If we use the $\pack$ function, to re-package the $X$ and $p$ values $t':= \pack\ X\ p$, we need to have that $t = t'$. In other words, if we unpack an existential type $(t: \exists X. P)$, and then re-pack it, we would like it to be equal to $t$.

  \define{\ExistsId}{ExistsId}{\texttt{ExistsId}}
  \begin{axiom} Let $(P: \U \to U)$. If we apply the pack function to the recursor, this yields the identity.
    \label{axiom:existsid}

    \begin{align*}
      \ExistsId \quad : \quad \RecExists\ (\exists X. P)\ \pack = \id{\exists X. P}
    \end{align*}
  \end{axiom}

  We shall use this axiom in \cref{stream_id}, where we also give some more details. We do not exclude that this axiom can be proven as a theorem by creating an improved variant of the existential type. This gives a good opportunity for further research.

  \section{Impredicative stream type}
  \label{impred:stream}
  \define{\Streamr}{Streamr}{\texttt{Stream}^*}

  Like before, we first give an impredicative definition of a stream type. We will show that this definition satisfies the computation rules ($\beta$-rules).

  \define{\hdr}{hd}{\texttt{hd}^*}
  \define{\tlr}{tl}{\texttt{tl}^*}
  \begin{definition} The \textbf{impredicative stream type}.

    \begin{align*}
      \Streamr & := \exists (X:\U). X \times (X \to E) \times (X \to X) \tag{formation}
    \end{align*}

    We first give the elimination rules for the stream type. The $\hdr$ function gives the head of the stream. The $\tlr$ function gives the tail of the stream. Both make use of the existential recursor $\RecExists$ to look inside the existential type. The tail function furthermore uses the $\pack$ function to create a new inhabitant of the existential type.

    \begin{align*}
      \hdr(s: \Streamr) & := \RecExists\ E\ (\lambda X. \lambda \tup{x, h, t}.\ h\ x)\ s                                        & : & \ E \tag{elimination} \\
      \tlr(s: \Streamr) & := \RecExists\ E\ (\lambda X. \lambda \tup{x, h, t}.\ \pack\ X\ \tup{t\ x, h,t})\ s \tag{elimination} & : & \ \Streamr
    \end{align*}
  \end{definition}

  A stream is build using some internal type $(X: \U)$. It consists of triples $\tup{x, h, t}$ where $x$ is some element to boot-strap the stream, $h$ is a function that computes the head of the stream and $t$ is a function that computes the tail of the stream. For example, we could have $\tup{0, \lambda n.n, \lambda n. n+1}$, where we use $\N$ both internally as our type $X$ and externally as our type $E$. To compute the head of this stream, we use the second element of the triple $(\lambda n. n)$ and apply it to the first element $0$ to obtain $(\lambda n.n)\ 0 \betar 0$. To obtain the tail of this stream, we use the third element $(\lambda n. n+1)$ to update the first element $((\lambda n. n+1)\ 0 \betar 1)$. To construct a stream, we define the corecursor. This is the dual notion of a recursor and is used not to destruct a type, but to construct one. We use the $\pack$ function to create an inhabitant of the existential type.

  \define{\CoRecStreamr}{CoRecStreamr}{\texttt{corec}^*_{\texttt{s}}}
  \begin{definition} The \textbf{impredicative stream corecursor}.

    \begin{align*}
      \CoRecStreamr             & : \prod (X : \U). (X \to E) \to (X \to X) \to X \to \Streamr \tag{introduction} \\
      \CoRecStreamr\ X\ h\ t\ x & := \pack\ X\ \tup{x, h, t}
    \end{align*}
  \end{definition}

  We can check that the introduction rule (the corecursor) and elemination rules (the head and tail functions) compute as one would expect.

  \begin{proposition} The stream corecursor satisfies the following $\beta$-rules. Let $(X: \U)$, $(h: X \to E)$, $(t: X \to X)$ and $(x: X)$.

    \begin{align*}
      \hdr (\CoRecStreamr\ X\ h\ t\ x) & \betar h\ x \tag{computation}                           \\
      \tlr (\CoRecStreamr\ X\ h\ t\ x) & \betar \CoRecStreamr\ X\ h\ t\ (t\ x) \tag{computation}
    \end{align*}

    \label{streamr_beta_rules}
    \begin{proof} We unfold the definitions of $\RecExists$, $\pack$, $\hdr$ and $\tlr$. To distinguish between between the outer variables $X$, $x$, $h$ and $t$, we rename the bounded inner variables to $A$, $a$, $b$ and $c$.

      \begin{align*}
        \hdr (\CoRecStreamr\ X\ h\ t\ x) & := \RecExists\ E\ (\lambda A. \lambda \tup{a, b, c}.\ b\ a)\ (\CoRecStreamr\ X\ h\ t\ x)       \\
                                         & := (\CoRecStreamr\ X\ h\ t\ x)\ E\ (\lambda X. \lambda \tup{a, b, c}.\ (b\ a))                 \\
                                         & := (\pack\ X\ \tup{x, h, t})\ E\ (\lambda X. \lambda \tup{a, b, c}.\ (b\ a))                   \\
                                         & := (\lambda Y. \lambda k. k\ X\ \tup{x, h, t})\ E\ (\lambda X. \lambda \tup{a, b, c}.\ (b\ a)) \\
                                         & \betar (\lambda X. \lambda \tup{a, b, c}.\ (b\ a))\ X\ \tup{x, h, t}                           \\
                                         & \betar h\ x
      \end{align*}
      \begin{align*}
        \tlr (\CoRecStreamr\ X\ h\ t\ x) & := \RecExists\ E\ (\lambda A. \lambda \tup{a, b, c}.\ \pack\ A\ \tup{c\ a, b, c})\ (\CoRecStreamr\ X\ h\ t\ x)      \\
                                         & :=  (\CoRecStreamr\ X\ h\ t\ x)\ E\ (\lambda A. \lambda \tup{a, b, c}.\ \pack\ A\ \tup{c\ a, b, c})                 \\
                                         & :=  (\pack\ X\ \tup{x, h, t})\ E\ (\lambda A. \lambda \tup{a, b, c}.\ \pack\ A\ \tup{c\ a, b, c})                   \\
                                         & :=  (\lambda Y. \lambda k. k\ X\ \tup{x, h, t})\ E\ (\lambda A. \lambda \tup{a, b, c}.\ \pack\ A\ \tup{c\ a, b, c}) \\
                                         & \betar (\lambda A. \lambda \tup{a, b, c}.\ \pack\ A\ \tup{c\ a, b, c})\ X\ \tup{x, h, t}                            \\
                                         & \betar \pack\ X\ \tup{t\ x, h, t}                                                                                   \\
                                         & =: \CoRecStreamr\ X\ h\ t\ (t\ x) \qedhere
      \end{align*}
    \end{proof}
  \end{proposition}

  We show a few examples of how to define streams using the corecursor.

  \define{\zeroes}{zeroes}{\texttt{zeroes}}
  \define{\ones}{ones}{\texttt{ones}}
  \define{\incr}{incr}{\texttt{incr}}
  \begin{example}
    Let $E := \N$. We can define the following streams:

    \begin{align*}
      \zeroes & := \CoRecStreamr\ \N\ (\lambda n. n)\ (\lambda n. 0)\ 0     \\
      \ones   & := \CoRecStreamr\ \N\ (\lambda n. n)\ (\lambda n. 1)\ 0     \\
      \incr   & := \CoRecStreamr\ \N\ (\lambda n. n)\ (\lambda n. n + 1)\ 0
    \end{align*}

    Here $\zeroes$ is an infinite stream of zeroes, $\ones$ is an infinite stream of ones and $\incr$ is an infinite stream of natural numbers starting at zero and incrementing by one each time. We can see that $\incr$ behaves as expected by evaluating the first few elements of the stream using the $\beta$-rules of \cref{streamr_beta_rules}.

    \begin{align*}
      \hdr\ \tlr\ \incr & := \hdr \Bigl(\tlr\ (\CoRecStreamr\ \N\ (\lambda n. n)\ (\lambda n. n + 1)\ 0)\bigr)                               \\
                        & \betar \hdr \Bigl( \CoRecStreamr\ \N\ (\lambda n. n)\ (\lambda n. n + 1)\ \bigl((\lambda n. n + 1)\ 0\bigr) \Bigr) \\
                        & \betar \hdr \Bigl( \CoRecStreamr\ \N\ (\lambda n. n)\ (\lambda n. n + 1)\ 1 \Bigr)                                 \\
                        & \betar (\lambda n. n)\ 1 \betar 1
    \end{align*}

  \end{example}

  We can also use $\CoRecStreamr$ to define functions on streams.

  \define{\squareStream}{squareStream}{\texttt{square}}
  \begin{example}
    Let $E := \N$. We define a function $\squareStream$ that takes a stream of natural numbers and returns a stream of the squares of those numbers.

    \begin{align*}
      \squareStream\ (s: \Streamr) & := \CoRecStreamr\ \Streamr\ (\lambda x. (\hdr\ x)*(\hdr\ x))\ (\lambda x. \tlr\ x)\ s
    \end{align*}

  \end{example}

  At a first glance, these stream types seem to be working quite well. We can construct them using the corecursor and destruct them using the head and tail functions. We have a way to compute using these streams because the $\beta$-rules are satisfied. However, we will see that the stream type does not satisfy the $\eta$-rule. If we construct two streams that \textit{behave} exactly the same, but use a different internal representation, we would like to have that these streams are \textit{propositionally} equal. Take for example $\zeroes$ and $\bigl(\CoRecStreamr\ \N\ (\lambda n. n * 2)\ (\lambda n. 0)\ 0\bigr)$. It is obvious that these streams will behave the same, therefore we would like to have an equality type between these different representations. This problem is usually solved using \textit{bisimulation}, sometimes called \textit{coinduction} \cite{Jacobs_Rutten_2011}. Using the bisimulation principle, two streams that \textit{behave} the same, actually \textit{are} the same (propositionally). In the remainder of this chapter, we construct an impredicative encoding that satisfies the $\eta$-rule and use this to prove the bisimulation principle.

  \section{Categorical streams}

  Within category theory, we can define streams as the final co-algebra of the endofunctor $\S$. We instantiate the definitions of \cref{coalgebra}. We again leave the category unspecified, but one can think of the category $\textbf{Set}$.

  \begin{align*}
    \S(X)          & := E \times X         \\
    \S(f: X \to Y) & := \tup{\text{id}, f}
  \end{align*}

  As shown in \cref{coalg_category}, we have a category of $\S$-coalgebras denoted as $\S$-\textbf{CoAlg} with objects $\tup{X, \tup{h,t}}$ where $X$ is a set and $(h: X \to E)$ and $(t: X \to X)$ are functions. Morphisms in $\S$-\textbf{CoAlg} satisfy the following commutative diagram:

  \begin{figure}[h]
    \begin{center}
      \begin{tikzcd}
        X & {E \times X} \\
        Y & {E \times Y}
        \arrow["{\tup{h, t}}", from=1-1, to=1-2]
        \arrow["f"', from=1-1, to=2-1]
        \arrow["{\S(f)}", from=1-2, to=2-2]
        \arrow["{=}"{description, pos=0.6}, shift left, draw=none, from=2-1, to=1-2]
        \arrow["{\tup{h',t'}}", from=2-1, to=2-2]
      \end{tikzcd}
    \end{center}
    \caption{Commutative diagram of $\S$-\textbf{CoAlg} morphisms.}
    \label{comm_coalg_stream}
  \end{figure}

  In formulas this comes down to the following:

  \begin{align*}
    h' \circ f = h \land t' \circ f = f \circ t
  \end{align*}

  The final object $\tup{F, \tup{\text{hd},\text{tl}}}$ of the category of $\S$-coalgebras $\S$-\textbf{CoAlg} has exectly one morphism $\bigl(u_X: \tup{X, \tup{h,t}} \to \tup{F, \tup{\text{hd},\text{tl}}}\bigr)$ for each object $\tup{X, \tup{h,t}}$ in the category $\S$-\textbf{CoAlg}. In addition to \cref{comm_coalg_stream}, these morphisms need to satisfy the uniqueness requirement of \cref{uniqueness_rules_co}. We make use of requirement $2)$. In formulas, this boils down to the following: for each $\S$-\textbf{CoAlg} morphism $(f: X \to Y)$, we have that $u_Y \circ f = u_X$.

  \begin{figure}[h]
    \begin{center}
      \[\begin{tikzcd}
          X & {E \times X} \\
          Y & {E \times Y} \\
          F & {E \times F}
          \arrow["{{\tup{h,t}}}", from=1-1, to=1-2]
          \arrow["f"', from=1-1, to=2-1]
          \arrow["{{=}}"{description}, draw=none, from=1-1, to=2-2]
          \arrow[""{name=0, anchor=center, inner sep=0}, "{{u_X}}"', bend right=55, from=1-1, to=3-1]
          \arrow["{{\S(f)}}", from=1-2, to=2-2]
          \arrow["{{\tup{h',t'}}}", from=2-1, to=2-2]
          \arrow["{{u_Y}}"', from=2-1, to=3-1]
          \arrow["{{=}}"{description}, draw=none, from=2-1, to=3-2]
          \arrow[from=2-2, to=3-2]
          \arrow["{{\tup{\text{hd},\text{tl}}}}", from=3-1, to=3-2]
          \arrow["{=}"{marking}, draw=none, from=2-1, to=0]
        \end{tikzcd}\]
    \end{center}
    \caption{Diagram of the final $\S$-coalgebra.}
    \label{final_coalg_stream}
  \end{figure}

  \section{Impredicative encoding of streams}

  \define{\MorphStream}{MorphStream}{\texttt{MorphStream}}

  As we saw before with lists, we shall now translate the categorical notion of final $\S$-coalgebra back to type theory. We define a predicate that states that $(f: X \to Y)$ forms an $\S$-coalgebra morphism.

  \begin{definition}
    Let $(X, Y: \U)$, $(h: X \to E)$, $(t: X \to X)$, $(h': Y \to E)$, $(t': Y \to Y)$ and $(f: X \to Y)$. We define the predicate $\MorphStream$ that states that $f$ forms an $\S$-coalgebra morphism. This predicate mirrors the commutative diagram from the categorical definition in \cref{comm_coalg_stream}.

    \begin{align*}
      \Bigl(\MorphStream\ X\ h\ t\ \quad Y\ h'\ t'\ \quad f: X \to Y :=  h' \circ f = h\ \land\ t' \circ f = f \circ t \Bigr)
    \end{align*}
  \end{definition}

  Let us take a moment to think about what we want to achieve. We want to define an encoding of $\Streamr$ that satisfies the $\eta$-rule. This $\eta$-rule is taken from the final coalgebra of $\S$ as shown in \cref{final_coalg_stream}. Given a morphism $(f: X \to Y)$, we thus want that $u_Y \circ f = u_X$ for morphisms $(u_X : X \to F)$ and $(u_Y : Y \to F)$. In the case of $\List$, the morphisms $u_X$ and $u_Y$ were given by the recursor. Taking the dual notion, we can see that these morphisms are created using the corecursor for $\Streamr$. We thus have $u_X := \CoRecStreamr\ X\ h\ t$ and $u_Y := \CoRecStreamr\ Y\ h'\ t'$.

  In the case of $\List$, we created a subtype of $\Listr$, by pairing each list with a proof that the list satisfies the $\eta$-rule. This made $\Listr$ an embedding of $\List$. In the case of streams, we take the dual notion of an embedding: a quotient. We explain the duality and the general constructions more in \cref{chap:generalization}.

  The quotient should equate two streams that are related by $u_Y \circ f = u_X$. To define this quotient, we create a relation $\texttt{CoLimStr}$. We relate two streams $\sigma$ and $\tau$ if they can be translated into each other by some $\S$-morphism $f$. The relation $\texttt{CoLimStr}$ is given below.

  \define{\CoLimStr}{CoLimStr}{\texttt{CoLimStr}}
  \begin{definition} We define the relation $\CoLimStr$ between streams $(\sigma, \tau : \Streamr)$ where $(\CoLimStr\ \sigma\ \tau)$ holds if $\sigma$ and $\tau$ are related by some $\S$-coalgebra morphism.
    \label{def:colimstr}

    \begin{gather}
      \CoLimStr\ \sigma\ \tau := \exists (X, Y: \U). \exists \substack{(h: X\to E)\\(h': Y\to E)}. \exists\substack{(t: X\to X)\\(t':Y\to Y)}. \exists (f: X\to Y). \exists (x: X)\\
      \bigl(\MorphStream\ X\ h\ t \quad Y\ h'\ t' \quad f\bigr)\ \land \\
      \sigma = \CoRecStreamr\ X\ h\ t\ x\ \land \\
      \tau = \CoRecStreamr\ Y\ h'\ t'\ (f\ x)
    \end{gather}
  \end{definition}

  \begin{notation}
    We use the infix notation $(\sigma\ \equiv\ \tau)$ to denote that $(\CoLimStr\ \sigma\ \tau)$ holds.
  \end{notation}

  \begin{lemma}
    The relation $\CoLimStr$ is neither \textit{symmetric} nor \textit{transitive}. By \cref{note:equivalence}, this does not hinder our further progress since the equality relation on $(\cls\ x)$ is an equivalence relation. When we speak about `the equivalence classes of $\CoLimStr$', we thus mean the equivalence classes of $(\cls\ x)$.
    \begin{proof}
      We elaborate more on why none of these properties hold in \cref{CoLimStr_not_eq}.
    \end{proof}
  \end{lemma}

  \define{\EqHd}{EqHd}{\texttt{EqHd}}
  \begin{lemma}[$\EqHd: (\EqCls\ \hdr\ \equiv)$] The head function $\hdr$ is constant on all equivalence classes of $\CoLimStr$.
    \label{eqhd}

    \begin{align*}
      \prod (\sigma, \tau: \Streamr).\ (\sigma \equiv \tau) \implies (\hdr\ \sigma = \hdr\ \tau)
    \end{align*}

    \begin{proof}
      Let $(\sigma, \tau: \Streamr)$ and suppose that $(\sigma \equiv \tau)$ holds. We then have that the following exists: $(X, Y: \U), \substack{(h: X\to E)\\(h': Y\to E)}, \substack{(t: X\to X)\\(t':Y\to Y)}, (f: X\to Y)$, and $(x: X)$ with

      \begin{align}
        \sigma     & = \CoRecStreamr\ X\ h\ t\ x\ \label{hdeq1}       \\
        \tau       & = \CoRecStreamr\ Y\ h'\ t'\ (f\ x) \label{hdeq2} \\
        h \circ f  & = h' \label{hdeq3}                               \\
        t' \circ f & = f \circ t \nonumber
      \end{align}

      We conclude the following:

      \begin{align*}
        \hdr \sigma \overset{\ref{hdeq1}} & {=} \hdr (\CoRecStreamr\ X\ h\ t\ x)            \\
                                          & \betar h\ x                                     \\
        \overset{\ref{hdeq3}}             & {=} h'\ (f\ x)                                  \\
                                          & \betar \hdr\ (\CoRecStreamr\ Y\ h'\ t'\ (f\ x)) \\
        \overset{\ref{hdeq2}}             & {=} \hdr\ \tau \qedhere
      \end{align*}
    \end{proof}
  \end{lemma}

  \define{\EqTl}{EqTl}{\texttt{EqTl}}
  \begin{lemma}[$\EqTl: (\EqCls\ (\cls \circ \tlr)\ \equiv)$] The lifted tail function $\cls \circ \tlr$ is constant on all equivalence classes of $\CoLimStr$.
    \label{eqtl}

    \begin{align*}
      \prod (\sigma, \tau: \Streamr).\  (\CoLimStr\ \sigma\ \tau) \ \to \ \cls\ (\tlr\ \sigma) = \cls\ (\tlr\ \tau)
    \end{align*}

    \begin{proof}
      Let $(\sigma, \tau: \Streamr)$ and suppose $(\CoLimStr\ \sigma\ \tau)$. We then have that the following exists: $(X, Y: \U), \substack{h: X\to E\\h': Y\to E}, \substack{t: X\to X\\t':Y\to Y}, (f: X\to Y)$ and $(x: Y)$ with:

      \begin{align}
        \sigma     & = \CoRecStreamr\ X\ h\ t\ x\ \label{tleq1}       \\
        \tau       & = \CoRecStreamr\ Y\ h'\ t'\ (f\ x) \label{tleq2} \\
        h \circ f  & = h' \nonumber                                   \\
        t' \circ f & = f \circ t \label{tleq4}
      \end{align}

      We  conclude the following:

      \begin{align*}
        \tlr \sigma \overset{\ref{tleq1}} & {=} \tlr (\CoRecStreamr\ X\ h\ t\ x)              \\
                                          & \betar \CoRecStreamr\ X\ h\ t\ (t\ x)             \\
                                          & \equiv \CoRecStreamr\ Y\ h'\ t'\ (f \circ (t\ x)) \\
        \overset{\ref{tleq4}}             & {=} \CoRecStreamr\ Y\ h'\ t'\ (t' \circ (f\ x))   \\
                                          & \betar \tlr\ (\CoRecStreamr\ Y\ h'\ t'\ (f\ x))   \\
        \overset{\ref{tleq2}}             & {=} \tlr\ \tau
      \end{align*}
      Because we have that $\bigl(\CoLimStr\ (\tlr\ \sigma)\ (\tlr\ \tau)\bigr)$, it follows from \cref{quotient_classes2} that $\cls \circ \tlr$ is constant on all equivalence classes of $\CoLimStr$.
    \end{proof}
  \end{lemma}

  We are now ready to define the impredicative encoding of coinductive streams as the quotient of the $\CoLimStr$ equivalence relation. We use our previously defined $\quot$ type.

  \define{\Stream}{Stream}{\texttt{Stream}}
  \define{\hd}{hd}{\texttt{hd}}
  \define{\tl}{tl}{\texttt{tl}}
  \define{\CoRecStream}{CoRecStream}{\texttt{corec}_{\texttt{s}}}

  \begin{definition}
    We define a \textbf{coinductive stream type} $\Stream$.

    \begin{align*}
      \Stream & := \quot\ \Streamr\ \CoLimStr
    \end{align*}

    We define the new head and tail functions $\hd$ and $\tl$ as the following lifted functions:

    \begin{align*}
      \hd & := \overline{\hdr}            & := & \quad \RecQuot\ E\ \hdr\ \EqHd                    \\
      \tl & := \overline{\cls \circ \tlr} & := & \quad \RecQuot\ \Stream\ (\cls \circ \tlr)\ \EqTl
    \end{align*}

    We define the coinductive constructor $\CoRecStream$ by lifting the corecursive constructor $\CoRecStreamr$:

    \begin{align}
      \CoRecStream & : \prod (X : \U). (X \to E) \to (X \to X) \to X \to \Stream \nonumber \\
      \CoRecStream & := \cls \circ \CoRecStreamr  \label{def:corec}
    \end{align}

  \end{definition}

  In the remainder of this section, we show that $\Stream$ is the final $\S$-coalgebra in the category $\S$-\textbf{CoAlg}. We first show that this new stream type still satisfies the $\beta$-rules.

  \begin{proposition}The new stream corecursor satisfies the following $\beta$-rules:

    \begin{align*}
      \hd\ (\CoRecStream\ X\ h\ t\ x) & \betar h\ x                          \\
      \tl\ (\CoRecStream\ X\ h\ t\ x) & \betar \CoRecStream\ X\ h\ t\ (t\ x)
    \end{align*}

    \label{stream_beta_rules}
    \begin{proof}
      The $\beta$ rules for $\Stream$ follow from the $\beta$-rules of $\Streamr$ and $\quot$.

      \begin{align*}
        \hd\ (\CoRecStream\ X\ h\ t\ x) & := \overline{\hdr}\ (\cls \circ \CoRecStreamr\ X\ h\ t\ x)              \\
        \overset{\beta-\quot}           & {=} \hdr (\CoRecStreamr\ X\ h\ t\ x)                                    \\
        \overset{\beta-\Streamr}        & {=} h\ x                                                                \\
        \tl (\CoRecStream\ X\ h\ t\ x)  & := (\overline{\cls \circ \tlr})\ (\cls \circ \CoRecStreamr\ X\ h\ t\ x) \\
        \overset{\beta-\quot}           & {=} (\cls \circ \tlr)\ (\CoRecStreamr\ X\ h\ t\ x)                      \\
        \overset{\beta-\Streamr}        & {=} \cls\ (\CoRecStreamr\ X\ h\ t\ (t\ x))                              \\
                                        & =: \CoRecStream\ X\ h\ t\ (t\ x) \qedhere
      \end{align*}
    \end{proof}
  \end{proposition}

  Next, we show that $(\CoRecStreamr\ X\ h\ t)$ and $\cls$ are $\S$-morphisms.

  \begin{lemma}
    Let $(X: \U)$, $(h: X \to E)$, $(t: X \to X)$ and $(x: X)$. We have that $(\CoRecStreamr\ X\ h\ t)$ forms an $\S$-morphism $(X \to \Streamr)$:
    \label{stream_morphism}

    \begin{align*}
      \Bigl(\MorphStream\ X\ h\ t \quad \Streamr\ \hdr\ \tlr \quad (\CoRecStreamr\ X\ h\ t)\Bigr)
    \end{align*}
    \begin{proof}
      The claim follows naturally from the $\beta$-rules of \cref{stream_beta_rules} and function extensionality.
    \end{proof}
  \end{lemma}

  \begin{lemma}
    \label{cls_morph}

    The class function $(\cls : \Streamr \to \Stream)$ forms an $\S$-morphism.

    \begin{align*}
      \Bigl(\MorphStream\ \Streamr\ \hdr\ \tlr \quad \Stream\ \hd\ \tl \quad \cls\Bigr)
    \end{align*}

    \begin{proof}
      We need to show $\hd = \hdr \circ \cls$ and that $\tl \circ \cls = \cls \circ \tlr$.
      Recall the  $\beta$-rule for $\RecQuot$, described in \cref{beta-quot}.
      We have that $\hd := \overline{\hdr} \overset{\beta}{=} \hdr \circ \cls$.
      Similarly, we have $\tl \circ \cls := \overline{\cls \circ \tlr} \circ \cls \overset{\beta}{=} \cls \circ \tlr$.
    \end{proof}
  \end{lemma}

  Before we show that the $\eta$-rule holds for this improved $\Stream$ type, we first need to prove the following helper lemmas.

\begin{lemma}
  \label{stream_cls_rel}
  Let $(\sigma: \Streamr)$ be a stream. Then $(\CoRecStreamr\ \Streamr\ \hdr\ \tlr\ (\cls\ \sigma))$ and $(\CoRecStreamr\ \Stream\ \hd\ \tl\ \sigma)$ are related by $\CoLimStr$.

  \begin{align}
    (\CoRecStreamr\ \Streamr\ \hdr\ \tlr\ \sigma) \equiv (\CoRecStreamr\ \Stream\ \hd\ \tl\ (\cls\ \sigma)) \label{stream_cls_eq}
  \end{align}

  \begin{proof}
    By \cref{cls_morph}, we have that $\cls$ is a morphism. Therefore, the $\CoLimStr$ predicate, instantiated using the obvious choices for $X, Y, h, h', t, t', f$ and $x$ follows directly.
  \end{proof}
\end{lemma}

\begin{lemma}
  \label{stream_id}
  Let $(\sigma: \Streamr)$ be a stream. Then $(\CoRecStreamr\ \Streamr\ \hdr\ \tlr\ \sigma)$ and $\sigma$ are related by $\CoLimStr$.

  \begin{align}
    (\CoRecStreamr\ \Streamr\ \hdr\ \tlr\ \sigma) \equiv \sigma \label{stream_id_eq}
  \end{align}
  \begin{proof}
    We shall make this proof as precise as possible to highlight the inner workings of the existential type used to define $\Streamr$. To show that the above relation holds, we need to find suitable instances of $X, Y, h, h', t, t', f$ and $x$ for \cref{def:colimstr}. We take $\sigma := \sigma$ and $\tau := (\CoRecStreamr\ \Streamr\ \hdr\ \tlr\ \sigma)$. We immediately see some good candidates for $Y := \Streamr$, $h' := \hdr$ and $t' := \tlr$. We can destruct the existential type $\sigma$ using $\RecExists$ to obtain $(X: \U)$, $(h: Y \to E)$, $(t: Y \to Y)$ and $(x: X)$. We are left to show the following three equations:

    \begin{gather}
      \MorphStream\ X\ h\ t \quad \Streamr\ \hdr\ \tlr \quad f \label{stream_id:1} \\
      \sigma = \CoRecStreamr\ X\ h\ t\ x \label{stream_id:2} \\
      \tau = (\CoRecStreamr\ \Streamr\ \hdr\ \tlr\ (f\ x)) \label{stream_id:3}
    \end{gather}

    Let's take a look at \cref{stream_id:2}. By destructing $\sigma$, we obtained $X, h, t$, and $x$. There is however a fundamental problem. We cannot relate these choices of $X$, $h$, $t$, and $x$ back to the original existential type $\sigma$. For this, we need \cref{axiom:existsid}. We apply the axiom below:

    \begin{align}
      \sigma \overset{\ExistsId}{=} \pack\ X\ \tup{x, h, t} =: \CoRecStreamr\ X\ h\ t\ x \label{stream_id:4}
    \end{align}

    We are now ready to look for a suitable candidate for $(f: X \to Y)$. We use $(f := \CoRecStreamr\ X\ h\ t)$. By \cref{stream_morphism} we know that $(\CoRecStreamr\ X\ h\ t)$ is a morphism. We thus have that \cref{stream_id:1} is satisfied. Additionally, by \cref{stream_id:4}, we have that $(f\ x) := (\CoRecStreamr\ X\ h\ t)\ x \overset{\ref{stream_id:4}}{=} \sigma$. Therefore, \cref{stream_id:3} follows trivially.
  \end{proof}
\end{lemma}

\begin{lemma} The stream recursor, together with the head function $\hd$ and the tail function $\tl$ form the identity.
  \label{stream_id2}

  \begin{align*}
    \CoRecStream\ \Stream\ \hd\ \tl = \id{\Stream}
  \end{align*}
  \begin{proof}
    By \cref{idLift}, we know that $\overline{\cls} = \id{\Stream}$, where $\overline{\cls} := \RecQuot\ \Stream\ \cls\ \EqCls2$. If we then apply the $\eta$-rule of the quotient type (\cref{eta-quot}), we see that, in order to show that $\CoRecStream\ \Stream\ \hd\ \tl = \id{\Stream} = \RecQuot\ \Stream\ \cls\ \EqCls2$, it is enough to show the following equation:

    \begin{gather}
      (\CoRecStream\ \Stream\ \hd\ \tl) \circ \cls = \cls \label{todoid}
    \end{gather}

    We reduce \cref{todoid} using function extensionality, the definition of $\CoRecStream$ and \cref{quotient_classes2}.

    \begin{align}
                                   & \CoRecStream\ \Stream\ \hd\ \tl \circ \cls = \cls \overset{\funext}                     & {\iff} \nonumber \\
      \forall (s: \Streamr) \qquad & \CoRecStream\ \Stream\ \hd\ \tl\ (\cls\ s) = \cls\ s \overset{\ref{def:corec}}          & {\iff} \nonumber \\
      \forall (s: \Streamr) \qquad & \cls\ \bigl(\CoRecStreamr\ \Stream\ \hd\ \tl\ (\cls\ s)\bigr) = \cls\ s \label{todoid2}
    \end{align}

    We shall prove \cref{todoid2}. Assume a stream $(s: \Streamr)$. We use the existential recursor $\RecExists$ and \cref{axiom:existsid} to destruct $s$. We thus destruct $(s = \CoRecStreamr\ X\ h\ t\ x)$ for some $(X: \U)$, $(h: X \to E)$, $(t: X \to X)$ and $(x: X)$. We use \cref{quotient_classes2} for the equational reasoning below:

    \begin{align*}
      \cls\ \bigl(\CoRecStreamr\ \Stream\ \hd\ \tl\ (\cls\ s)\bigr) \overset{\ref{stream_cls_rel}} & {=} \cls\ \bigl(\CoRecStreamr\ \Streamr\ \hdr\ \tlr\ s\bigr) \\
      \overset{\ref{stream_id}}                                                                    & {=} \cls\ s \qedhere
    \end{align*}
  \end{proof}
\end{lemma}

We are now ready to show that the recursor is unique.

\begin{theorem}
  For all $(X: \U), (h: X \to E), (t: X \to X)$ and $(f: X \to \Stream)$ we have:
  \label{eta_stream}

  \begin{gather*}
    \bigl(\MorphStream\ X\ h\ t \quad \Stream\ \hd\ \tl\ \quad f\bigr)\implies (f = \CoRecStream\ X\ h\ t)
  \end{gather*}

  \begin{proof}
    Assume the following:

    \begin{align*}
       & (X: \U)                                                             \\
       & (h: X \to E)                                                        \\
       & (t: X \to X)                                                        \\
       & (f: X \to \Stream)                                                  \\
       & \bigl(\MorphStream\ X\ h\ t\ \quad \Stream\ \hd\ \tl\ \quad f\bigr)
    \end{align*}

    By function extensionality, it is enough to prove $(f\ x = \CoRecStream\ X\ h\ t\ x)$ for all $(x: X)$. We can apply the result from \cref{stream_id2} and the fact that $(\CoRecStreamr\ X\ h\ t\ x)$ and $(\CoRecStreamr\ \Stream\ \hd\ \tl\ (f\ x))$ are related by $\CoLimStr$.

    \begin{align*}
      f\ x\quad \overset{\ref{stream_id2}} & = \CoRecStreamr\ \Stream\ \hd\ \tl\ (f\ x) \\
                                           & = \CoRecStreamr\ X\ h\ t\ x \qedhere
    \end{align*}
  \end{proof}

\end{theorem}

\[\begin{tikzcd}
    \Stream & {E \times \Stream} \\
    \\
    \Streamr & {E \times \Streamr} \\
    \\
    \\
    X & {E\times X}
    \arrow["{{\tup{\hd,\tl}}}", from=1-1, to=1-2]
    \arrow["\cls", from=3-1, to=1-1]
    \arrow["{{\tup{\hdr,\tlr}}}", from=3-1, to=3-2]
    \arrow["{{\tup{\id{E}, \cls}}}", from=3-2, to=1-2]
    \arrow["{\CoRecStream\ X\ h\ t}", bend left=30, from=6-1, to=3-1]
    \arrow["f"', bend right=30, from=6-1, to=3-1]
    \arrow["\equiv"{description}, draw=none, from=6-1, to=3-1]
    \arrow["{{\tup{h,t}}}", from=6-1, to=6-2]
    \arrow["{{\tup{\id{E}, f}}}", from=6-2, to=3-2]
  \end{tikzcd}\]
\section{Stream bisimulation principle}

In this section, we show that the coinduction principle, also known as the bisimulation principle, holds for the newly defined stream type $\Stream$. The bisimulation principle states that two streams are equal if they have the same behavior, i.e. if they are related by a bisimulation relation. We define a bisimulation as a relation $\texttt{BiSimStream}$ and show that two streams are equal if and only if they are related by this relation.

\define{\BiSim}{BiSim}{\texttt{BiSim}}
\define{\IsBiSim}{IsBiSim}{\texttt{IsBiSim}}
\begin{definition}
  We define the bisimulation relation $\BiSim$ that relates two streams if they can be related via some bisimulation $R$. A relation $R$ is a bisimulation if the predicate $(\IsBiSim\ R)$ holds. A bisimulation between streams $(\sigma, \tau: \Stream)$ is a relation that implies the heads of both streams are equal and the tails are related.

  We define the bisimulation relation that relates streams with the same behavior. Plainly, two streams are bisimilar when their heads are equal and their tails are bisimilar:

  \begin{align*}
    \BiSim\ \sigma\ \tau & := \sum (R: \Stream \to \Stream \to \U). \IsBiSim\ R \land R\ \sigma\ \tau                                        \\
    \IsBiSim\ R          & := \prod (\sigma, \tau: \Stream). R\ \sigma\ \tau \to \hd\ \sigma = \hd\ \tau \land R\ (\tl\ \sigma)\ (\tl\ \tau)
  \end{align*}

\end{definition}

\begin{notation}
  We denote the relation $\BiSim$ using the infix symbol $\sim$, so instead of writing $(\BiSim\ \sigma\ \tau)$ we write $(\sigma \sim \tau)$.
\end{notation}

\begin{lemma}
  The propositional equality relation on streams is a bisimulation on streams. So for all $(\sigma, \tau: \Stream)$ we have $(\sigma = \tau) \implies (\sigma \sim \tau)$. We thus have that the predicate $\IsBiSim$ holds for $=$.
  \label{equal_bisim}

  \begin{align}
    \IsBiSim\ (\lambda \sigma. \lambda \tau. \sigma = \tau)
  \end{align}

  \begin{proof}
    Define $R := (\lambda \sigma. \lambda \tau.\ \sigma = \tau)$. Let $(\sigma, \tau: \Stream)$ and suppose that $(R\ \sigma\ \tau)$. By definition of $R$ we have that $\sigma = \tau$. It follows that $\hd\ \sigma = \hd\ \tau$. We also have that $\tl\ \sigma = \tl\ \tau$, by which we have that $R\ (\tl\ \sigma)\ (\tl\ \tau)$. We conclude that $\hd\ \sigma = \hd\ \tau \land R\ (\tl\ \sigma)\ (\tl\ \tau)$.
  \end{proof}
\end{lemma}

\define{\CoIndStream}{CoIndStream}{\texttt{CoIndStream}}
\newcommand{\bisimquot}[0]{\Stream/\hspace{-0.3em}\sim}
\define{\hdt}{hdt}{\overline{\texttt{hd}}}
\define{\clst}{clst}{\cls_{\sim}}
\define{\tlt}{tlt}{\overline{\clst\circ\texttt{tl}}}
\begin{theorem}
  We have the following \textit{coinduction} proof principle $\CoIndStream$.

  \begin{align*}
    \CoIndStream\ \sigma\ \tau := \sigma\sim\tau \iff \sigma = \tau
  \end{align*}

  \begin{proof}
    ($\impliedby$) This follows from \cref{equal_bisim}.

    ($\implies$)
    Let $(\sigma, \tau: \Stream)$. Assume we have $\sigma\sim\tau$.

    We define the following quotient:

    \begin{align*}
      \bisimquot & := \quot\ \Stream\ \BiSim
    \end{align*}

    The proofs that $\hdt$ and $\tlt$ are well-defined are very similar to \cref{eqhd} and \cref{eqtl} and have been omitted. We show that the following diagram commutes.

    \begin{figure}[H]
      \centering
      \begin{tikzcd}
        \Stream \arrow[dddd, "\id{\Stream}"'] \arrow[rdd, "\cls_{\sim}"] \arrow[dddd, "=", phantom, bend left] & \\
        &                                                                 \\
        & \bisimquot \arrow[ldd, "\CoRecStream\ (\Stream / \sim)\ \hdt\ \tlt"] \\
        &                                                                 \\
        \Stream &
      \end{tikzcd}
    \end{figure}

    By $\sigma \sim \tau$ there exists a bisimulation $(R : \Stream \to \Stream \to \U)$ with $(R\ \sigma\ \tau)$ and $(\IsBiSim\ R)$. So we have $\hd\ \sigma = \hd\ \tau$ and $R\ (\tl\ \sigma)\ (\tl\ \tau)$.

    Define $f := \bigl(\CoRecStream\ (\bisimquot)\ \hdt\ \tlt\bigr) \circ \clst$. We prove that $f$ forms an $\S$-morphism by showing that the predicate $\bigl(\MorphStream\ \Stream\ \hd\ \tl\ \quad \Stream\ \hd\ \tl \quad f \bigr)$ holds.

    \begin{align*}
      \hd\ (f\ \sigma)          & := \hd\ (\CoRecStream\ (\bisimquot)\ \hdt\ \tlt\ (\clst\ \sigma))      \\
                                & \betar \hdt\ (\clst\ \sigma)                                           \\
      \overset{\ref{beta-quot}} & {=} \hd\ \sigma                                                        \\
      \\
      \tl\ (f\ \sigma)          & := \tl\ (\CoRecStream\ (\bisimquot)\ \hdt\ \tlt\ (\clst\ \sigma))      \\
                                & \betar \CoRecStream\ (\bisimquot)\ \hdt\ \tlt\ (\tlt\ (\clst\ \sigma)) \\
      \overset{\ref{beta-quot}} & {=} \CoRecStream\ (\bisimquot)\ \hdt\ \tlt\ (\clst\ (\tl\ \sigma))     \\
                                & \betar f\ (\tl\ \sigma)
    \end{align*}

    By the application of the $\eta$-rule for streams (\cref{eta_stream}), we find that $f = \CoRecStream\ \Stream\ \hd\ \tl$.
    We thus have for all streams $(\sigma: \Stream)$, that the following holds:

    \begin{align}
      f\ \sigma := (\CoRecStream\ (\bisimquot)\ \hdt\ \tlt\ (\clst\ \sigma)) = (\CoRecStream\ \Stream\ \hd\ \tl\ \sigma)\label{equiv_bisim}
    \end{align}

    Because $(\sigma \sim \tau)$, it follows from \cref{quotient_classes}, that:

    \begin{align}
      \clst\ \sigma = \clst\ \tau \label{sim_clst}
    \end{align}

    We can now derive the following:

    \begin{align*}
      \sigma \overset{\ref{stream_id2}} & {=} \CoRecStream\ \Stream\ \hd\ \tl\ \sigma                 \\
      \overset{\ref{equiv_bisim}}       & {=} \CoRecStream\ (\bisimquot)\ \hdt\ \tlt\ (\clst\ \sigma) \\
      \overset{\ref{sim_clst}}          & {=} \CoRecStream\ (\bisimquot)\ \hdt\ \tlt\ (\clst\ \tau)   \\
      \overset{\ref{equiv_bisim}}       & {=} \CoRecStream\ \Stream\ \hd\ \tl\ \tau                   \\
      \overset{\ref{stream_id2}}        & {=} \tau
    \end{align*}

    We conclude that $\sigma = \tau$.
  \end{proof}
\end{theorem}

\chapter{Generalization}
\label{chap:generalization}

In this chapter, we zoom out and look at the general construction that was used to define $\List$ and $\Stream$. We first look at the general construction of creating inductive and coinductive types in System F. We then look at the generic requirements of the initiality and finality of the $\L$-algebra and $\S$-coalgebra respectively. We shall see that the encodings are defined as well-chosen limit and colimit respectively.

Let's summarize the constructions of the previous chapters. We defined $\List$ as the initial algebra of the functor $\L(X) := 1+E\times X$ and $\Stream$ as the final coalgebra of the functor $\S(X) := E\times X$. Both already satisfied the expected $\beta$-rules, (which made them weakly initial and weakly final respectively). We had to encode the $\eta$-rules so the uniqueness of the defining morphisms was guaranteed. We used a subtype to encode this for $\List$ and a quotient type to encode this for $\Stream$. For $\List$ this gave us the induction principle and for $\Stream$ this gave us the coinduction principle (often called the bisimulation principle).

\section{Impredicative definition}

The first question that might arise is: given a functor $F$, how can we find the right impredicative definition? The answer can be found in a general construction of free data structures as described in \cite[11.4. Representation of free structure]{Girard_Taylor_Lafont_1993}. We can describe any data structure given by a positive endofunctor $F$ using the following System F type:

\begin{align}
  \prod (X: \U) X^{F(X)} \to X \label{algebra_system_f}
\end{align}

Let's look at how we can obtain the definition of $\Listr$ from \cref{algebra_system_f}. Before we continue, we need the following currying isomorphism:

\begin{align}
  \bigl(\sum (a: A). B(a) \to X\bigr) \to X \cong
  \prod (a: A). (B(a) \to X) \to X \label{currying}
\end{align}

This isomorphism holds in a very general setting \cite{nlab:hom-functor_preserves_limits}. It is essentially the same idea that lets us consider a function that takes two arguments to be isomorphic to a function that takes one argument and returns a function that takes the second argument, also known as currying  \cite[Section 4.6]{rijke2022introductionhomotopytypetheory}.

Using this definition, we can describe inductive types. Let us look at the definition of $\List$ with the functor $\L(X) := 1+E\times X$.

\begin{align*}
  \prod (X: \U) X^{\L(X)} \to X & := \prod (X: \U). X^{1+E\times X} \to X                          \\
                                & := \prod (X: \U). \bigl((1+E\times X) \to X\bigr) \to X          \\
                                & \cong \prod (X: \U). \bigl(X\times (E\times X \to X)\bigr) \to X \\
  \overset{\ref{currying}}      & {\cong} \prod (X: \U). X\to (E\to X \to X) \to X                 \\
                                & =: \Listr
\end{align*}

For coinductive data type, the story is reversed. We can describe any data structure given by a positive endofunctor $F$ using the following System F type \cite[4.2]{ChurchScottDataTypes}:

\begin{align}
  \exists (X: \U). X \times (F(X))^X \label{eq:coalg}
\end{align}

Using this definition, we can describe coinductive types. Let us look at the definition of $\Stream$ with the functor $\S(X) := E\times X$. We can describe the type of streams as follows:

\begin{align*}
  \exists (X: \U). X \times (\S(X))^X & := \exists (X: \U). X \times (E\times X)^X                 \\
                                      & := \exists (X: \U). X \times \bigl(X \to (E\times X)\bigr) \\
                                      & \cong \exists (X: \U). X \times (X \to E) \times (X \to X) \\
                                      & =: \Streamr
\end{align*}

\section{Ensuring the $\eta$-rules}

For $\List$ we wanted to guarantee the uniqueness of $(\RecList \ X\ x\ g)$. We ensured this by requiring that the following equality holds for $\L$-morphisms $f$:

\begin{align*}
  f \circ (\RecList\ X\ x\ g) = \RecList\ Y\ y\ h
\end{align*}

For $\Stream$ we wanted to guarantee the uniqueness of $(\CoRecStream\ X\ h\ t)$. We ensured this by requiring that the following equality holds for $\S$-morphisms $f$:

\begin{align*}
  (\CoRecStream\ Y\ h'\ t') \circ f = \CoRecStream\ X\ h\ t
\end{align*}

We can describe these constructions using equalizers (\cref{diagrams}(c)) and coequalizers (\cref{diagrams}(d)). To obtain the initial object of $\L$-\textbf{Alg}, we take the equalizer of the two morphisms $f\ \circ\ (\RecList\ X\ x\ g)$ and $\RecList\ Y\ y\ h$, which is $\tup{\List, \pr1}$. To obtain the final object of $\S$-\textbf{CoAlg}, we take the coequalizer of the two morphisms $\CoRecStream\ Y\ h'\ t' \circ f$ and $\CoRecStream\ X\ h\ t$, which is $\tup{\Stream, \cls}$.

To be precise, we actually need a (co)equalizer for each free variable that appears in the diagram (for each of the quantifiers in the definition of $\LimList$ and $\CoLimStr$). This means that $\List$ and $\Stream$ are the (co)limits of some big commutative diagrams. Since (co)limits are nothing more than a product of (co)equalizers, we can focus on the given (co)equalizers and leave the quantification implicit. This significantly reduces the complexity of the diagrams below.

\begin{figure}[H]
  \centering
  \begin{subfigure}[b]{0.475\textwidth}
    \begin{tikzcd}
      1+(E\times\Listr) \arrow[rr, "{[\nilr,\consr]}"] \arrow[dd] &  & \Listr \arrow[dd, "\RecList\ X\ x\ g"'] \arrow[dddd, "\RecList\ Y\ y\ h", bend left] \\
      &  &                                                                                      \\
      1+(E\times X) \arrow[rr, "{[x,g]}"] \arrow[dd]              &  & X \arrow[dd, "f"']                                                                   \\
      &  &                                                                                      \\
      1+(E\times Y) \arrow[rr, "{[y, h]}"]                        &  & Y
    \end{tikzcd}
    \caption{Commutative diagram of $\Listr$}
  \end{subfigure}
  \hfill
  \begin{subfigure}[b]{0.475\textwidth}
    \begin{tikzcd}
      E\times \Streamr &  & \Streamr \arrow[ll, "{\tup{\hdr,\tlr}}"']                                                          \\
      &  &                                                                                      \\
      E \times Y \arrow[uu] &  & Y \arrow[ll, "{\tup{h', t'}}"'] \arrow[uu, "\CoRecStream\ Y\ h'\ t'"]                                          \\
      &  &                                                                                      \\
      E \times X \arrow[uu] &  & X \arrow[uu, "f"] \arrow[ll, "{\tup{h, t}}"'] \arrow[uuuu, "\CoRecStream\ X\ h\ t"', bend right]
    \end{tikzcd}
    \caption{Commutative diagram of $\Streamr$}
  \end{subfigure}
  \vskip\baselineskip
  \begin{subfigure}[b]{0.475\textwidth}
    \centering
    \begin{tikzcd}
      \List \arrow[r, "\pr{1}"] & \Listr \arrow[r, "\RecList\ Y\ y\ h", bend left] \arrow[r, "f\ \circ\ (\RecList\ X\ x\ g)"', bend right] & Y
    \end{tikzcd}
    \caption{Equalizer diagram of $\List$}
  \end{subfigure}
  \hfill
  \begin{subfigure}[b]{0.475\textwidth}
    \centering
    \begin{tikzcd}
      \Stream & \Streamr \arrow[l, "\cls"'] &  X \arrow[l, "(\CoRecStream\ Y\ h'\ t')\ \circ\ f", bend left] \arrow[l, "\CoRecStream\ X\ h\ t"', bend right]
    \end{tikzcd}
    \caption{Co-equalizer diagram of $\Stream$}
  \end{subfigure}
  \caption{Diagrams of $\List$ and $\Stream$}
  \label{diagrams}
\end{figure}

\section{Alternative predicates}

In this thesis, we used the limit and colimit to encode the (co)inductive data types similar to \cite{encodings}. It is also possible to take different predicates/relations. One can for example directly encode the induction principle within the embedding, by creating a predicate \texttt{Ind} as was done in \cite{Xavier}. It also seems possible to directly encode the $\eta$-rule in a way that is closer to Requirement 1) of \cref{uniqueness_rules} and Requirement 1) of \cref{uniqueness_rules_co} using a predicate \texttt{Unq}. Such a predicate would closely match the statement of the various $\eta$-rules in this thesis. In the table below, we summarized the different encodings and how they relate to the different theorems/definitions in this thesis.

\begin{table}[H]
  \centering
  \begin{tabular}{ll|ll}
    \texttt{Lim} & \cref{def:w-type-enc} & \texttt{CoLim} & \cref{def:m-type-enc} \\
    \texttt{Unq} & \cref{w-eta}          & \texttt{Unq}   & \cref{m-eta}          \\
    \texttt{Ind} & \cref{w-ind}          & \texttt{CoInd} & \cref{m-ind}
  \end{tabular}
\end{table}
Regardless of the exact choice of the embedding or quotient; the principle idea remains the same: we encode the (co)limit/uniqueness/(co)induction principles within the definition of the type.

In the next chapter, we use this knowledge about the general construction to create an embedding of a W-type. This shows that the technique of \cite{encodings} can be generalized to arbitrary (non-higher) inductive types.

\chapter{W-types}
\label{chap:wtypes}

In this chapter, we introduce W-types. W-types generalize inductive types such as natural numbers, lists, and trees. They were originally introduced by Martin-Löf \cite[Wellorderings]{MartinLof} as a way to define inductive data structures within type theory. In this chapter, we give an impredicative definition of W-types. This impredicative definition yields the expected $\beta$-rule, just like we saw for the $\Listr$ type in \cref{chap:list}. We again create an impredicative encoding that ensures the $\eta$-rule also holds. We finish by using this $\eta$-rule to prove the induction principle for W-types.

\newcommand{\WABr}{W^*_{AB}}
\newcommand{\WAB}{W_{AB}}
\newcommand{\W}{\mathcal{W}}

\section{Introduction to W-types}
\label{intro-w-types}

Before we give a formal definition of W-types, we first sketch the intuitive idea. For further reading, we recommend \cite[5.3 W-Types]{Hott}.

The essential property of inductive data structures is of course their self-referential nature. Natural numbers are created by repeatedly taking the successors of natural numbers and lists are created by appending elements to lists. This self-referential property is quite explicit within functional programming languages like Haskell. Take for example a Haskell definition of natural numbers and lists:

\begin{lstlisting}[language=Haskell,deletekeywords={List}]
  data Nat = Z | S Nat
  data List = Nil | Cons E List
  \end{lstlisting}

In this context, the explicit usage of \texttt{Nat} and \texttt{List} on both sides of the definition highlights the inherent recursive nature of inductive types. It also highlights that inductive types are created using several constructors. For example, natural numbers have two constructors: $\texttt{Z}$ and $\texttt{S}$. Each of these constructors has an arity. This is the number of arguments that the constructor requires. The constructor $\texttt{Z}$ has arity $0$ since it does not require any arguments. The constructor $S$ has arity $1$ because we require one argument. For the list type, we could say that it also has two constructors. One can also consider \texttt{Cons} as a \textit{family} of constructors: one for each $(e: E)$. So if $E$ has infinitely many inhabitants, then the list type has infinitely many constructors. Each of these constructors combines the arguments in some way to produce an element of the inductive type. An inductive type can thus be defined using a set of constructors that each have an arity.

The concept of W-types captures this idea. We have a type $(A:\U)$ of \textit{labels}. Each label represents one of the constructors of the inductive type. For each label $(a: A)$, we have a type $(B(a): \U)$ that describes the arity of the constructor.

\begin{notation}
  To denote the W-type with label type $(A: \U)$ and arity type $(B: A \to \U)$, we write $W(a:A).B(a)$, or $\WAB$.
\end{notation}

Note that the existence of W-types is often postulated as an axiom. The beauty of the following approach is that we get W-types as a constructable type. Let us take a look at the W-types for natural numbers and lists.

\begin{example}
  We can define the natural numbers as a W-type. We need to find suitable instances of $A$ and $B$. As we have seen, the natural numbers are made up of two constructors. Therefore, we define our label type $A$ as a type that has two inhabitants. We use $(A := \textbf{1} + \textbf{1})$. For each of the labels ($(\inl\ *) : \textbf{1} + \textbf{1}$) and ($(\inr\ *): \textbf{1} + \textbf{1}$) we define the arity. We construct the arity type $(a: A \vdash B(a))$ using the recursor of $+$. We map $(\inl\ *)$ to $\textbf{0}$ and $(\inr\ *)$ to $\textbf{1}$. Our type $B(a)$ then becomes $(B := [\lambda x.\textbf{0}, \lambda x. \textbf{1}])$. We thus define the $W$-variant of natural numbers as follows:
  \begin{align*}
    \N^W := W(a: \textbf{1} + \textbf{1}). [\lambda x.\textbf{0}, \lambda x.\textbf{1}]\ a
  \end{align*}
\end{example}

\begin{example}
  Let us also define lists using W-types. We need a suitable label type that captures that we either create an empty list or append some element $(e: E)$ to the list. For this purpose we define $(A := \textbf{1} + E)$. The empty list does not require any recursive arguments. The cons constructor requires one. The resulting W-type thus becomes:

  \begin{align*}
    \texttt{List}^W := W(a: \textbf{1} + E). [\lambda x.\textbf{0}, \lambda e.\textbf{1}]\ a
  \end{align*}
\end{example}

Next, we take a look at how we construct elements of W-types. We have one constructor called $\texttt{sup}$. This function simply takes one of the constructors, and its arguments, and creates an element of the W-type. The basic idea is as follows: In the case of $\N^W$, we create $0^W$ using $\texttt{sup(Z, *)}$ and $1^W$ using $\texttt{sup(S, sup(Z,*))}$. Likewise, we create the list $[1,2]^W$ as follows:

\begin{align*}
  [1,2]^W := \texttt{sup(Cons1, sup(Cons2, sup(Nil, *)))}
\end{align*}

The \texttt{sup} constructor is nothing more than an interface for the original constructors.

\section{Impredicative W-type}
\label{impr-w}
In this section, we give an impredicative definition of W-types. Before we begin, we would like to remind the reader that, when we write $(\sum (a: A). B(a) \to X)$, the binding is as follows: $(\sum (a: A). (B(a) \to X))$. In other words, arrows bind stronger than sigma-types.

In the literature, e.g. \cite[5.3 W-Types]{Hott}, we see that W-types can be defined as the initial algebra of the functor $\W$ below.

\begin{definition}\label{w-functor}
  We define the endofunctor $(\W : \C \to \C)$. We leave the category $\C$ unspecified.

  \begin{align*}
    \W(X) & := \sum (a: A). B(a) \to X    \\
    \W(f) & := \tup{\pi_1, f \circ \pi_2}
  \end{align*}
\end{definition}

Of course, we aim at \textit{defining} W-types, not \textit{postulating} them. To that effect, we follow the same approach as we did for $\Listr$ in \cref{chap:list} and detailed in \cref{chap:generalization}. We first define the type $W^*(a:A).B(a)$, which serves as our impredicative basis. We again need the currying isomorphism of \cref{currying}. We instantiate \cref{algebra_system_f} using the functor $\W$ and apply \cref{currying}.

\begin{align*}
  \prod (X: \U). X^{\W(X)} \to X & := \prod (X: \U). X^{\sum (a: A). B(a) \to X} \to X                           \\
                                 & := \prod (X: \U). \Bigl(\bigl(\sum (a: A). B(a) \to X\bigr) \to X\Bigr) \to X \\
  \overset{\ref{currying}}       & {\cong} \prod (X: \U). \Bigl(\prod (a: A). (B(a) \to X) \to X\Bigr) \to X
\end{align*}

We are able to give an impredicative definition of W-types.

\define{\supr}{supr}{\texttt{sup}^*}
\begin{definition}\label{def-w-impr}
  We define \textbf{impredicative W-types}. Let $(A: \U)$ be a label type and $(B: A \to \U)$ be a type describing the arities of these labels.

  \begin{align*}
    W^*(a:A).B(a) & := \prod (X: \U). \Bigl(\prod (a: A). (B(a) \to X) \to X\Bigr) \to X \tag{formation}
  \end{align*}

  We define the $\supr$ constructor.

  \begin{align*}
    \supr       & :\ \;\prod (a:A). \prod (r: B(a) \to \WABr) \to \WABr \tag{introduction}    \\
    \supr\ a\ r & := \lambda (X:\U). \lambda \bigl(g: \prod (a: A). (B(a) \to X) \to X\bigr). \\
                & \hspace{1.8em} g\ a\ \bigl(\lambda b. r\ b\ X\ g\bigr)
  \end{align*}

\end{definition}

\begin{example}
  Using the formally defined $\supr$ function, we define $0^{W^*}$ and $1^{W^*}$ and $[1,2]^{W^*}$. First, we would like to remind the reader that we can always create an inhabitant of the type $\zero \to X$, using $\RecZ$, as defined in \cref{datatypes:systemf}.

  \begin{align*}
    0^{W^*}     & := \supr\ (\inl\ *)\ (\lambda (b: \textbf{0}). \RecZ\ \N^{W^*}\ b)                                                                                                                                      \\
    1^{W^*}     & := \supr\ (\inr\ *)\ (\lambda (b: \textbf{1}). 0^{W^*})                                                                                                                                                 \\
    1^{W^*}     & := \supr\ (\inr\ *)\ (\lambda (b: \textbf{1}). 1^{W^*})                                                                                                                                                 \\
    [1,2]^{W^*} & := \supr\ (\inr\ 1)\ \biggl(\lambda (b: \textbf{1}). \supr\ (\inr\ 2)\ \Bigl(\lambda (b: \textbf{1}). \supr\ (\inl\ *)\ \bigl(\lambda (b: \textbf{0}). \RecZ\ \texttt{List}^{W^*}\ b\bigr)\Bigr)\biggr)
  \end{align*}

\end{example}

It is time to create a way to compute using the W-types by giving the recursion principle.

\define{\RecWr}{RecWr}{\texttt{rec}^*_{\texttt{W}}}
\begin{definition}
  We define the \textbf{$W$-recursor}.

  \begin{align*}
    \RecWr          & : \prod (X: \U). \Bigl(\prod (a: A). (B(a) \to X) \to X\Bigr) \to \WABr \to X \tag{elimination} \\
    \RecWr\ X\ g\ w & := w\ X\ g
  \end{align*}
\end{definition}

This recursion principle behaves as one would expect. It thus satisfies the $\beta$-rule below.

\begin{lemma} The following $\beta$-rule holds for $\WABr$. Let $(X: \U)$, $(a:A)$ and $(r: B(a) \to \WABr)$ be given. Suppose $\bigl(g: \prod (a: A). (B(a) \to X) \to X\bigr)$.

  \begin{align*}
    \RecWr\ X\ g\ (\supr\ a\ r) & \betar g\ a\ ((\RecWr\ X\ g) \circ r) \tag{computation}
  \end{align*}

  \begin{proof}
    \begin{align*}
      \RecWr\ X\ g\ (\supr\ a\ r) & := (\supr\ a\ r)\ X\ g                          \\
                                  & \betar g\ a\ (\lambda b.\ r\ b\ X\ g)           \\
                                  & \betar g\ a\ (\lambda b.\ \RecWr\ X\ g\ (r\ b)) \\
                                  & \betar g\ a\ ((\RecWr\ X\ g) \circ r) \qedhere
    \end{align*}
  \end{proof}
\end{lemma}

\begin{example} We define a function \texttt{doubleW} that doubles a natural number. The recursive schema is as follows:

  \begin{align*}
    \texttt{0}    & \longmapsto \texttt{0}                  \\
    \texttt{S}\ n & \longmapsto \texttt{S}\ (\texttt{S}\ n)
  \end{align*}

  The function is defined using the recursor for W-types:

  \begin{align*}
    \texttt{doubleW}    & : \N^{W^*} \to \N^{W^*}   \\
    \texttt{doubleW}\ n & := \RecWr\ \N^{W^*}\ g\ n
  \end{align*}

  We have to define $g$. We first look at the type of $g$. We named the inner function $t$ for convenience.

  \begin{align*}
    g: \prod (a: \textbf{1} + \textbf{1}). \prod \Bigl(t: \bigl([\lambda x. \textbf{0}, \lambda x. \textbf{1}]\ a\bigr) \to \N^{W^*}\Bigr) \to \N^{W^*}
  \end{align*}

  There are two cases, either $a = (\inl\ *)$ or $a = (\inr\ *)$.

  \begin{itemize}
    \item If $a = (\inl\ *)$, we have $(t : \textbf{0} \to \N^{W^*})$. This $t$ is thus empty. In this base case, we want to return $0^{W^*}$.
    \item If $a = (\inr\ *)$, we have $(t : \textbf{1} \to \N^{W^*})$. In other words, we have $\bigl((t\ \star) : \N^{W^*}\bigr)$ and can obtain a value of $\N^{W^*}$. This is the recursive call. We want to map this value to its double. In other words, we map $(t\ \star)$ to $\supr\ (\inr\ *)\ \bigl(\lambda b. \supr\ (\inr\ *)\ (\lambda b. t\ \star)\bigr)$.
  \end{itemize}

  All together, we have the following definition of $g$:

  \begin{align*}
    g := \lambda a. [ &                                                                                                                                          \\
                      & \qquad \lambda (s: \star). \lambda (t: \textbf{0} \to \N^{W^*}). 0^{W^*},                                                                \\
                      & \qquad \lambda (s: \star). \lambda (t: \textbf{1} \to \N^{W^*}). \supr\ (\inr\ *)\ (\lambda b. \supr\ (\inr\ *)\ (\lambda b. (t\ \star)) \\
                      & ]\ a
  \end{align*}

  Let us finish by computing the value of $(\texttt{doubleW}\ 1^{W^*})$.

  \begin{align*}
    \texttt{doubleW}\ 1^{W^*} & := \RecWr\ \N^{W^*}\ g\ 1^{W^*}                                                                                                          \\
                              & := \RecWr\ \N^{W^*}\ g\ \bigl(\supr\ (\inr\ *)\ (\lambda b. 0^{W^*})\bigr)                                                               \\
                              & \betar g\ (\inr\ *)\ \bigl((\RecWr\ \N^{W^*}\ g) \circ (\lambda b. 0^{W^*}) \bigr)                                                       \\
                              & = g\ (\inr\ *)\ \bigl(\lambda (b: \star).\RecWr\ \N^{W^*}\ g\ ((\lambda b. 0^{W^*})\ b) \bigr)                                           \\
                              & \betar g\ (\inr\ *)\ \bigl(\lambda b.\RecWr\ \N^{W^*}\ g\ 0^{W^*} \bigr)                                                                 \\
                              & \betar \supr\ (\inr\ *)\ \Bigl(\lambda b. \supr\ (\inr\ *)\ \bigl(\lambda b. (\lambda b.\RecWr\ \N^{W^*}\ g\ 0^{W^*})\ \star\bigr)\Bigr) \\
                              & \betar \supr\ (\inr\ *)\ \Bigl(\lambda b. \supr\ (\inr\ *)\ \bigl(\lambda b. \RecWr\ \N^{W^*}\ g\ 0^{W^*}\bigr)\Bigr)                    \\
                              & \betar \supr\ (\inr\ *)\ \bigl((\lambda b. \supr\ (\inr\ *)\ (\lambda b. 0^{W^*})\bigr)                                                  \\
                              & =: 2^{W^*}
  \end{align*}

\end{example}

In this section, we have seen that we can create W-types using an impredicative definition. Even though this definition looks quite complicated, this is mostly due to the nature of formal type-theoretic definitions rather than an inherent complexity of W-types. As mentioned in \cref{intro-w-types}, W-types are usually defined as the initial algebra of the $\W$-endofunctor. This initiality property, and in particular the uniqueness requirement of initiality, allows us to prove the induction principle of W-types. We shall again continue by taking a look at the categorical definition of W-types. We then encode the uniqueness requirement within the definition of an inductive W-type. We finish by showing that the induction principle of W-types holds for this improved definition.

\section{Categorical W-type}

In this section, we take a look at the categorical definition of W-types as the initial algebra of the $\W$-functor of \cref{w-functor}. We first look at the commutative diagram of $\W$-morphisms.

\begin{figure}[h]
  \centering
  \begin{tikzcd}
    {\sum(a:A).B(a) \to X} && X \\
    \\
    {\sum(a:A). B(a) \to Y} && Y
    \arrow["f", dashed, from=1-3, to=3-3]
    \arrow["{\tup{\pi_1, f \circ \pi_2}}"', dashed, from=1-1, to=3-1]
    \arrow["g", from=1-1, to=1-3]
    \arrow["{g'}"', from=3-1, to=3-3]
    \arrow["{=}"{description}, draw=none, from=1-1, to=3-3]
  \end{tikzcd}
  \caption{Commutative diagram of $\W$-morphisms.}
\end{figure}

In formulas, a function $(f: X \to Y)$ is a $\W$-morphism if the following holds:

\begin{gather}
  f \circ g = g' \circ \tup{\pi_1, f \circ \pi_2} \nonumber \qquad \iff\\
  \forall (a: A), (t: B(a) \to X).\qquad f \circ g \circ \tup{a, t} = g' \circ \tup{a, f \circ t} \label{w-morph-form}
\end{gather}

As we saw before, we write out the uniqueness requirement of the initial $\W$-algebra. In formulas, this $\eta$-rule states that for each $\W$-morphism $(f: X \to Y)$, we have $f \circ u_X = u_Y$. In the next section, we encode this requirement within the definition of an inductive W-type.

\begin{figure}[h]
  \centering
  \begin{tikzcd}
    {\sum (a: A). B(a) \to I} & I \\
    {\sum (a: A).B(a) \to X} & X \\
    {\sum (a: A). B(a)\to Y} & Y
    \arrow[from=1-1, to=1-2]
    \arrow[from=1-1, to=2-1]
    \arrow["{=}"{description}, draw=none, from=1-1, to=2-2]
    \arrow["{u_X}"', from=1-2, to=2-2]
    \arrow[""{name=0, anchor=center, inner sep=0}, "{u_Y}", bend left=49, from=1-2, to=3-2]
    \arrow["g", from=2-1, to=2-2]
    \arrow[from=2-1, to=3-1]
    \arrow["{=}"{description}, draw=none, from=2-1, to=3-2]
    \arrow["f"', from=2-2, to=3-2]
    \arrow["{g'}", from=3-1, to=3-2]
    \arrow["{=}"{marking, allow upside down}, draw=none, from=2-2, to=0]
  \end{tikzcd}
  \caption{Uniqueness requirement of the initial $\W$-algebra.}
  \label{w-uniqu-diagram}
\end{figure}

\section{Impredicative encoding of W-types}

\define{\MorphW}{MorphW}{\texttt{MorphW}}
In this section we give an impredicative encoding of the W-type that we defined in \cref{impr-w}. To start, we give a predicate $\MorphW$ that states that a function $f$ is a $\W$-morphism.

\begin{definition} Let $(X, Y: \U)$, $\bigl(g: \prod (a: A). (B(a) \to X) \to X\bigr)$ and $\bigl(g': \prod (a: A). (B(a) \to Y) \to Y\bigr)$. We define a predicate $\MorphW$ that states that $(f: X \to Y)$ forms a $\W$-morphism based on \cref{w-morph-form}.

  \begin{align*}
    \MorphW\ X\ g\quad Y\ g'\quad f & :=  \prod (a: A). \prod (t: B(a) \to X). f\ (g\ a\ t) = g'\ a\ (f \circ t)
  \end{align*}
\end{definition}

We are now ready to define an inductive W-type. This is an embedding of the previously defined impredicative definition \cref{def-w-impr}. We thus have that $W(a:A). B(a) \hookrightarrow W^*(a:A). B(a)$. We use a $\Sigma$-type to encode a proof term of the uniqueness requirement of \cref{w-uniqu-diagram}.

\define{\LimW}{LimW}{\texttt{LimW}}
\define{\RecW}{RecW}{\texttt{rec}_{\texttt{W}}}
\begin{definition} \label{def:w-type-enc}
  We define an \textbf{inductive W-type}. Let $(A: \U)$ be a label type and let $(B: A \to \U)$ be the type of arities.

  \begin{align*}
    W(a:A). B(a) := & \sum \bigl(w :W^*(a:A).B(a)\bigr). \LimW\ w \tag{formation}                \\
    \LimW\ w :=     & \prod (X, Y: \U)                                                           \\
                    & \prod \Bigl(g: (\prod (a: A). B(a) \to X) \to X\Bigr)                      \\
                    & \prod \Bigl(g': (\prod (a: A). B(a) \to Y) \to Y\Bigr).                    \\
                    & \prod (f: X \to Y).                                                        \\
                    & (\MorphW\ X\ g\ Y\ g'\ f) \implies f\ (\RecWr\ X\ g\ w) = \RecWr\ Y\ g'\ w
  \end{align*}

  We can create a recursor for this new W-type.

  \begin{align*}
    \RecW          & : \prod (X: \U). \bigl(\prod (a: A). (B(a) \to X) \to X\bigr) \to W(a:A). B(a) \to X \tag{elimination} \\
    \RecW\ X\ g\ w & := \RecWr\ X\ g\ (\pr1\ w)
  \end{align*}

\end{definition}

We have to show that elements of the old W-type satisfy the $\LimW$ predicate. Because we have just one introduction rule, namely $\supr$, we just have to check that $\LimW$ holds for this constructor.

\define{\LimWSup}{LimWsup}{\texttt{LimWSup}}
\begin{lemma}[$\LimWSup$]
  For $(a: A)$ and $\bigl(r: B(a) \to \WAB \bigr)$ we have that the $\supr$ constructor satisfies the $\LimW$ predicate.

  \begin{align*}
    \LimW\ (\supr\ a\ (\pr1\ \circ r))
  \end{align*}

  \begin{proof}
    Let $(X, Y: \U)$, $(g: \prod (a: A). (B(a) \to X) \to X)$, $(g': \prod (a: A). (B(a) \to Y) \to Y)$ and $(f: X \to Y)$ be given. Suppose $(\MorphW\ X\ g\ Y\ g'\ f)$ holds. We need to show that $f\ (\RecWr\ X\ g\ \supr(a,r)) = \RecWr\ Y\ g'\ \supr(a,r)$ holds. Note that by instantiating \\
    $(\MorphW\ X\ g\ Y\ g'\ f)$ with $a$ and $(g\ a\ (\RecWr(X, g) \circ r))$ we obtain the following equation:

    \begin{align}
      f\ \bigl(g\ a\ ((\RecWr\ X\ g) \circ r)\bigr) = g'\ a\ (f \circ (\RecWr\ X\ g) \circ r) \label{LimWsup:morph}
    \end{align}

    For all $(a: A)$ and $(b: B(a))$ we have that $(r\ b)$ is of the type $\WAB$. Thus, we can use the proof obtained from $(\pr2\ (r\ b))$. In other words, we have that $(\LimW\ (\pr1\ (r\ b)))$ holds. We instantiate this with the variables $X$, $Y$, $g$, $g'$ and $f$ to obtain the following equation:

    \begin{align}
      \prod (a: A). \prod (b: B(a)). f\ \bigl(\RecWr\ X\ g\ (\pr1\ (r\ b))\bigr) & = \RecWr\ Y\ g'\ (\pr1\ (r\ b)) \iff \notag                 \\
      f \circ (\RecWr\ X\ g) \circ (\pr1\ \circ r)                               & = (\RecWr\ Y\ g') \circ (\pr1\ \circ r) \label{LimWsup:lim}
    \end{align}

    We conclude:

    \begin{align*}
      f\ \bigl(\RecWr\ X\ g\ (\supr\ a\ (\pr1\ \circ r))\bigr) & \betar\ f\ \bigl(g\ a\ ((\RecWr\ X\ g) \circ (\pr1\ \circ r))\bigr) \\
      \overset{\eqref{LimWsup:morph}}                          & {=} g'\ a\ (f \circ (\RecWr\ X\ g)) \circ (\pr1\ \circ r))          \\
      \overset{\eqref{LimWsup:lim}}                            & {=} g'\ a\ (\RecWr\ Y\ g') \circ (\pr1\ \circ r))                   \\
                                                               & =: \RecWr\ Y\ g'\ (\supr\ a\ (\pr1\ \circ r)) \qedhere
    \end{align*}

  \end{proof}
\end{lemma}

\define{\supp}{supp}{\texttt{sup}}

\begin{definition} We define the introduction rule for the new inductive W-type.

  \begin{align*}
    \supp       & : \prod (a:A). \prod (r: B(a) \to \WAB) \to \WAB \tag{introduction} \\
    \supp\ a\ r & := \tup{\supr\ a\ (\pr1\ \circ r),\ \LimWSup\ a\ (\pr1\ \circ r)}
  \end{align*}
\end{definition}

\begin{lemma} The $\beta$-rule hold for $W(a:A). B(a)$. Let $(X: \U)$, $(a:A)$ and $(r: B(a) \to \WAB)$ be given. Suppose $\bigl(g: \prod (a: A). (B(a) \to X) \to X\bigr)$. We have the following equation:
  \label{beta2-W}

  \begin{align*}
    \RecW\ X\ g\ (\supp\ a\ r) \betar g\ a\ ((\RecW\ X\ g) \circ r) \tag{computation}
  \end{align*}

  \begin{proof}
    \begin{align*}
      \RecW\ X\ g\ (\supp\ a\ r) & := \RecWr\ X\ g\ (\pr1(\supp\ a\ r))                \\
                                 & := \RecWr\ X\ g\ (\supr\ a\ (\pr1\ \circ r))        \\
                                 & \betar g\ a\ ((\RecWr\ X\ g) \circ (\pr1\ \circ r)) \\
                                 & = g\ a\ (\lambda b.\ \RecWr\ X\ g\ (\pr1\ (r\ b)))  \\
                                 & =: g\ a\ (\lambda b.\ \RecW\ X\ g\  (r\ b))         \\
                                 & = g\ a\ ((\RecW\ X\ g) \circ r) \qedhere
    \end{align*}
  \end{proof}
\end{lemma}

Now that we have shown the formation, introduction, elimination and computation rules for the inductive W-type, we are ready to show that this type satisfies the uniqueness or $\eta$-rule. We start by showing that $(\RecW\ X\ g)$ is a $\W$-morphism. Like before, these recursors shall play the role of the unique morphisms $u_X$ of the initial object of the $\W$-algebra.

\begin{lemma} Let $(X: \U)$ and $\bigl(g: \prod (a: A). (B(a) \to X) \to X\bigr)$. Then we have:
  \label{RecMorphW}

  \begin{align*}
    \MorphW\ \ \WAB\ \supp\quad X\ g \quad (\RecW\ X\ g)
  \end{align*}

  \begin{proof}
    Let $(X: \U)$ and $(g: \prod (a: A). (B(a) \to X) \to X)$. By filling $\MorphW$ with the right arguments, we see that the required equation is equal to the $\beta$-rule for $\WAB$ as shown in \cref{beta2-W}.

    \begin{align*}
      \RecW\ X\ g\ (\supp\ a\ r) & \betar g\ a\ ((\RecW\ X\ g) \circ r) \qedhere
    \end{align*}
  \end{proof}
\end{lemma}

To prove the complete $\eta$-rule, we first show a special case.

\define{\RecWId}{RecWId}{\texttt{RecWId}}
\begin{lemma}[\RecWId] The following holds:
  \label{RecWId}

  \begin{align*}
    \RecW\ \WAB\ \supp = \id{\WAB}
  \end{align*}

  \begin{proof}
    Let $(w: \WAB)$. We first reduce the statement using function extensionality and $\beta$-reduction. In the $(\hookrightarrow)$ step, we use the fact that $W^*(a:A).B(a)^* \hookrightarrow \WAB$ is an embedding.

    \begin{align}
                                 & \RecW\ \WAB\ \supp = \id{\WAB} \overset{\funext}                    & {\iff} \nonumber \\
      \forall\ (w: \WAB). \qquad & \RecW\ \WAB\ \supp\ w = \id{\WAB}\ w \overset{\beta}                & {\iff} \nonumber \\
      \forall\ (w: \WAB). \qquad & \RecW\ \WAB\ \supp\ w = w \overset{\hookrightarrow}                 & {\iff} \nonumber \\
      \forall\ (w: \WAB). \qquad & \pr1\ \bigl(\RecW\ \WAB\ \supp\ w\bigr) = \pr1\ w \overset{\funext} & {\iff} \nonumber \\
      \forall\ \substack{(w: \WAB), (X:\U),                                                                               \\ (g: \prod (a: A). (B(a) \to X) \to X)}. \qquad &\Bigl(\pr1\ \bigl(\RecW\ \WAB\ \supp\ w\bigr)\Bigr)\ X\ g = (\pr1\ w)\ X\ g \label{RedWIdRed}
    \end{align}

    By \cref{RecMorphW} we known that $\bigl((\RecW\ X\ g): \WAB \to X\bigr)$ is a $\W$-morphism. By looking at $(p:= \pr2\ w)$, which has type $\LimW(\pr1\ w)$, we obtain the following equation:

    \begin{align}
      (\RecW\ X\ g)\ (\RecWr\ \WAB\ \supp\ (\pr1\ w)) = (\RecWr\ X\ g\ (\pr1\ w)) \label{LimWPr}
    \end{align}

    We conclude the proof by showing \cref{RedWIdRed}. Let $(w: \WAB), (X:\U)$, and $\bigl(g: \prod (a: A). (B(a) \to X) \to X\bigr)$.

    \begin{align*}
      \Bigl(\pr1\ \bigl(\RecW\ \WAB\ \supp\ w\bigr)\Bigr)\ X\ g & =: \RecWr\ X\ g\ \Bigl(\pr1\ \bigl(\RecW\ \WAB\ \supp\ w\bigr)\Bigr) \\
                                                                & =: \RecW\ X\ g\ \bigl(\RecW\ \WAB\ \supp\ w\bigr)                    \\
                                                                & =: \RecW\ X\ g\ \bigl(\RecWr\ \WAB\ \supp\ (\pr1\ w)\bigr)           \\
      \overset{\eqref{LimWPr}}                                  & {=} \RecWr\ X\ g\ (\pr1\ w)                                          \\
                                                                & = (\pr1\ w)\ X\ g \qedhere
    \end{align*}

  \end{proof}
\end{lemma}

We are now ready to show the $\eta$-rule for the inductive W-type. This rule states that the recursors $(\RecW\ X\ g)$ are unique.

\begin{theorem} \label{w-eta}
  The $\eta$-rule holds. Let $(X: \U)$, $\bigl(g: \prod (a: A). (B(a) \to X) \to X\bigr)$ and $(f: \WAB \to X)$ be given. We have the following equation:

  \begin{align*}
    \MorphW(\WAB,\ \supp,\quad X,\ g,\quad f) \implies f = \RecW\ X\ g
  \end{align*}

  \begin{proof}
    Let $(X: \U)$, $\bigl(g: \prod (a: A). (B(a) \to X) \to X\bigr)$ and $(f: \WAB \to X)$. Suppose $\MorphW(\WAB,\ \supp,\ X,\ g,\ f)$ holds. Fix $(w: \WAB)$. By looking at $(p:= \pr2\ w)$, which has type $\LimW(\pr1\ w)$, we obtain the following equation:

    \begin{align}
      f\ (\RecWr\ \WAB\ \supp\ (\pr1\ w)) = \RecWr\ X\ g\ (\pr1\ w) \label{LimWPr2}
    \end{align}

    We derive the following:

    \begin{align*}
      f\ w \overset{\RecWId}    & {=} f\ \bigl(\RecW\ \WAB\ \supp\ w\bigr)          \\
                                & =:\ f\ \bigl(\RecWr\ \WAB\ \supp\ (\pr1\ w)\bigr) \\
      \overset{\eqref{LimWPr2}} & {=}\ \RecWr\ X\ g\ (\pr1\ w)                      \\
                                & =:\ \RecW\ X\ g\ w
    \end{align*}

    By function extensionality we conclude that $(f = \RecW\ X\ g)$.
  \end{proof}

\end{theorem}

\section{W-induction principle}

In this section, we show that we have an induction principle on our impredicative encoding. We follow the same approach as in the proof for list induction in \cref{thm:indlist}. Before we begin, we first take a moment to investigate what an induction principle for W-types would look like.
\define{\HW}{HW}{\texttt{Hyp}_{W}}

Induction is used to prove properties about \textit{all} members of some type. We thus assume some predicate $(P: \WAB \to \U)$ that we want to show. The end result of the induction principle is thus a proof that this predicate holds for all members: $\bigl(\prod (w: \WAB). P\ w\bigr)$. What requirements do we need to satisfy in order to prove this? We know that W-types are constructed using just one constructor: $(\supp\ a\ r)$. The induction hypothesis is therefore as follows:

\begin{definition} Induction \textbf{hypothesis for $W$-induction}. Let $(a: A)$, and $(r: (B\ a) \to \WAB)$.
  \label{def:indhypw}

  \begin{align*}
    \HW\ a\ r := \prod\Bigl(h: \prod (b: B\ a). P\ (r\ b)\Bigr) \implies P\ (\supp\ a\ r)
  \end{align*}
\end{definition}

The combined induction principle for W-types is quite straightforward:
\define{\IndW}{IndW}{\texttt{Ind}_{W}}
\begin{definition} We define the \textbf{induction principle of W-types}.

  \begin{align*}
    \IndW := \Bigl(\prod(a:A). \prod\bigl(r: (B\ a) \to \WAB\bigr).\ \HW\ a\ r\Bigr) \implies \prod \bigl(w: \WAB\bigr). P\ w
  \end{align*}
\end{definition}

To show that this induction principle holds, we make use of the $\eta$-rule of \cref{w-eta}. We first formulate our plan of attack. Note the similarity to what we have seen in \cref{thm:indlist}.

\define{\hh}{hh}{\texttt{h}}

\begin{minipage}{.66\textwidth}
  \begin{enumerate}
    \item We show that $(T := \sum(w: \WAB). P\ w)$ together with some function $\hh$ forms a $\W$-algebra.
    \item We have that $\pr1$ is a $\W$-morphism $(\pr1: T \to \WAB)$.
    \item We have the unique $\W$-morphism $(\RecW\ T\ \hh: \WAB \to T)$.
    \item We have that $(\pr1 \circ (\RecW\ T\ \hh))$ is a $\W$-morphsim \\
          $\bigl(\pr1 \circ (\RecW\ T\ \hh) : \WAB \to \WAB\bigr)$.
    \item We have that $\id{\List}$ is also an $\mathcal{L}$-morphism from $\List$ to $\List$.
    \item By the $\eta$-rule we conclude that that $\pr1 \circ (\RecW\ T\ \hh) = \id{\WAB}$.
    \item For all $(w: \WAB)$ we have that $\pr1\ (\RecW\ T\ \hh\ w) = w$ and thus conclude that $(\pr2\ (\RecW\ T\ \hh\ w): P\ w)$.
  \end{enumerate}
\end{minipage}
\begin{minipage}{.33\textwidth}
  \[
    \begin{tikzcd}
      {\W(\WAB)} & \WAB \\
      {\W(T)} & T \\
      {\W(\WAB)} & \WAB
      \arrow["\supp", from=1-1, to=1-2]
      \arrow["{\W(\RecW\ T\ \hh)}"', from=1-1, to=2-1]
      \arrow["{=}"{description, pos=0.4}, draw=none, from=1-1, to=2-2]
      \arrow["{\RecW\ T\ \hh}", from=1-2, to=2-2]
      \arrow["{\id{\WAB}}", bend left=100, from=1-2, to=3-2]
      \arrow["\hh", from=2-1, to=2-2]
      \arrow["{\W(\pr1)}"', from=2-1, to=3-1]
      \arrow["{=}"{description}, draw=none, from=2-1, to=3-2]
      \arrow["\pr1", from=2-2, to=3-2]
      \arrow[from=3-1, to=3-2]
    \end{tikzcd}
  \]
\end{minipage}

\begin{lemma} Assume a predicate $(P: \WAB \to \U)$. We assume the $W$-induction hypothesis of \cref{def:indhypw} holds. Then, e have that $(T := \sum(w: \WAB). P\ w)$ together with some function $\hh$ forms a $\W$-algebra.
  \label{TalgW}
  \begin{proof}
    We define $(T := \sum(w: \WAB). P\ w)$. To make this a $W$-algebra, we need a function $(\hh: \bigl(\sum(a: A). B(a) \to T\bigr) \to T)$. We define $\hh$ as follows:

    \begin{align*}
      \hh & : \Bigl(\sum(a: A). B(a) \to \sum(w: \WAB). P\ w\Bigr) \to \sum(w: \WAB). P\ w                  \\
      \hh & := \lambda \tup{a, r}. \tup{\supp\ a\ (\pr1\ \circ r),\ H\ a\ (\pr1\ \circ r)\ (\pr2\ \circ r)}
    \end{align*}

    Let's take a moment to decipher this definition. First, we destruct the left-most sigma type. This yields a tuple $\tup{a, r}$ where $(a: A)$ and $(r:  B(a) \to \sum(w: \WAB). P\ w)$. Next, we need to create a value of type $(\sum(w: \WAB). P\ w)$. We thus need to create a tuple. The left element needs to be of type $\WAB$. We use $\supp$ to construct the required W-type. Note that we need to take the first projection of $r$.

    For the right element of the tuple, we need a proof of type $(P\ w)$. Because we are working with a dependent sum type, this $w$ is the value we just gave for the left element of the tuple. We thus need to give a proof of $(P\ (\supp\ a\ (\pr1\ \circ r)))$. We use the induction hypothesis. We first initialize $H$ with the right types:

    \begin{align*}
      (H\ a\ (\pr1\ \circ r)) : \prod\Bigl(k: \prod (b: B\ a). P\ ((\pr1\ \circ r)\ b)\Bigr) \implies P\ (\supp\ a\ (\pr1\ \circ r))
    \end{align*}
    If we find a suitable value for $k$, the required result follows directly. We are thus left with finding a suitable value of $k$.

    \begin{align*}
      k: \prod (b: B\ a). P\ ((\pr1\ \circ r)\ b)
    \end{align*}

    Assume $(b: B\ a)$. Note that $\bigl((r\ b) : \sum (w: \WAB). P\ w\bigr)$, where $\bigl((\pr1\ (r\ b)) : \WAB\bigr)$ and therefore $\bigl((\pr2\ (r\ b)) : P\ (\pr1\ (r\ b))\bigr)$. We thus see that $(\pr2\ \circ r)$ gives the desired value for $k$.
  \end{proof}

\end{lemma}

\begin{lemma} We have that $(\pr1: T \to \WAB)$ is a $\W$-morphism. Formally, the following holds:
  \label{pr1-w-morph}

  \begin{align*}
    \bigl(\MorphW\ T\ \hh \quad \WAB\ \supp \quad \pr1\bigr)
  \end{align*}

  \begin{proof}
    Let $(a: A)$ and $(t: (B\ a) \to T)$. We have to show that $\pr1\ (\hh\ a\ t) = \supp\ a\ (\pr1\ \circ t)$. This follows directly by looking at (the first projection of) the definition of $(\hh\ a\ t)$.
  \end{proof}

\end{lemma}

\begin{theorem} \label{w-ind}
  Induction holds for $\WAB$.

  Suppose $(P: \WAB \to \U)$ is a predicate on $\WAB$. Suppose we have for all $(a: A)$ and $\bigl(r: (B\ a) \to \WAB\bigr)$ that $(\HW\ a\ r)$ holds, then we conclude $\prod \bigl(w: \WAB\bigr). P\ w$.

  \begin{proof}
    Let us assume a predicate $(P: \WAB \to \U)$. We assume that the induction hypothesis $(\HW\ a\ r)$ holds for all $(a: A)$ and $\bigl(r: (B\ a) \to \WAB\bigr)$. This means that we have a proof $(H\ a\ r : (\HW\ a\ r))$ for all $(a: A)$ and $\bigl(r: (B\ a) \to \WAB\bigr)$.

    \begin{enumerate}
      \item By \cref{TalgW} we have that $(T := \sum(w: \WAB). P\ w)$ forms a $\W$-algebra.
      \item By \cref{pr1-w-morph} we have that $(\pr1: T \to \WAB)$ formas a $\W$-morphism.
      \item By \cref{RecMorphW} have a $\W$-morphism $((\RecW\ T\ \hh): \WAB \to T)$, which is unique by the $\eta$-rule of \cref{w-eta}.
      \item By a straightforward adaptation of \cref{comp_morph} we have that $\bigl((\pr1 \circ (\RecW\ T\ \hh)): \WAB \to \WAB \bigr)$ is a $\W$-morphism.
      \item By a straightforward adaptation of \cref{id_morph} we have that $\id{\List}$ is a $\W$-morphism.
      \item By \cref{w-eta} we have that $(\pr1 \circ (\RecW\ T\ \hh)) = (\RecW\ \WAB\ \sup)$. By \cref{RecWId} we have that $(\RecW\ \WAB\ \sup) = \id{\WAB}$. We combine these to obtain that $\pr1\ (\RecW\ T\ \hh) = \id{\List}$.
      \item Let $(w : \WAB)$ be any member of $\WAB$. We want to show that $(P\ w)$ holds. We just saw that $(\pr1 \circ (\RecW\ T\ \hh))\ w = w$. We can thus obtain a proof of $(P\ l)$ by looking at the second projection $\bigl((\pr2\ (\RecW\ T\ \hh))\ w: P\ w\bigr)$. \qedhere
    \end{enumerate}
  \end{proof}

\end{theorem}

\chapter{M-types}
\label{chap:mtypes}

\newcommand{\MABr}{M^*_{AB}}
\newcommand{\MAB}{M_{AB}}
\newcommand{\M}{\mathcal{M}}

In this chapter, we introduce M-types. M-types are the dual notion of W-types and generalize coinductive types such as streams,  possibly infinite lists, and non-wellfounded trees. In this chapter, we give an impredicative definition of M-types. This definition yields the expected $\beta$-rules just like we saw for $\Streamr$ in \cref{chap:stream}. We again create an impredicative encoding that ensures that the $\eta$-rule also holds. We finish by using this $\eta$-rule to prove the coinduction principle (also called bisimulation) for M-types.

\section{Introduction to M-types}

Before we give a formal definition, we first sketch the idea behind M-types. Coinductive data types are structures that are possibly infinite. Take for example the type of streams. These streams are infinite in the sense that one can take the tail of the stream any amount of times without depleting the stream. It is also possible to create streams that might terminate at some point. They are often called possibly infinite lists. Inhabitants of this type might be infinite or might terminate at some point. Another example are non-wellfounded trees, in other words, trees that might go on forever.

Just like a W-type has a number of constructors, an M-type has a number of ways it can destruct. We saw that streams have two destructors: $(\hd : \Stream \to E)$ and $(\tl : \Stream \to \Stream)$. We can also combine these into a new destructor: $(\texttt{destr}: \Stream \to (E \times \Stream))$ which simultaneously takes the first element of the stream and returns the tail of the stream. Note that we can define $\hd$ and $\tl$ in terms of $\texttt{destr}$ by taking the first and second projections respectively. Just like we could describe $\cons$ as a family of constructors (one for each $(e: E)$), we can also describe $\texttt{destr}$ as a family of destructors $\texttt{destr}_E$. For each particular $(e: E)$, we have a destructor $(\texttt{destr}_E : \Stream \to \Stream)$.

To define M-types, we take some type $(A: \U)$ of $\textit{labels}$. Each label represents one of the destructors of the coinductive type. For each label $(a: A)$, we have a type $(B(a) : \U)$ that describes the arity of the destructor.

\begin{notation}
  To denote the M-type with label type $(A: \U)$ and arity type $(B: A \to \U)$, we write $M(a: A). B(a)$ or $\MAB$.
\end{notation}

\begin{example}
  We define infinite streams as an M-type. Define $A := E$ and $B := \lambda e. \textbf{1}$.

  \begin{align*}
    \texttt{Stream}^M := M(e: E). (\lambda e. \textbf{1})\ e = M(e: E). \textbf{1}
  \end{align*}

  For each M-type, we have an elimination principle $\texttt{elimM}$ that destructs that M-type. The type definition of this elimination principle is as follows. Note that the scope of the sigma type extends all the way to the right.

  \begin{align*}
    \texttt{elimM} & : \quad \MAB \to \sum (a: A). B(a) \to \MAB                                 \\
    \intertext{In the case of $\texttt{Stream}^M$, we thus have the following type definition.}  \\
    \texttt{elimM} & : \quad \texttt{Stream}^M \to \sum (e: E). \textbf{1} \to \texttt{Stream}^M \\
    \intertext{Because $B$ does not depend on $(e: E)$, we equivalently have:}                   \\
    \texttt{elimM} & : \quad \texttt{Stream}^M \to (E \times \texttt{Stream}^M)
  \end{align*}

  We thus see that at least the types of the destructors of $\Stream$ and $\texttt{Stream}^M$ coincide.
\end{example}

\begin{example}
  Let us give another example where the type $B$ does depend on $(a: A)$. We define possibly infinite lists over some element type $E$. This means that if we destruct the possibly infinite list, we either find that it continues, or that it terminates. We define $A := E + \textbf{1}$ such that the list continues if it destructs to the left $(\inl \ e)$, and it terminates if it destructs to the right $(\inr \ \star)$. If the list continues, we again obtain an element of the possibly infinite list from the destructor. If it terminates, we obtain nothing. We define $(B\ (\inl \ e)) := \textbf{1}$ and $(B\ (\inr\ \star)) := \textbf{0}$. We can thus define the type of possibly infinite lists as follows:

  \begin{align*}
    \texttt{PIList}^M                                                  & := M(a: E + \textbf{1}). [\lambda e. \textbf{1}, \lambda x. \textbf{0}]\ a  \\
    \intertext{We take a look at the elimination principle:}                                                                                         \\
    \texttt{elimM}                                                     & : \quad \texttt{PIList}^M \to \sum (e: E). \textbf{1} \to \texttt{PIList}^M \\
    \intertext{If we abuse notation, we see that there are two cases:}                                                                               \\
    \texttt{elimM}\ (l: \texttt{PIList}^M)\ \tup{e: E, -}              & : \textbf{1} \to \texttt{PIList}^M                                          \\
    \texttt{elimM}\ (l: \texttt{PIList}^M)\ \tup{\star: \textbf{1}, -} & : \textbf{0} \to \texttt{PIList}^M                                          \\
    \intertext{In other words, if we destruct $l$, we either find that it continues, or that it terminates:}                                         \\
    \texttt{elimM}\ (l: \texttt{PIList}^M)\ \tup{e: E, -}              & : \texttt{PIList}^M                                                         \\
    \texttt{elimM}\ (l: \texttt{PIList}^M)\ \tup{\star: \textbf{1}, -} & : \textbf{1}
  \end{align*}

\end{example}

\section{Impredicative M-types}
\label{impred-m-type}

In this section, we give an impredicative definition of M-types. M-types can be defined as the final coalgebra of the functor $\M$ below \cite{otten2020mtypes}.

\begin{definition}\label{m-functor}
  We define the endofunctor $(\M: \C \to \C)$. We leave the category $\C$ unspecified.

  \begin{align*}
    \M(X) & := \sum (a: A). B(a) \to X    \\
    \M(f) & := \tup{\pi_1, f \circ \pi_2}
  \end{align*}
\end{definition}

To define an impredicative M-type, we instantiate \cref{eq:coalg} using the functor $\M$. Just like the coinductive type of streams, we also define M-types using existential types.

\begin{align*}
  \exists (X: \U). X \times (\M(X))^X & := \exists (X: \U). X \times \bigl(\sum (a: A). B(a) \to X\bigr)^X      \\
                                      & =\ \exists (X: \U). X \times \bigl( X \to \sum (a: A). B(a) \to X\bigr)
\end{align*}

We are able to give an impredicative definition of M-types.

\define{\CoRecMr}{CoRecMr}{\texttt{corec}^*_{\texttt{M}}}
\define{\ElimMr}{ElimMr}{\texttt{elimM}^*}

\begin{definition}\label{def-m-impr}
  We define \textbf{impredicative M-types}. Let $(A: \U)$ be a label type and $(B: A \to \U)$ be a type describing the arities of these labels.

  \begin{align*}
    M^*(a: A). B(a) & := \exists (X: \U). X \times \bigl(X \to \sum (a: A). B(a) \to X\bigr) \tag{formation}
  \end{align*}

  The following corecursor serves as our introduction rule.

  \begin{align*}
    \CoRecMr          & : \prod (X: \U). \bigl(X \to \sum (a: A). B(a) \to X\bigr) \to X \to \MABr \tag{introduction} \\
    \CoRecMr\ X\ r\ x & := \pack\ X\ \tup{x, r}
  \end{align*}

\end{definition}

As we outlined before, we need a destructor, or elimination principle. Because we are working with existential types, we need to use the $\RecExists$ recursor of the existential type to operate on the contents of the M-type.

\begin{definition}
  For each M-type $\MAB$, we define the \textbf{M-eliminator} as follows:

  \begin{align*}
    \ElimMr    & : \MABr \to \sum (a: A). B(a) \to \MABr \tag{elimination}                                                                                    \\
    \ElimMr\ m & := \RecExists\ \bigl(\sum (a: A). B(a) \to \MABr\bigr)                                                                                       \\
               & \quad\ \Bigl(\lambda X. \lambda \tup{x, f}.\ \bigl\langle \pr1\ (f\ x),\ \lambda b. \pack\ X\ \tup{\pr2\ (f\ x)\ b, f}\ \bigr\rangle\ \Bigr)
  \end{align*}

  Let us dissect this definition. We want to create a function that takes an M-type $(m: \MABr)$ and produces a value of type $\bigl(\sum (a: A). B(a) \to \MABr\bigr)$. We use the existential recursor $\RecExists$ with output type $(\sum (a: A). B(a) \to \MABr)$. We can thus create a function where we get some type $(X: \U)$ and a tuple $\tup{x, f}$, where $(x: X)$ and $\bigl(f: X \to \sum (a: A). B(a) \to X\bigr)$ and return some value $(\tup{l,r} : \sum (a: A). B(a) \to \MABr)$:

  \begin{align*}
    \ElimMr\ m & := \RecExists\ \bigl(\sum (a: A). B(a) \to \MABr\bigr)                                                                                                                                                                                                    \\
               & \quad\ \Bigl(\lambda X. \lambda \tup{x, f}.\ \bigl\langle \underbrace{\pr1\ (f\ x)}_{l\ :\ A},\ \underbrace{\lambda b. \pack\ X\ \tup{(\ \overbrace{\vspace{-1em}\pr2\ (f\ x)}^{:\ B(a) \to \WAB}\ )\ b, f}}_{r\ :\ B(a) \to \MABr}\ \bigr\rangle\ \Bigr)
  \end{align*}

  For the left-hand side $l$, we extract the value of `$a$' from the first projection of $(f\ x)$. To create a value for the right-hand side $r$, we need to create a function that takes a value of type $(b: B(a))$ and yields a value of type $\MABr$. To this effect, we use the right-hand projection of $(f\ x)$ and apply $b$.
\end{definition}

Now that we have both an introduction and an elimination principle, we can check if the computation rule or $\beta$-rule holds.

\begin{lemma} \label{beta-mr}
  The $\beta$-rule holds for $\MABr$. Let $(X: \U)$, $(r: X \to \sum (a: A). B(a) \to X)$ and $(x: X)$. We have that the following equation holds:

  \begin{gather*}
    \ElimMr\ (\CoRecMr\ X\ r\ x) \betar \bigl\langle \pr1\ (r\ x),\ (\CoRecMr\ X\ r) \circ (\pr2\ (r\ x)) \bigr\rangle \tag{computation}
  \end{gather*}

  \begin{proof}
    Let $X$, $r$, and $x$ be as described. The desired result follows by folding and unfolding the definitions and using \cref{axiom:existsid}.

    \begin{align*}
      \ElimMr\ (\CoRecMr\ X\ r\ x) & := \ElimMr\ (\pack\ X\ \tup{x, r})                                                             \\
      \overset{\ExistsId}          & {=} \bigl\langle \pr1\ (r\ x), \lambda b. \pack\ X\ \tup{\pr2\ (r\ x)\ b, r} \bigr\rangle      \\
                                   & =: \bigl\langle \pr1\ (r\ x), \lambda b. \CoRecMr\ X\ r\ (\pr2\ (r\ x)\ b) \bigr\rangle        \\
                                   & =: \bigl\langle \pr1\ (r\ x),\ (\CoRecMr\ X\ r) \circ (\pr2\ (r\ x)) \bigr\rangle     \qedhere
    \end{align*}
  \end{proof}

\end{lemma}

Now that we have an impredicative definition of M-types, we can follow the approach outlined in \cref{chap:generalization} to create an impredicative encoding. We first take a look at the categorical definitions.

\section{Categorical M-type}

In this section, we take a look at the categorical definition of M-types as the final coalgebra of the $\M$-functor of \cref{m-functor}. We first look at the commutative diagram of $\M$-morphisms.

\begin{figure}[h]
  \centering
  \begin{tikzcd}
    X && {\sum (a: A). B(a)\to X} \\
    \\
    Y && {\sum (a:A). B(a) \to Y}
    \arrow["g", from=1-1, to=1-3]
    \arrow["f"', from=1-1, to=3-1]
    \arrow["{=}"{description}, draw=none, from=1-1, to=3-3]
    \arrow["{\tup{\pi_1, f \circ \pi_2}}", from=1-3, to=3-3]
    \arrow["{g'}", from=3-1, to=3-3]
  \end{tikzcd}
  \caption{Commutative diagram of $\M$-morphisms.}
\end{figure}

In formulas, a function $(f: X \to Y)$ is an $\M$-morphism, if the following holds:

\begin{gather}
  g' \circ f = \tup{\pi_1, f \circ \pi_2} \circ g \iff \nonumber \\
  \forall (x: X). \qquad g'\ (f\ x) =  \tup{\pi_1(g\ x), f \circ (\pi_2\ (g\ x))} \iff \nonumber\\
  \forall (x: X). \qquad \pi_1 (g'\ (f\ x)) =  \pi_1(g\ x) \quad \land \quad \pi_2 (g'\ (f\ x)) = f \circ (\pi_2\ (g\ x)) \label{m-morph-form}
\end{gather}

We take a look at the uniqueness requirement of the final $\M$-coalgebra. In formulas, this $\eta$-rule states that for each $\M$-morphism $(f: X \to Y)$, we have $u_X = f \circ u_Y$.

\begin{figure}[h]
  \centering
  \begin{tikzcd}
    X & {\sum (a: A). B(a)\to X} \\
    Y & {\sum (a:A). B(a) \to Y} \\
    F & {\sum (a: A). B(a) \to F}
    \arrow["g", from=1-1, to=1-2]
    \arrow["f"', from=1-1, to=2-1]
    \arrow["{=}"{description}, draw=none, from=1-1, to=2-2]
    \arrow[""{name=0, anchor=center, inner sep=0}, "{u_X}"', bend right=55, from=1-1, to=3-1]
    \arrow[from=1-2, to=2-2]
    \arrow["{g'}", from=2-1, to=2-2]
    \arrow["{u_Y}"', from=2-1, to=3-1]
    \arrow["{=}"{description}, draw=none, from=2-1, to=3-2]
    \arrow[from=2-2, to=3-2]
    \arrow[from=3-1, to=3-2]
    \arrow["{=}"{description}, draw=none, from=0, to=2-1]
  \end{tikzcd}
  \caption{Uniqueness requirement of the final $\M$-coalgebra.}
\end{figure}

\section{Impredicative encoding of M-types}

\define{\MorphM}{MorphM}{\texttt{MorphM}}
In this section, we give an impredicative encoding of the M-type that we defined in \cref{impred-m-type}. We first give a predicate $\MorphM$ that states that a given function $(f: X \to Y)$ is an $\M$-morphism.

\begin{definition}
  Let $(X, Y: \U)$, $(g: X \to \sum (a: A). B(a) \to X)$, $(g': Y \to \sum (a: A). B(a) \to Y)$ and $(f: X \to Y)$. We define a predicate $\MorphM$ that states that $(f: X \to Y)$ forms an $\M$-morphism based on \cref{m-morph-form}.

  \begin{align*}
    \MorphM\ X\ g\quad Y\ g'\quad f := \prod (x: X).\quad & \pr1\ (g'\ (f\ x)) = \pr1\ (g\ x) \quad \land \\
                                                          & \pr2\ (g'\ (f\ x)) = f \circ (\pr2\ (g\ x))
  \end{align*}
\end{definition}

\define{\CoLimM}{CoLimM}{\texttt{CoLimM}}

As we saw with $\Streamr$, we shall define a colimit relation $\CoLimM$ that relates two M-types if they are related by an $\M$-morphism. We then create a quotient using this relation such that two M-types that are related by $\CoLimM$ are equated. We finally prove that this quotient satisfies the $\eta$-rule.

\begin{definition} \label{def:m-type-enc}
  We define the relation $\CoLimM$. Let $(m, n: \MAB)$. The predicate $(\CoLimM\ m\ n)$ holds if $m$ and $n$ are related by some $\M$-morphism.

  \begin{gather*}
    \CoLimM\ m\ n := \exists (X, Y: \U). \exists \substack{g: X \to \sum (a: A). B(a) \to X\\ g': Y \to \sum (a: A). B(a) \to Y}. \exists (f: X \to Y). \exists (x: X).\\
    (\MorphM\ X\ g\quad Y\ g'\quad f)\ \land \\
    m = \CoRecMr\ X\ g\ x\ \land \\
    n = \CoRecMr\ Y\ g'\ (f\ x)
  \end{gather*}

\end{definition}

\begin{notation}
  We use the infix notation $(m \equiv n)$ to denote that $(\CoLimM\ m\ n)$ holds.
\end{notation}

Even though this relation is not an equivalence relation (see note \cref{note:equivalence}), we shall speak about the equivalence classes in the sense of the elements that are equal when taking the quotient $\quot$.

\define{\CoRecM}{CoRecM}{\texttt{corec}_{\texttt{M}}}
\begin{definition}
  We define the \textbf{coinductive M-type} $\MAB$.

  \begin{align}
    M(a: A). B(a) := \quot\ (M^*(a: A). B(a))\ \CoLimM \tag{formation}
  \end{align}

  We define the corecursor as follows:

  \begin{align*}
    \CoRecM          & : \prod (X: \U). \bigl(X \to \sum (a: A). B(a) \to X\bigr) \to X \to \MAB \tag{introduction} \\
    \CoRecM\ X\ r\ x & := \cls\ (\CoRecMr\ X\ r\ x)
  \end{align*}

\end{definition}

In order to lift the $\ElimMr$ function, we need to show that it is constant under the equivalence classes of $\CoLimM$. We show this in two parts: first, we show that the left projection is constant under equivalence and then we show that the right projection is constant under $\cls$.

\define{\ElimEqo}{ElimEqo}{\texttt{ElimEq1}}
\begin{lemma}[$\ElimEqo : (\EqCls\ (\pr1 \circ \ElimMr)\ \equiv)$] \label{eqmlem1}
  The left projection of the M-eliminator $\ElimMr$ is constant on all equivalence classes of $\CoLimM$.

  \begin{align*}
    \prod (m, n: \MAB). m \equiv n \implies (\pr1\ (\ElimMr\ m)) = (\pr1\ (\ElimMr\ n))
  \end{align*}

  \begin{proof}
    Let $(m, n: \MAB)$ be M-types. Suppose that $(\CoLimM\ m\ n)$ holds.
    We thus have some $(X, Y: \U)$, $(g: X \to \sum (a: A). B(a) \to X)$, $(g': Y \to \sum (a: A). B(a) \to Y)$, $(x: X)$ and $(f: X \to Y)$ such that the following hold:

    \begin{align}
      \pr1\ (g'\ (f\ x)) & = \pr1\ (g\ x) \label{elimeq1:l1}            \\
      \pr2\ (g'\ (f\ x)) & = f \circ (\pr2\ (g\ x)) \label{elimeq1:l2}  \\
      m                  & = \CoRecMr\ X\ g\ x \label{elimeq1:l3}       \\
      n                  & = \CoRecMr\ Y\ g'\ (f\ x) \label{elimeq1:l4}
    \end{align}

    We derive the following:

    \begin{align*}
      \pr1\ (\ElimMr\ m) \overset{\ref{elimeq1:l3}} & {=} \pr1\ (\ElimMr\ (\CoRecMr\ X\ g\ x))                                                                        \\
                                                    & \betar \pr1\ \bigl\langle \pr1\ (g\ x),\ \lambda b. \CoRecMr\ X\ g\ (\pr2\ (g\ x)\ b) \bigr\rangle              \\
                                                    & \betar \pr1\ (g\ x)                                                                                             \\
      \overset{\ref{elimeq1:l1}}                    & {=} \pr1\ (g'\ (f\ x))                                                                                          \\
                                                    & \betar \pr1\ \bigl\langle \pr1\ (g'\ (f\ x)),\ \lambda b. \CoRecMr\ Y\ g'\ (\pr2\ (g'\ (f\ x))\ b) \bigr\rangle \\
                                                    & \betar \pr1\ (\ElimMr\ (\CoRecMr\ Y\ g'\ (f\ x)))                                                               \\
      \overset{\ref{elimeq1:l4}}                    & {=} \pr1\ (\ElimMr\ n) \qedhere
    \end{align*}
  \end{proof}
\end{lemma}

\define{\ElimEqt}{ElimEqt}{\texttt{ElimEq2}}
\begin{lemma}[$\ElimEqt : (\EqCls\ (\cls \circ \pr2 \circ \ElimMr)\ \equiv)$] \label{eqmlem2}
  The right projection of the M-eliminator $\ElimMr$ is constant on all equivalence classes of $\CoLimM$.

  \begin{align*}
    \prod (m, n: \MAB). m \equiv n \implies (\cls\ (\pr2\ (\ElimMr\ m))) = (\cls\ (\pr2\ (\ElimMr\ n)))
  \end{align*}

  \begin{proof}
    Let $(m, n: \MAB)$ be M-types. Suppose that $(\CoLimM\ m\ n)$ holds.
    We thus have some $(X, Y: \U)$, $(g: X \to \sum (a: A). B(a) \to X)$, $(g': Y \to \sum (a: A). B(a) \to Y)$, $(x: X)$, and $(f: X \to Y)$ such that the following hold:

    \begin{align}
      \pr1\ (g'\ (f\ x)) & = \pr1\ (g\ x) \label{l1}            \\
      \pr2\ (g'\ (f\ x)) & = f \circ (\pr2\ (g\ x)) \label{l2}  \\
      m                  & = \CoRecMr\ X\ g\ x \label{l3}       \\
      n                  & = \CoRecMr\ Y\ g'\ \label{l4} (f\ x)
    \end{align}

    We derive the following. Note that we have left open a step at $(*)$.

    \begin{align*}
      \cls\ (\pr2\ (\ElimMr\ m)) \overset{\ref{l2}} & {=} \cls\ (\pr2\ (\ElimMr\ (\CoRecMr\ X\ g\ x)))                                                                        \\
      \overset{\ref{beta-mr}}                       & {=} \cls\ (\pr2\ \bigl\langle \pr1\ (g\ x),\ \lambda b. \CoRecMr\ X\ g\ (\pr2\ (g\ x)\ b) \bigr\rangle)                 \\
                                                    & \betar \cls\ (\lambda b. \CoRecMr\ X\ g\ (\pr2\ (g\ x)\ b))                                                             \\
      \overset{(*)}                                 & {=} \cls\ (\lambda b. \CoRecMr\ Y\ g'\ (f\ (\pr2\ (g\ x)\ b)))                                                          \\
      \overset{\ref{l2}}                            & {=} \cls\ (\lambda b. \CoRecMr\ Y\ g'\ (\pr2\ (g'\ (f\ x))))                                                            \\
                                                    & \betar \cls\ (\pr2\ \bigl\langle \pr1\ (g'\ (f\ x)),\ \lambda b. \CoRecMr\ Y\ g'\ (\pr2\ (g'\ (f\ x))\ b) \bigr\rangle) \\
      \overset{\ref{beta-mr}}                       & {=} \cls\ (\pr2\ (\ElimMr\ (\CoRecMr\ Y\ g'\ (f\ x))))                                                                  \\
      \overset{\ref{l4}}                            & {=} \cls\ (\pr2\ (\ElimMr\ n))
    \end{align*}

    We shall now fill in the step $(*)$. Let $(b: B(a))$. It suffices to show the following:

    \begin{gather}
      \CoRecMr\ X\ g\ (\pr2\ (g\ x)\ b) \equiv \CoRecMr\ Y\ g'\ (f\ (\pr2\ (g\ x)\ b)) \tag{$*$}
    \end{gather}

    This clearly holds since $f$ is indeed a morphism by assumption.
  \end{proof}
\end{lemma}

By combining \cref{eqmlem1} and \cref{eqmlem2}, we can lift $\ElimMr$.

\define{\ElimM}{ElimM}{\texttt{elimM}}
\begin{definition}
  We define the lifted version of $\ElimMr$:

  \begin{align*}
    \ElimM\  & : \prod (m: \MAB) \to \MAB \to \sum (a: A). B(a) \to \MAB \tag{elimination} \\
    \ElimM\  & := \overline{\tup{\pr1, \cls \circ \pr2} \circ \ElimMr}
  \end{align*}
\end{definition}

We show that this improved M-type satisfies the expected computation rule.

\begin{proposition} \label{beta-m}
  The $\beta$-rule holds for $\MAB$. Let $(X: \U)$, $(r: X \to \sum (a: A). B(a) \to X)$ and $(x: X)$. We have that the following equation holds:

  \begin{align*}
    \ElimM\ (\CoRecM\ X\ r\ x) \betar \bigl\langle \pr1\ (r\ x),\ \CoRecM\ X\ r\ (\pr2\ (r\ x)) \bigr\rangle \tag{computation}
  \end{align*}

  \begin{proof}
    We derive the statement below:

    \begin{align*}
      \ElimM\ (\CoRecM\ X\ r\ x)  & := \overline{\tup{\pr1, \cls \circ \pr2} \circ \ElimMr}\ (\CoRecM\ X\ r\ x)                                     \\
                                  & := \overline{\tup{\pr1, \cls \circ \pr2} \circ \ElimMr}\ (\cls\ (\CoRecMr\ X\ r\ x))                            \\
      \overset{\beta\text{-quot}} & {=} (\tup{\pr1, \cls \circ \pr2} \circ \ElimMr)\ (\CoRecMr\ X\ r\ x)                                            \\
                                  & = \tup{\pr1, \cls \circ \pr2}\ (\ElimMr\ (\CoRecMr\ X\ r\ x))                                                   \\
      \overset{\ref{beta-mr}}     & {=} \tup{\pr1, \cls \circ \pr2}\ \bigl\langle \pr1\ (r\ x),\ (\CoRecMr\ X\ r) \circ (\pr2\ (r\ x)) \bigr\rangle \\
                                  & \betar \bigl\langle \pr1\ (r\ x),\ \cls \circ (\CoRecMr\ X\ r) \circ (\pr2\ (r\ x)) \bigr\rangle                \\
                                  & =: \bigl\langle \pr1\ (r\ x),\ (\CoRecM\ X\ r) \circ (\pr2\ (r\ x)) \bigr\rangle \qedhere
    \end{align*}
  \end{proof}
\end{proposition}

In the remainder of this section, we shall prove that the $\eta$-rule holds. We first show that $(\CoRecMr\ X\ g)$ and $\cls$ are $\S$-morphisms.

\begin{lemma} \label{corec-morphm}
  Let $(X: \U)$, and $(g: X \to \sum (a: A). B(a) \to X)$. We have that $\bigl((\CoRecMr\ X\ g) : X \to \MABr\bigr)$ forms an $\M$-morphism.

  \begin{align*}
    \Bigl( \MorphM\ X\ g\quad \MABr\ \ElimMr\quad (\CoRecMr\ X\ g) \Bigr)
  \end{align*}

  \begin{proof}
    Let $(x: X)$. We need to show the following equations:

    \begin{align}
      \pr1\ \bigl(\ElimMr\ (\CoRecMr\ X\ g\ x)\bigr) & = \pr1\ (g\ x)                          \\
      \pr2\ \bigl(\ElimMr\ (\CoRecMr\ X\ g\ x)\bigr) & = (\CoRecMr\ X\ g) \circ (\pr2\ (g\ x))
    \end{align}

    Both follow immediately from the $\beta$-rules of \cref{beta-mr}.
  \end{proof}
\end{lemma}

\begin{lemma} \label{cls-morphm}
  The class function $(\cls : \MABr \to \MAB)$ forms an $\M$-morphism.

  \begin{align*}
    \Bigl( \MorphM\ \MABr\ \ElimMr \quad \MAB\ \ElimM\quad \cls \Bigr)
  \end{align*}

  \begin{proof}
    Let $(m: \MABr)$. We need to show the following equations:

    \begin{align}
      \pr1\ (\ElimM\ (\cls\ m)) & = \pr1\ (\ElimMr\ m)      \label{cls-morph-e1}          \\
      \pr2\ (\ElimM\ (\cls\ m)) & = \cls \circ (\pr2\ (\ElimMr\ m))  \label{cls-morph-e2}
    \end{align}

    These equations follow directly from the definition of $\ElimM$ and the $\beta$-rule of quotients from \cref{beta-quot}. We first show \cref{cls-morph-e1}.

    \begin{align*}
      \pr1\ (\ElimM\ (\cls\ m)) & := \pr1\ (\overline{\tup{\pr1, \cls \circ \pr2} \circ \ElimMr}\ (\cls\ m)) \\
      \overset{\ref{beta-quot}} & {=} \pr1\ ((\tup{\pr1, \cls \circ \pr2} \circ \ElimMr)\ m)                 \\
                                & \betar \pr1\ (\ElimMr\ m)
    \end{align*}

    To conclude, we show that \cref{cls-morph-e2} holds.

    \begin{align*}
      \pr1\ (\ElimM\ (\cls\ m)) & :=  \pr2\ (\overline{\tup{\pr1, \cls \circ \pr2} \circ \ElimMr}\ (\cls\ m)) \\
      \overset{\ref{beta-quot}} & {=} \pr2\ ((\tup{\pr1, \cls \circ \pr2} \circ \ElimMr)\ m)                  \\
                                & \betar \cls \circ (\pr2\ (\ElimMr\ m)) \qedhere
    \end{align*}
  \end{proof}
\end{lemma}

\begin{lemma} \label{cls-rel}
  Let $(m: \MABr)$ be an M-type. Then we have that the following are related by $\CoLimM$:

  \begin{align*}
    \bigl(\CoRecMr\ \MABr\ \ElimMr\ (\cls\ m)\bigr) \equiv \bigl( \CoRecMr\ \MAB\ \ElimM\ m \bigr)
  \end{align*}

  \begin{proof}
    By \cref{cls-morphm} we know that $\cls$ forms an $\M$-morphism from which it follows that $\CoLimM$ holds.
  \end{proof}
\end{lemma}

\begin{lemma} \label{id-colimm}
  Let $(m: \MABr)$ be an M-type. We have the following identity:

  \begin{align*}
    m \equiv (\CoRecMr\ \MABr\ \ElimMr\ m)
  \end{align*}

  \begin{proof}
    We destruct $m$ to obtain $(X: \U)$, $(x: X)$ and $(g: X \to \sum (a: A). B(a) \to X)$ such that $m = \pack\ X\ \tup{x, g} =: \CoRecMr\ X\ g\ x$ by using \cref{axiom:existsid}. By \cref{corec-morphm} we know that $(\CoRecMr\ X\ g)$ forms an $\M$-morphism. We thus have that:

    \begin{align*}
      m                             & = \CoRecMr\ X\ g\ x\ \land                        \\
      (\CoRecMr\ \MABr\ \ElimMr\ m) & = \CoRecMr\ \MABr\ \ElimMr\ ((\CoRecMr\ X\ g)\ x)
    \end{align*}

    We thus satisfied the $\CoLimM$ predicate which relates $m$ to $(\CoRecMr\ \MABr\ \ElimMr\ m)$.
  \end{proof}
\end{lemma}

\begin{lemma} \label{id-corecm}
  We have the following identity:

  \begin{align*}
    \CoRecM\ \MAB\ \ElimM = \id{\MAB}
  \end{align*}

  \begin{proof}
    It is enough to show the following equation (c.f. \cref{stream_id2}):

    \begin{align}
      (\CoRecM\ \MAB\ \ElimM) \circ \cls = \cls \label{xxxx}
    \end{align}

    Let $(m: \MABr)$ be an M-type. By function extensionality and the definition of $\CoRecM$ we can reduce \cref{xxxx} to the following:

    \begin{align}
      \cls\ (\CoRecMr\ \MAB\ \ElimM\ (\cls\ m)) = \cls\ m
    \end{align}

    We destruct $m = \CoRecMr\ X\ g\ x$ for some $(X: \U)$, $(g: X \to \sum (a: A). B(a) \to X)$ and $(x: X)$. We conclude the following.

    \begin{align*}
      \cls\ (\CoRecMr\ \MAB\ \ElimM\ (\cls\ m)) \overset{\ref{cls-rel}} & {=} \cls\ (\CoRecMr\ \MABr\ \ElimMr\ m) \\
      \overset{\ref{id-colimm}}                                         & {=} \cls\ m \qedhere
    \end{align*}
  \end{proof}
\end{lemma}

\begin{theorem}
  \label{m-eta}
  For all $(X: \U)$, $(g: X \to \sum (a: A). B(a) \to X)$ and $(f: X \to \MAB)$ we have:

  \begin{align}
    \bigl( \MorphM\ X\ g\quad \MAB\ \ElimM\quad f \bigr) \implies (f = \CoRecM\ X\ g)
  \end{align}

  \begin{proof}
    Let $X, g$, and $f$ be as described. Assume $\bigl( \MorphM\ X\ g\quad \MAB\ \ElimM\quad f \bigr)$ holds. We show the statement by function extensionality by fixing some $(x: X)$. Note that $(\CoRecM\ X\ g\ x)$ and $(\CoRecM\ \MAB\ \ElimM\ (f\ x))$ are related by $\CoLimM$. This yields the following conclusion.

    \begin{align*}
      f\ x \overset{\ref{id-corecm}} & {=} \CoRecM\ \MAB\ \ElimM\ (f\ x) \\
                                     & = \CoRecM\ X\ g\ x \qedhere
    \end{align*}
  \end{proof}
\end{theorem}

\section{Coinduction}

\define{\BiSimM}{BiSimM}{\texttt{BiSimM}}
\define{\IsBisimM}{IsBisimM}{\texttt{IsBisimM}}

In this final section, we shall show that the newly defined M-types have a coinduction principle. This principle states that two M-types are equal if they have the same behavior, i.e. if they are related by a bisimulation relation. We define the bisimulation relation $\BiSimM$ and show that two M-types are equal if and only if they are related by this relation.

\begin{definition}
  We define the bisimulation relation between M-types. Two M-types are bisimilar if there exists a bisimulation relation. Let $(m, n: \MAB)$ be M-types.

  \begin{align*}
    \BiSimM\ m\ n := \sum (R: \MAB \to \MAB \to \U).     & \IsBisimM\ R\ \land R\ m\ n                           \\
    \IsBisimM\ R := \prod (m, n: \MAB). R\ m\ n \implies & \pr1\ (\ElimM\ m) = \pr1\ (\ElimM\ n) \land           \\
                                                         & R\ ((\pr2\ (\ElimM\ m))\ b)\ ((\pr2\ (\ElimM\ n))\ b)
  \end{align*}
\end{definition}

\begin{notation}
  We denote the relation $\BiSimM$ using the infix symbol $\sim$, so instead of writing $(\BiSimM\ m\ n)$ we write $(m \sim n)$.
\end{notation}

\begin{lemma} \label{equality-bisimm}
  The propositional equality relation on M-types is a bisimulation. So for all $(m, n: \MAB)$ we have $(m = n) \implies (m \sim n)$. In other words, we have that the $\IsBisimM$ predicate holds for $=$.

  \begin{align}
    \IsBisimM\ (\lambda m. \lambda n. m = n)
  \end{align}

  \begin{proof}
    Define $R := (\lambda m. \lambda n. m = n)$. Let $(m, n: \MAB)$ and suppose $(R\ m\ n)$. From $m = n$, if follows that $(\ElimM\ m) = (\ElimM\ n)$. The statement follows trivially.
  \end{proof}
\end{lemma}

\define{\CoIndM}{CoIndM}{\texttt{CoIndM}}
\define{\ElimX}{ElimX}{\texttt{elimX}}
\begin{theorem} \label{m-ind}
  We have the following coinduction proof principle $\CoIndM$. Let $(m, n: \MAB)$.

  \begin{align*}
    \CoIndM\ m\ n := m \sim n \iff m = n
  \end{align*}

  \begin{proof}
    $(\impliedby)$ This follows from \cref{equality-bisimm}.\\
    $(\implies)$ Let $(m, n: \MAB)$ be M-types and assume $(m \sim n)$.
    We define the following quotient:

    \begin{align*}
      \MAB / \sim := \quot\ \MAB\ \BiSimM
    \end{align*}

    By $(m \sim n)$ there exists an bisimulation $(R : \MAB \to \MAB \to \U)$. Let $R$ be that bisimulation.
    We lift the M-destructor and define $\ElimX := \overline{\tup{\pr1, \cls_\sim \circ \pr2} \circ \ElimM}$. This construction is well-defined by the definition of $\IsBisimM$ and a proof is almost identitcal to the (lengthy though not complicated) proofs of \cref{eqmlem1} and \cref{eqmlem2}.

    We show that the following diagram commutes:

    \begin{figure}[H]
      \centering
      \begin{tikzcd}
        \MAB \\
        \\
        & {\MAB / \sim} \\
        \\
        \MAB
        \arrow["{\cls_{\sim}}", from=1-1, to=3-2]
        \arrow["{\id{\MAB}}"', from=1-1, to=5-1]
        \arrow["{=}"{description}, bend left, draw=none, from=1-1, to=5-1]
        \arrow["{\CoRecM\ (\MAB / \sim)\ \ElimM}", from=3-2, to=5-1]
      \end{tikzcd}
    \end{figure}

    We define $f := (\CoRecM\ (\MAB / \sim)\ \ElimX) \circ \cls_{\sim}$.
    We have that $(f : \MAB \to \MAB)$ forms an $\M$-morphism. To show this, we prove the following equations for $(x: X)$.

    \begin{align}
      \pr1\ (\ElimM\ (f\ x)) = \pr1\ (\ElimM\ x) \quad \label{m-ind-e1} \\
      \pr2\ (\ElimM\ (f\ x)) = f \circ (\pr2\ (\ElimM\ x)) \label{m-ind-e2}
    \end{align}

    We first show \cref{m-ind-e1} using the $\beta$-rule for quotients and the $\beta$-rule for M-types.

    \begin{align*}
      \pr1\ (\ElimM\ (f\ x))    & := \pr1\ \Bigl(\ElimM\ \bigl(\CoRecM\ (\MAB / \sim)\ \ElimX\ (\cls_{\sim}\ x)\bigr)\Bigr) \\
      \overset{\ref{beta-m}}    & {=} \pr1\ (\ElimX\ (\cls_{\sim}\ x))\                                                     \\
      \overset{\ref{beta-quot}} & {=} \pr1\ \bigl((\tup{\pr1, \cls_\sim \circ \pr2} \circ \ElimM)\ x\bigr)                  \\
                                & = \pr1\ (\ElimM\ x)
    \end{align*}

    Next, we show \cref{m-ind-e2}. Again using the $\beta$-rule for quotients and the $\beta$-rule for M-types.

    \begin{align*}
      \pr2\ (\ElimM\ (f\ x))    & := \pr2\ \Bigl(\ElimM\ \bigl(\CoRecM\ (\MAB / \sim)\ \ElimX\ (\cls_{\sim}\ x)\bigr)\Bigr)                  \\
      \overset{\ref{beta-m}}    & {=} \CoRecM\ (\MAB / \sim)\ \ElimX\ \bigl(\pr2\ (\ElimX\ (\cls_\sim\ x))\bigr)                             \\
      \overset{\ref{beta-quot}} & {=} \CoRecM\ (\MAB / \sim)\ \ElimX\ \bigl(\pr2\ ((\tup{\pr1, \cls_\sim \circ \pr2} \circ \ElimM)\ x)\bigr) \\
                                & = \bigl((\CoRecM\ (\MAB / \sim)\ \ElimX) \circ \cls_{\sim}\bigr)\ (\cls_{\sim}\ (\pr2\ (\ElimM\ x)))       \\
                                & =: f \circ (\pr2\ (\ElimM\ x))
    \end{align*}

    From the $\eta$-rule of M-types (\cref{m-eta}), we have that:

    \begin{align}
      f := (\CoRecM\ \MAB\ \ElimX) \circ \cls_\sim = (\CoRecM\ \MAB\ \ElimM) \label{here}
    \end{align}

    Because $(m \sim n)$ we also have that:

    \begin{align}
      \cls_\sim\ m = \cls_\sim n \label{here2}
    \end{align}

    We derive the following:

    \begin{align*}
      m \overset{\ref{id-corecm}} & {=} \CoRecM\ \MAB\ \ElimM\ m                       \\
      \overset{\ref{here}}        & {=} \CoRecM\ (\MAB / \sim)\ \ElimX\ (\cls_\sim\ m) \\
      \overset{\ref{here2}}       & {=} \CoRecM\ (\MAB / \sim)\ \ElimX\ (\cls_\sim\ n) \\
      \overset{\ref{here}}        & {=} \CoRecM\ \MAB\ \ElimM\ n                       \\
      \overset{\ref{id-corecm}}   & {=} n
    \end{align*}

    We conclude that $m = n$.
  \end{proof}
\end{theorem}

\chapter{Conclusion}

In this thesis, we extended the technique of \cite{encodings} of encoding $\eta$-rules within the definition of System F style data types and applied this to lists, quotients, streams, W-types, and M-types. We showed how to define System F style impredicative data types and that each definition satisfied the expected formation, introduction, elimination, and computation rules.

For the inductive lists and W-types, we looked at the categorical definition using the initial algebra of the functors $\L(X) = 1 + E \times X$ and $\W(X) = \sum(a: A). B(a) \to X$ respectively. We distilled the uniqueness requirement of these initial objects and encoded them as a subtype using sigma types. We showed that these improved types satisfied the $\eta$-rules and proved the induction principles.

We also showed how to dualize the technique. We first created an encoding of the inductive quotient type and showed that it satisfied the $\eta$-rule and the induction principle. We then created a stream type and M-type using existential types by looking at the final coalgebra of the functors $\S(X) = E \times X$ and $\mathcal{M}(X) = \sum(a: A). B(a) \to X$ respectively.  We distilled the uniqueness requirement of these final objects and encoded them using our quotients. We showed that these improved types satisfied the $\eta$-rules and proved the bisimulation principles.

Finally, we looked at the technique in general, closing the gap between the concrete implementations in this thesis and the abstract generic definitions in the paper \cite{encodings}.

\chapter{Future work}

In this final chapter, we list some possible future topics one can look at.

\begin{itemize}
  \item First, the entirety of this thesis can be formalized in a proof assistant such as Coq or Lean. Some attempts have been made to formalize parts of the paper \cite{encodings} by Awodey \url{https://github.com/awodey/Impredicative}.
  \item We have not looked at higher inductive types. In this case, one should drop the \UIP\ axiom and use a notion of \texttt{Set}. The same goes for higher coinductive types.
  \item An interesting avenue is to look at the differences between the inductive definitions in Coq and the impredicative encodings. Perhaps inductive types in Coq can be translated to these impredicative encodings.
  \item We used the equality type from Homotopy Type Theory. Perhaps the Leibnitz equality `$a = b$' $ := \prod (P: A \to \U). P\ a \iff P\ b$ will also suffice. This would make the system smaller.
  \item It might be interesting to look at the exact difference or similarity between \texttt{Lim}, \texttt{Ind} and \texttt{Unq} mentioned in \cref{chap:generalization}.
  \item We have not looked at the exact circumstances under which the initial and final objects exist in this system. This has something to do with the ``positivity'' of the functor $F$ and the existence of limits/colimits.
\end{itemize}

\appendix
\counterwithout{theorems}{section}
\counterwithin{theorems}{chapter}

\chapter{System F typing rules}
\label{systemf}

We define the syntax of System F using a BNF grammar. Let `$X$' be a set of type variables and `$x$' a set of term variables.

\begin{align*}
  A, B & ::= X \mid A \to B \mid \prod (X: \U). A \tag{types}                                 \\
  t, u & ::= x \mid \lambda (x: T). t \mid t\ u \mid \lambda (X: \U). t \mid t\ A \tag{terms}
\end{align*}

We write $A[X]$ to denote a type with a free type variable $X$ and $t[x]$ to denote a term with a free term variable $x$. We write $A[X := B]$ for the substitution of $X$ with the type $B$. Similarly, we write $t[x := u]$ for the substitution of $x$ with the term $u$.

Next, we define the typing rules of System F.

\begin{gather*}
  \infer[\text{id}]{\Gamma,\, x: A \vdash x : A}{} \\
  \begin{aligned}
    \infer[\to\text{-intro}]{\Gamma \vdash \lambda (x : A). t : A \to B}{\Gamma,\, x: A \vdash t : B}              & \hspace{5em}
    \infer[\to\text{-elim}]{\Gamma \vdash t\ u : B}{\Gamma \vdash t : A \to B                                      & \Gamma \vdash u : A} \\
    \infer[\Pi\text{-intro}^*]{\Gamma \vdash \lambda (X : \U). t : \prod (X: \U). A}{\Gamma,\, X: \U \vdash t : A} & \hspace{5em}
    \infer[\Pi\text{-elim}]{\Gamma \vdash t\ A : B[X := A]}{\Gamma \vdash t : \prod (X: \U). A}
  \end{aligned}\\
  ^* \text{where } X \text{ is not free in } \Gamma
\end{gather*}

Finally, we define $\beta$-reduction for System F.

\begin{align*}
  (\lambda (x. A). t[x]) \ u \betar t[x := u] \\
  (\lambda (X: \U). t[X]) \ A \betar t[X := A]
\end{align*}

\chapter{Data types in System F}
\label{datatypes:systemf}

\begin{definition} We define a System F \textbf{unit} type.

  \begin{align*}
    \textbf{1}^*                & := \prod (X: \U). X \to X                 \tag{formation}                                \\
    \texttt{tt}^*               & := \lambda (X: \U). \lambda (x: X). x     \tag{introduction}                             \\
    \texttt{rec}_{\textbf{1}}^* & := \lambda (X: \U). \lambda (x: X). \lambda (u: \textbf{1}^*). u\ X\ x \tag{elimination}
  \end{align*}

\end{definition}

\begin{definition} We define a System F \textbf{empty} type.

  \begin{align*}
    \textbf{0}^*                & := \prod (X: \U). X                               \tag{formation}   \\
    \texttt{rec}_{\textbf{0}}^* & := \lambda (X: \U). \lambda (z: \textbf{0}). z\ X \tag{elimination}
  \end{align*}

\end{definition}

\begin{definition} We define a System F \textbf{coproduct} type.

  \begin{align*}
    A +^* B                  & := \prod (X: \U). (A \to X) \to (B \to X) \to X \tag{formation}                                                    \\
    \texttt{inl}^*           & := \lambda (a: A). \lambda (X: \U). \lambda (f: A \to X). \lambda (g: B \to X). f\ a \tag{introduction}            \\
    \texttt{inr}^*           & := \lambda (b: B). \lambda (X: \U). \lambda (f: A \to X). \lambda (g: B \to X). g\ b \tag{introduction}            \\
    \texttt{rec}_{A +^* B}^* & := \lambda (X: \U). \lambda (f: A \to X). \lambda (g: B \to X). \lambda (t: A +^* B). t\ X\ f\ g \tag{elimination}
  \end{align*}

  We denote $[f, g] := \texttt{rec}_{A +^* B}^*\ X\ f\ g$ when $X$, $A$ and $B$ are clear from the context.

\end{definition}

\begin{definition} We define a System F \textbf{product} type.

  \begin{align*}
    A \times^* B    & := \prod (X: \U). (A \to B \to X) \to X \tag{formation}                                                    \\
    \texttt{pair}^* & := \lambda (a: A). \lambda (b: B). \lambda (X: \U). \lambda (f: A \to B \to X). f\ a\ b \tag{introduction} \\
    \texttt{fst}^*  & := \lambda (p: A \times^* B). p\ A\ (\lambda (a: A). \lambda (b: B). a) \tag{elimination}                  \\
    \texttt{snd}^*  & := \lambda (p: A \times^* B). p\ B\ (\lambda (a: A). \lambda (b: B). b) \tag{elimination}
  \end{align*}
\end{definition}

\chapter{Inference rules of the system}
\label{chap:complete_system}

These rules are partly taken from Appendix A of the master's thesis of Sam Speight \cite{speight} and are equal to those of HoTT \cite{Hott}.

\section*{Inference rules for product types}
\vspace{1em}
\begin{gather*}
  \begin{aligned}
    \infer[\Pi\text{-form}_1]{\Gamma \vdash \prod (a: A). B : \U}{%
      \Gamma \vdash A : \U_i
     &
      \Gamma, a: A \vdash B : \U
    }
    \qquad
     &
    \infer[\Pi\text{-form}_2]{\Gamma \vdash \prod (a: A). B : \U_i}{%
      \Gamma \vdash A : \U_i
     &
      \Gamma, a: A \vdash B : \U_i
    }
    \\\\
    \infer[\Pi\text{-intro}]{\Gamma \vdash \lambda (a: A). b : \prod (a: A). B}{%
      \Gamma, a: A \vdash b : B
    }
    \qquad
     &
    \infer[\Pi\text{-elim}]{\Gamma \vdash f\ a : B[x:=a]}{%
      \Gamma \vdash f : \prod (x: A). B
     &
      \Gamma \vdash a: A
    }
    \\\\
    \infer[\Pi\text{-}\beta]{\Gamma \vdash (\lambda x. b)\ a \betar b[x:=a] : B[x := a] }{%
      \Gamma, x: A \vdash b: B
     &
      \Gamma \vdash a: A
    }
    \qquad
     &
    \infer[\Pi\text{-}\eta]{\Gamma \vdash f = (\lambda x. f\ x) : \prod (x: A). B }{%
      \Gamma \vdash f : \prod (x: A). B
    }
  \end{aligned}
\end{gather*}

\section*{Inference rules for sigma types}
\vspace{1em}

\begin{gather*}
  \begin{aligned}
    \infer[\Sigma\text{-form}_1]{\Gamma \vdash \sum (a:A). B : \U}{%
      \Gamma \vdash A : \U
     &
      \Gamma, a: A \vdash B : \U
    }
    \ \ \
     &
    \infer[\Sigma\text{-form}_1]{\Gamma \vdash \sum (a:A). B : \U_i}{%
      \Gamma \vdash A : \U_i
     &
      \Gamma, a: A \vdash B : \U_i
    }
  \end{aligned}
  \\\\
  \infer[\Sigma\text{-intro}]{\Gamma \vdash \tup{a, b} : \sum (x: A). B}{%
    \Gamma, x: A \vdash B : \U_i
    &
    \Gamma \vdash a : A
    &
    \Gamma \vdash b : B[x := a]
  }
  \\\\
  \begin{aligned}
    \infer[\Sigma\text{-elim}]{\Gamma \vdash \pr1\ p : A}{%
      \Gamma \vdash p : \sum (x: A). B
    }
    \qquad
     &
    \infer[\Sigma\text{-elim}]{\Gamma \vdash \pr2\ p : B[x := \pr1\ p]}{%
      \Gamma \vdash p : \sum (x: A). B
    }
    \\\\
    \infer[\Sigma\text{-}\beta]{\Gamma \vdash \pr1\ \tup{a, b} \betar a : A}{%
      \Gamma \vdash a: A
     &
      \Gamma, a: A \vdash b: B
    }
    \qquad
     &
    \infer[\Sigma\text{-}\beta]{\Gamma \vdash \pr2\ \tup{a, b} \betar b : B[x:=a]}{%
      \Gamma \vdash a: A
     &
      \Gamma, x: A \vdash b: B
    }
  \end{aligned}
  \\\\
  \infer[\Sigma\text{-}\eta]{\Gamma \vdash p = \tup{\pr1\ p,\ \pr2\ p} : \sum (x: A). B}{%
    \Gamma \vdash p : \sum (x: A). B
  }
\end{gather*}

\newpage
\newtheorem{innercustomthm}{Lemma}
\newenvironment{customthm}[1]
{\renewcommand\theinnercustomthm{#1}\innercustomthm}
{\endinnercustomthm}

\begin{customthm}{3.2.3} We have the following sigma injection principle.

  \begin{align*}
    (\hookrightarrow) : \prod (X:\ \U). \prod (P: X \to \U). \prod \bigl(y, y': \sum (x: X). P\ x\bigr). y = y' \iff \pr1\ y = \pr1\ y'
  \end{align*}

  \begin{proof}
    Assume $(X:\ \U)$, $(P: X \to \U)$ and $\bigl(y, y': \sum (x: X). P\ x\bigr)$. \\
    $(\implies)$ Assume we know that $y = y'$. By the $\Sigma\text{-}\eta$ rule, we know $y = \tup{\pr1\ y,\ \pr2\ y}$ and $y' = \tup{\pr1\ y',\ \pr2\ y'}$. Because $y = y'$, we know that $\tup{\pr1\ y,\ \pr2\ y} = \tup{\pr1\ y',\ \pr2\ y'}$. By taking the first projection of both sides, we see that $\pr1\ y = \pr1\ y'$.\\
    $(\impliedby)$ Assume we know that $\pr1\ y = \pr1\ y'$. Because $y$ and $y'$ are well-formed, we know that $(\pr2\ y : P\ (\pr1\ y))$ and $(\pr2\ y' : P\ (\pr1\ y'))$. Since $\pr1\ y = \pr1\ y'$ we obtain by $\UIP$ that $P\ (\pr1\ y) = P\ (\pr1\ y')$. We thus we have that $\pr1\ y = \pr1\ y'$ and that $\pr2\ y = \pr2\ y'$ and conclude that $y = y'$.
  \end{proof}
\end{customthm}

\section*{Inference rules for equality types}
\vspace{1em}

\begin{gather*}
  \begin{aligned}
    \infer[\text{=-form}_1]{\Gamma \vdash (a =_A b) : \U}{%
      \Gamma \vdash A : \U
     &
      \Gamma \vdash a : A
     &
      \Gamma \vdash b : A
    }
     &
    \ \ \
    \infer[\text{=-form}_1]{\Gamma \vdash (a =_A b) : \U_i}{%
      \Gamma \vdash A : \U_i
     &
      \Gamma \vdash a : A
     &
      \Gamma \vdash b : A
    }
  \end{aligned}
  \\\\
  \infer[\text{=-intro}]{\Gamma \vdash \texttt{refl} : a =_A a}{%
    \Gamma \vdash A : \U_i
    &
    \Gamma \vdash a : A
  }
  \\\\
  \infer[\text{=-elim}]{\Gamma \vdash \texttt{ind}_= (c,a,b,q) : C[a, b, x:=q, y, p]}{%
    \begin{gathered}
      \Gamma \vdash a : A \qquad
      \Gamma \vdash b : B \qquad
      \Gamma \vdash p : a =_A b
      \\
      \Gamma, x: A, y: A, p: a =_A b \vdash C : \U_i \qquad
      \Gamma, z: A \vdash c : C[z, z, x := \texttt{refl}_z, y, p]
    \end{gathered}
  }
  \\\\
  \infer[\text{=-}\beta]{\Gamma \vdash \texttt{ind}_= (c,a,a,\texttt{refl}_a) \betar c[z := a]: C[a, a, x := \texttt{refl}_a, y, p]}{%
    \Gamma \vdash a : A \qquad
    \Gamma, x: A, y: A, p: a =_A b \vdash C : \U_i \qquad
    \Gamma, z: A \vdash c : C[z, z, x := \texttt{refl}_z, y, p]
  }
\end{gather*}

\section*{Axioms}
\vspace{1em}

\begin{align*}
  \funext \quad   & : \quad \Bigl( \prod(x: X). f\ x = g\ x \Bigr) \implies f = g    \\\\
  \UIP \quad      & : \quad \prod (X: \U) \prod (x, y: X) \prod (p, q: x = y). p = q \\\\
  \ExistsId \quad & : \quad \RecExists\ (\exists X. P)\ \pack = \id{\exists X. P}
\end{align*}

\chapter{CoLimStr is no equivalence relation}
\label{CoLimStr_not_eq}

In this appendix, we elaborate on why the relation $\CoLimStr$ of \cref{def:colimstr} is not an equivalence relation. In fact, it is neither \textit{symmetric} nor \textit{transitive}. The relation is reflexive.

\subsection*{Reflexivity}

Suppose $(\sigma : \Streamr)$. Does $(\CoLimStr\ \sigma\ \sigma)$ hold? We use the recursor $\RecExists$, we obtain some $(X: \U)$, $(h: X \to E)$, $(t: X \to X)$ and $(x: X)$. Together with $\ExistsId$, we know that $\sigma = \CoRecStream\ X\ h\ t\ x$. It is easy to see that the identity function forms a morphism. Therefore we have $\CoLimStr\ \sigma\ \sigma$. Note that without $\ExistsId$, we are unable to prove that $\sigma = \CoRecStream\ X\ h\ t\ x$.

\subsection*{Symmetry}

Suppose $(\sigma, \tau : \Streamr)$ and $(\CoLimStr\ \sigma\ \tau)$. Does $(\CoLimStr\ \tau\ \sigma)$ hold? By $(\CoLimStr\ \sigma\ \tau)$ we have $(X: \U)$, $(h: X \to E)$, $(t: X \to X), (x: X)$ and $(Y: \U)$, $(h': Y \to E)$, $(t': Y \to Y)$ and $(f: X \to Y)$ such that $\sigma = \CoRecStream\ X\ h\ t\ x$, $\tau = \CoRecStream\ Y\ h'\ t'\ (f\ x)$ and $(\MorphStream\ X\ h\ t\quad Y\ h'\ t\quad f)$ hold. In this type system, we cannot simply inverse $f$ to obtain a function $(g: Y \to X)$.
\subsection*{Transitivity}

Suppose $(\sigma, \tau, \upsilon : \Streamr)$, $(\CoLimStr\ \sigma\ \tau)$ and $(\CoLimStr\ \tau\ \upsilon)$. From $(\CoLimStr\ \sigma\ \tau)$ we obtain the left column, from $(\CoLimStr\ \tau\ \upsilon)$, we obtain the right column:

{\setstretch{-2.5}
\begin{align*}
  X: \U \qquad                                       & \qquad \widetilde{Y}: \U                                                                   \\
  h: X \to E \qquad                                  & \qquad h: \widetilde{Y} \to E                                                              \\
  t: X \to X \qquad                                  & \qquad t: \widetilde{Y} \to \widetilde{Y}                                                  \\
  x: X \qquad                                        & \qquad y: \widetilde{Y}                                                                    \\
  Y: \U \qquad                                       & \qquad Z: \U                                                                               \\
  h': Y \to E \qquad                                 & \qquad h'': Z \to E                                                                        \\
  t': Y \to Y \qquad                                 & \qquad t'': Z \to Z                                                                        \\
  f: X \to Y \qquad                                  & \qquad g: \widetilde{Y} \to Z                                                              \\
  \MorphStream\ X\ h\ t\quad Y\ h'\ t'\quad f \qquad & \qquad \MorphStream\ \widetilde{Y}\ \widetilde{h'}\ \widetilde{t'}\quad Z\ h''\ t''\quad g \\
  \sigma = \CoRecStream\ X\ h\ t\ x \qquad           & \qquad \sigma = \CoRecStream\ \widetilde{Y}\ \widetilde{h'}\ \widetilde{t'}\ \widetilde{y} \\
  \tau = \CoRecStream\ Y\ h'\ t'\ (f\ x) \qquad      & \qquad \tau = \CoRecStream\ Z\ h''\ t''\ (g\ y)                                            \\
\end{align*}
}

We are unable to prove that $Y = \widetilde{Y}$, since the exact implementation of $\tau$ is hidden within the existential type. Neither can we obtain a proof of $Y = \widetilde{Y}$ because this proof lives in type universe $\U_1$ since $(Y : \U)$.

\nocite{*}

\chapter{References}
\printbibliography[heading=none]
\printindex

\end{document}